\documentclass[11pt]{article}
\usepackage{eurosym}
\usepackage{amssymb}
\usepackage{graphicx}
\usepackage{amsmath}
\usepackage{makeidx}
\usepackage{indentfirst}
\usepackage[T1]{fontenc}
\usepackage[utf8]{inputenc}
\usepackage{authblk}

\newcounter{resultnum}[section]
\newcounter{conclusionnum}[section]
\newcounter{conditionnum}[section]
\newcounter{conjecturenum}[section]
\newcounter{examplenum}[section]
\newcounter{exercisenum}[section]
\newcounter{lemmanum}[section]
\newcounter{notationnum}[section]
\newcounter{theoremnum}[section]
\newcounter{definitionnum}[section]
\newcounter{corollarynum}[section]
\newcounter{remarknum}[section]
\newcounter{propositionnum}[section]
\newcounter{acknowledgementnum}[section]
\newcounter{algorithmnum}[section]
\newcounter{axiomnum}[section]
\newcounter{casenum}[section]
\newcounter{claimnum}[section]
\newcounter{summarynum}[section]
\newcounter{problemnum}[section]
\begin{document}

\title{On Supersymmetric Geometric Flows and \\
$\mathcal{R}^2$ Inflation From Scale Invariant Supergravity}
\date{June 28, 2017}

\author{ %${}$ \\
\vspace{.1 in}{ \textbf{Subhash Rajpoot}}\\
{\small \textit{California State University at Long Beach, }}%
{\small \textit{Long Beach, California, USA }}\\
{\small \textit{email: Subhash.Rajpoot@csulb.edu ${}$ }}\\
\vspace{.1 in} \textbf{Sergiu I. Vacaru}\footnote{{\it Address for post correspondence:\ } 
67 Lloyds Street South, Manchester, M14 7LF, the UK; \newline the USA affiliation is for a visiting position }\\
{\small \textit{Physics Department, California State University Fresno, Fresno, CA 93740, USA}} \\
{\small and } \ {\small \textit{ Project IDEI, University "Al. I. Cuza" Ia\c si, Romania}}\\
{\small \textit{emails: sergiu.vacaru@gmail.com  and dougs@csufrenso.edu}}}

\maketitle

\begin{abstract}
Models of geometric flows pertaining to $\mathcal{R}^2$ scale invariant
(super) gravity theories coupled to conformally invariant matter fields are
investigated. Related to this work are supersymmetric scalar manifolds that
are isomorphic to the K\"{a}hlerian spaces $\mathcal{M}_n=SU(1,1+k)/U(1)%
\times SU(1+k)$ as generalizations of the non--supersymmetric analogs with $%
SO(1,1+k)/SO(1+k)$ manifolds. For curved superspaces with geometric
evolution of physical objects, a complete supersymmetric theory has to be
elaborated on nonholonomic (super) manifolds and bundles determined by
non--integrable superdistributions with additional constraints on (super)
field dynamics and geometric evolution equations. We also consider
generalizations of Perelman's functionals using such nonholonomic variables
which result in the decoupling of geometric flow equations and Ricci soliton
equations with supergravity modifications of the $R^2$ gravity theory. As
such, it is possible to construct exact non--homogeneous and locally
anisotropic cosmological solutions for various types of (super) gravity
theories modelled as modified Ricci soliton configurations. Such solutions
are defined by employing the general ansatz encompassing coefficients of
generic off--diagonal metrics and generalized connections that depend
generically on all spacetime coordinates. We consider nonholonomic
constraints resulting in diagonal homogeneous configurations encoding
contributions from possible nonlinear parametric geometric evolution
scenarios, off--diagonal interactions and anisotropic polarization/
modification of physical constants. In particular, we analyze small
parametric deformations when the underlying scale symmetry is preserved and
the nontrivial anisotropic vacuum corresponds to generalized de Sitter
spaces. Such configurations may mimic quantum effects whenever transitions
to flat space are possible. Our approach allows us to generate solutions
with scale violating terms induced by geometric flows, off--diagonal
interactions and supersymmetric modifications of effective potentials. We
study reconstructing the formalism for (super) geometric flows and modified
gravity cosmological scenarios. We also analyze the conditions under which
such modified mimetic type theories and solutions reproduce the Starobinsky
inflationary models in the double scalar approach.

\end{abstract}

\tableofcontents

\section{Introduction}

Physical models with geometric evolution are well known in quantum field
theory (QFT) since the late 70's, due to research on gauge and nonlinear
sigma theories inspired by A. M. Polyakov and D. Friedan \cite%
{polyakov,friedan1,friedan2,friedan3}. Certain new directions in modern
mathematics, with important achievements in geometric analysis and topology,
were elaborated upon, beginning with the works of R. Hamilton \cite%
{hamilt1,hamilt2,hamilt3}. It induced a great social interest in the field
after the famous proofs of Thurston's geometrization program and the Poincar%
\'{e} conjecture by G. Perelman \cite{perelman1,perelman2,perelman3} (for
reviews of mathematical results see \cite{monogrrf1,monogrrf2,monogrrf3}).
Such significant ideas and results intrigued many physicists to pursue
research in the field, with the hope that the new geometric methods would
offer surprising and far reaching perspectives in mathematical particle
physics, QFT, string theory, (modified) gravity, geometric mechanics and
thermodynamics etc. Here we call attention to certain important applications
of the Ricci flow theory in the study of nonlinear sigma models \cite%
{nitta,tsey,oliy,carfora}, research on geometric flow evolution of modified
(non) holonomic commutative and noncommutative gravity theories \cite%
{bakaslust,vnhrf,vnrflnc,streets} and exact solutions for (modified) gravity and geometric flows,
Ricci solitons  \cite{vlf,vvrfthbh,muen01,carfora1,carstea,vrfijmpa,vacvis}.

In spite of the many constructions elaborated upon by physicists that have
not yet passed the censorship of mathematicians (who request proofs of
theorems on hundred pages using sophisticate geometric analysis methods),
the field offers new physical ideas and results, constituting a source of
inspiration for new interesting developments both in mathematics and
physics. In this paper, we neither concentrate on QFT issues nor consider
how certain geometric models of the space of probability measures over
Riemannian metric measure spaces provide a natural connection between
nonlinear sigma models and Ricci flow theory \cite{carfora}. Our goals and
methods are very different from the ones employed in "pure" mathematics
elaborated in geometric analysis with relations to quantum theories.
Instead, we follow a series of our recent results relating to exact
solutions for nonlinear off--diagonal systems and gravity. Such geometric
constructions and methods manifest that via geometric flows with Ricci
soliton configurations we can model modified gravity theories, MGTs, and
general relativity, GR, using new classes of physically important solutions
in modern gravity and cosmology:

\begin{enumerate}
\item Nonholonomic constraints on geometric flows may result in the
evolution of a (pseudo) Riemannian geometry, for instance, into a
generalized Lagrange-Finsler type, or metric--affine geometry, almost K\"{a}%
hler configurations, noncommutative geometries, Ho\v{r}ava--Lifshits gravity
theories, $f(R)$ and $R^2$ modified gravity theories etc., see \cite%
{vnhrf,vnrflnc,bakaslust,vvrfthbh,muen01}. Complex manifolds and (in various
different approaches) supermanifolds are defined by certain respective
classes of nonholonomic distributions. It is expected that the method of
nonholonomic deformations of fundamental functionals, actions and derived
field equations and physical/geometric objects in supergravity and
superstring theories will result in the elaborating of various models of
supergeometric flows and (super) Ricci soliton spaces.

\item For self--similar fixed configurations, the (non) holonomic geometric
flow equations transform into modified Ricci soliton equations which (for
well defined conditions) are equivalent to gravitational field equations in
MGTs, or GR. In a certain sense, the bulk of MGTs can be realized as some
generalized Ricci soliton geometries arising in corresponding fixed points
of parametric geometric flow evolution of certain classes of fundamental
geometric/physical geometric objects on (modified) spacetime manifolds.

\item Mathematically, it has not been elaborated upon as yet a well--defined
geometric analysis formalism which allows to study rigorously the geometric
evolution of metrics of pseudo--Euclidean signature and to find exact
solutions of Ricci flow equations for generic off--diagonal metrics. There
are a number of conceptual and fundamental issues which have not been
addressed by mathematicians in order to develop methods of nonlinear
functional analysis and the geometry of Lorentz manifolds for (modified)
gravity theories. For physical models of relativistic spacetimes and
gravity, the Ricci tensor in 4-d does not approximate in the limit of weak
gravitational interactions to the Laplace (diffusion) operator but to the
D'Alambert (wave) operator. The standard geometric evolution paradigm
proposed initially in the Hamilton--Perelman program (with nonlinear
evolution equations and associated entropy type functionals, all defined by
Riemannian metrics) must be modified in order to study realistic geometric
flow evolution models related to physical theories. For instance, further
research on relativistic Ricci flows is possible, for instance, following
methods on nonlinear relativistic hydrodynamics generalized in certain forms
which allows to consider nonlinear geometric "heat" transfers, propagating
relativistically entropy and temperature. Such issues go back to the
fundamental works on relativistic definition of temperature and heat in the
framework of relativistic kinetics and thermodynamics in special and general
relativity theories (see details and references in \cite{vvrfthbh}; on
Einstein-Ott-Plank thermodynamics, Einstein-Vlasov kinetic theory,
relativistic diffusion and hydrodynamics etc.).

\item In our research, we have found that it is possible to introduce such
nonholonomic variables when the geometric flow evolution and Ricci soliton
equations can be decoupled in certain general forms. This allows us to
construct various classes of exact solutions with generic off--diagonal
metrics and generalized connections. The coefficients of such objects depend
on all space-time coordinates via generating and integration functions and
various types of commutative and noncommutative parameters and
integration/physical constants. Such geometric techniques of constructing
exact solutions in geometric flow evolution and MGTs have been elaborated
following the so--called anholonomic frame deformation method, AFDM. For
details, examples of exact solutions, and various applications, we cite \cite%
{sv2001,vsingl,svvvey,tgovsv} and references therein. Having generated certain
classes of parametric solutions describing modifications by geometric flows
on certain hypersurfaces, with 3+1 splitting, or with effective temperature
and entropy, for instance, of certain black hole, or cosmological,
solutions, we can analyze in explicit form possible physical consequences.

\item Another very important direction for research is connected to
generalized geometric flow theories via corresponding modifications of
Perelman's functionals. In his original work \cite{perelman1}, G. Perelman
introduced two Lyapunov type functionals called F- entropy and W-entropy,
which were crucial to the proof of the Poincar\'{e} conjecture. It allowed
to elaborate an associated statistical thermodynamics model characterizing
geometric flows. Such functionals can be formally generalized for various
types of commutative and noncommutative geometric models also encoding
interactions of matter fields \cite{vnhrf,vnrflnc,vvrfthbh,streets,svvvey}.
For relativistic geometric flows, the (modified) F- and W--functionals do
not have a simple interpretation as entropy types and do not result in
well--defined nonlinear diffusion type evolution equations. Nevertheless,
redefining and adapting the geometric constructions to certain nonholonomic
double 2+2 and 3+1 splitting, we can provide a thermodynamic
characterization of MGTs and the corresponding classes of exact solutions.
Such an approach is more general, being elaborated on different fundamental
principles, than that considered for black holes using the
Hawking-Bekenstein entropy (derived for 2--d hypersurface gravity). In
general, Perelman's W-entropy can be computed, for instance, for general
cosmological inhomogeneous and cosmological solutions even in supergravity
theories (as we prove in this work). For such constructions, we can
speculate on how Ricci flow conjugated initial data sets can be related to
new concepts of entropy and produce cosmological averaging data \cite%
{carfora1}.
\end{enumerate}

The goal of this work is to elaborate on certain most important principles
and methods for formulating supergeometric flow evolution models in order to
encode modified (super) gravity theories which are compatible with the
modern accelerating cosmology paradigm, see \cite{kounnas3,kehagias,muen01}
and references therein. It should be noted that the first supersymmetric
geometric flow constructions were considered for two dimensional K\"{a}%
hler--Ricci flows and solitons \cite{gunara,gunara1}. In our approach, we
develop a different approach orientated to supersymmetric geometric flow
generalizations of MGTs and GR. We shall construct exact solutions with
total metrics which in general are not K\"{a}hler, contain pseudo-Riemann
bosonic parts of necessary finite dimension, and describe certain equivalent
almost K\"{a}hler configurations. This can be important for research in
classical and quantum gravity, string theory, deformation quantization,
geometric flows with algebroid structure etc. \cite{vwitten,vjp,vdq1,vmedit}%
. The present work can be considered as a supersymmetric / superflow partner
of our recent works \cite{muen01,vvrfthbh,tgovsv,bubv} where the direction is
developed for supergeometric flow models and exact solutions in superstring
MGTs, in particular, related to $R^{2}$ theories and cosmology. We construct
new classes of generalized inhomogeneous cosmological and geometric flow
solutions for modified Ricci soliton equations (such solutions can be
extended to higher dimensions as in \cite{vtamsuper,vrajpootjet} using
noncommutative and/or supersymmetric variables).

The last twenty years has witnessed the changing of GR cosmological paradigm
to other ones related to MGTs and performed in order to include
observational data (beginning late 90's) on late--time acceleration of our
Universe \cite{acelun}. For reviews of results and some insightful studies,
we cite \cite{2a,capozzello,2b,3}, when it is important to consult the works
on early--time inflation \cite{4a,4b,4c}, modified $F(R)$ gravity \cite{5}
with certain attempts to address the early-time and late-time acceleration
within the same theoretical framework \cite{2a,capozzello,2b,6}. The
late--time acceleration is attributed to dark energy modelled, for instance,
as a negative pressure perfect fluid which dominates our Universe's energy
density at the percentage $\Omega _{DE}\sim 68$. The rest of energy is
controlled by cold dark matter, $\Omega _{DE}\sim 27$, and ordinary matter $%
\Omega _{m}\sim 5$. There is no up to date experimentally and/or
observationally confirmed theory which predicts the nature of dark matter
and dark energy. There have been developed a number of MGTs and cosmological
scenarios, all approached by different constructions and methods, but
addressing the same fundamental problems of modern cosmology.

In our recent papers, we elaborated on two main ideas related to MGTs and
cosmology: The cosmological acceleration scenarios may involve 1) certain
geometric spacetime flow evolution of 2) generic off--diagonal modified
Ricci soliton configurations both resulting in modification of standard
model with effective running and/or locally anisotropic polarization of
physical constants. In a certain sense, such constructions can be modelled
by generic off--diagonal solutions of MGTs and GR, which (for applications
in cosmology) can be treated equivalently in the framework of (modified)
mimetic gravity theories \cite{mimet1,mimet2,odints2}. In order to study
possible supergravity contributions to geometric flows and cosmology, the
reconstruction method with $F(R)$ modifications \cite{bamba,much3,odintsrec}
is more convenient. We shall develop such approaches in order to include
geometric flows and off-diagonal configurations.

The article is organized as follows:\ In section \ref{sec2} we briefly
recall the geometry of nonholonomic manifolds enabled with nonlinear
connection, N--connection, 2+2 splitting and introduce nonholonomic
variables with deformation of the linear connection and frame structures. We
define modified Perelman's functionals in nonholonomic variables and show
how geometric relativistic flow equations can be derived. For self-similar
and fixed parameter configurations, such equations define Ricci soliton
spacetimes modelling (for corresponding classes of nonholonomic constraints)
MGTs and generic off-diagonal Einstein spacetimes. Important examples of
nonholonomic geometric evolution of and Ricci solitons corresponding to $%
R^{2}$ gravity with conformally coupled matter are also presented. Also
considered are $SO(1,1+k)$ generalizations of geometric flow and pure $R^{2}$
gravity models .

Section \ref{sec3} is devoted to supersymmetric extensions of Perelman's
functionals and nonholonomic Ricci solitons for $R^{2}$ gravity with $%
SU(1,1+k)$ modifications. This is possible due to using the constructions
elaborated for multi--field and $SO(1,1+k)$ configurations. There is
considered the proof of geometric flow equations in nonhlonomic
supersymmetric variables and analysis of possible configurations of the
vacuum structure of MGTs determined by effective scalar potentials and
associated cosmological constants. We speculate on actions which are
associated with such supersymmetric modifications of Ricci solitons and
related $R^{2}$ gravity models. Such effective modified potentials are
important for constructing exact solutions and elaborating on reconstruction
modified models in the following sections.

In section \ref{sec4}, we develop the AFDM for generating cosmological
solutions of supersymmetric modifications of Ricci flows and $R^2$ gravity,
Following such geometric methods, we prove that nonlinear systems of PDEs
with explicit flow parameter and time coordinate dependencies can be
decoupled in general forms for generic off--diagonal metrics, generalized
connections with nonholonomically induced torsions and (effective) matter
field sources encoding contributions from supergravity models. There are
constructed in general forms the corresponding classes of inhomogeneous and
locally anisotropic Ricci soliton metrics corresponding to gravitational
field equations in $R^2$ and GR theories. We perform a generalization of
such systems for geometric flows with factorized dependence of geometric
evolution parameter. We show also that the AFDM allows to generate exact
cosmological solutions of geometric flow equations encoding nontrivial
vacuum configurations with general flows (non-factorized) on evolution
parameter. We also show how to constrain such cosmological solutions for
extracting torsionless configurations in GR. In order to be able to provide
certain physical interpretation of new classes of geometric flow and Ricci
soliton solutions, we formulate a procedure of small deformations on a $%
\varepsilon$ parameter which allows us to study generic off--diagonal
deformations and geometric evolution of cosmologically important metrics in
MGTs.

Finally, in section \ref{sec5}, we study how cosmological geometric flow and
Ricci soliton effects can be connected to modified gravity scenarios and
observable cosmological data. We follow a reconstruction method which allows
us to construct effective mimetic potential and Lagrange multiplier
determined by generalized supersymmetric and Starobinsky type potentials in
modified double scalar models. Also are computed and analyzed possible
contributions of running of physical constants and locally anisotropic
interactions on observational indices. The approach is elaborated for
generalized mimetic $F(A,\mathcal{R}^2)$ theories with flexibility for
constraints to $\mathcal{R}^2$ and GR configurations. Discussions and
conclusions are presented at the end of the article in section \ref{sec6}.

\section{Nonholonomic Deformations of Perelman's Functionals and the $%
\mathcal{R}^{2}$ Gravity Theory}

\label{sec2}There are a few works on nonholonomic variables in supergravity
and superstring theories. The geometric constructions for supersymmetric
Finsler models on tangent superbundles \cite{vnpfinsl} can be re-defined to
noholonomic Einstein configurations and generalized for fibered superspaces
\cite{vtamsuper}) and superysmmetric K\"{a}hler--Ricci flows and solitons
\cite{gunara,gunara1}. By the same token, there is extensive work on
supergravity generalizations and modifications, see \cite%
{kounnas3,lah,ant,fer1,fer2,dal}, of the Starobinsky model \cite%
{star1,star2,much}. Our approach provides a set up for topological and
geometric flow evolution models and scale invariant theories in a de Sitter
background, extending the $\mathcal{R}^{2}$ plus conformally invariant
matter theories, in the framework of $N=1$ supergravity and exact
cosmological solutions for the accelerated expansion cosmology.

\subsection{Geometric objects adapted to nonlinear connection splitting}

In focusing on exact solutions for geometric flows, Ricci solitons and MGTs
and GR, we must bypass three obstructions (two technical and one physical):
the first one is to decouple such systems of nonlinear partial differential
equations, PDEs; the second one is to integrate (find solutions) the
equations in general form; the third one is to provide physical
interpretations of such new classes of solutions. The first two mathematical
tasks can be solved by using the so--called nonholonomic variables with
deformations of the frame, metric and nonlinear and linear connections
structures. We outline the key ideas and notations on nonholonomic variables
in (super) gravity theories, elaborated and reviewed in our previous works
\cite{svvvey,tgovsv,vtamsuper,vnpfinsl}.

Let us consider a (pseudo) Riemannian manifold $\mathbf{V}$ (we can consider
any dimension $n+m,$ with $n,m\geq 2$ like in \cite{tgovsv,bubv};\ for
simplicity, we shall formulate the main results and discuss with some
examples for $n=m=2$).\footnote{%
The local coordinates on $V$ are labelled $u^{\mu }=(x^{i},y^{a}),$ (in
brief, we set $u=(x,y)$), where indices respectively take values of type $%
i,j,...=1,2...,n$ and $a,b,...=n+1,n+2,...,n+m.$ The cumulative small Greek
indices run with values $\alpha ,\beta ,...=1,2,n+m.$ In four dimensions, we
consider $u^{4}=y^{4}=t$ as a time like coordinate.} We can introduce a
conventional 2+2 (or $n+m$) splitting into horizontal (h) and vertical (v)
components defined by a Whitney sum
\begin{equation}
\mathbf{N}:\ T\mathbf{V}=h\mathbf{\mathbf{V\oplus }}v\mathbf{V},
\label{ncon}
\end{equation}%
where $T\mathbf{V}$ is the tangent bundle on a spacetime manifold $\mathbf{V}
$ (in 4-d, we can consider a Lorentzian manifold with local pseudo-Euclidean
signature $(+++-)$). \ Any N--connection structure (\ref{ncon}) is
determined locally by a corresponding set of coefficients $N_{i}^{a},$ when $%
\mathbf{N}=N_{i}^{a}(u)dx^{i}\otimes \partial _{a}.$

There are structures, elongated linearly on N--coefficients, respectively,
of partial derivatives and differentials, $\mathbf{e}_{\nu }=(\mathbf{e}%
_{i},e_{a}),$ and cobases, $\mathbf{e}^{\mu }=(e^{i},\mathbf{e}^{a}),$ when
\begin{eqnarray}
\mathbf{e}_{\nu } &=&(\mathbf{e}_{i}=\partial /\partial x^{i}-\
N_{i}^{a}\partial /\partial y^{a},\ e_{a}=\partial _{a}=\partial /\partial
y^{a}),\mbox{ and }  \label{nader} \\
\mathbf{e}^{\mu } &=&(e^{i}=dx^{i},\mathbf{e}^{a}=dy^{a}+\ N_{i}^{a}dx^{i}).
\label{nadif}
\end{eqnarray}%
Such N--adapted bases (local frames) are nonholonomic because, in general,
there are satisfied relations of the type
\begin{equation}
\lbrack \mathbf{e}_{\alpha },\mathbf{e}_{\beta }]=\mathbf{e}_{\alpha }%
\mathbf{e}_{\beta }-\mathbf{e}_{\beta }\mathbf{e}_{\alpha }=W_{\alpha \beta
}^{\gamma }\mathbf{e}_{\gamma },  \label{anhcoef}
\end{equation}%
with nontrivial anholonomy coefficients $W_{ia}^{b}=\partial
_{a}N_{i}^{b},W_{ji}^{a}=\Omega _{ij}^{a}=\mathbf{e}_{j}\left(
N_{i}^{a}\right) -\mathbf{e}_{i}(N_{j}^{a}).$ We obtain holonomic
(integrable) bases if and only if $W_{\alpha \beta }^{\gamma }=0.$ In a more
general context, we can consider an arbitrary local basis $e^{\alpha
}=(e^{i},e^{a})$ and the corresponding dual one, co-basis, $e_{\beta
}=(e_{j},e_{b}).$

A local coordinate bases is denoted by $\partial _{\alpha ^{\prime
}}=(\partial _{i^{\prime }},\partial _{a^{\prime }})$ [for instance, $%
\partial _{i^{\prime }}=\partial /\partial x^{i^{\prime }}],$ and the
respective dual cobasis is written as $du^{\alpha ^{\prime }}=(dx^{i^{\prime
}},dy^{a^{\prime }}).$ The frame (vierbein) transforms can be written in the
form $e_{\beta }=A_{\beta }^{\ \beta ^{\prime }}(u)\partial _{\beta ^{\prime
}}$ and $e^{\alpha }=A_{\ \alpha ^{\prime }}^{\alpha }(u)du^{\alpha ^{\prime
}}$ (in particular, we can parameterize the $A$--matrices via
N--coefficients in (\ref{nader}) and (\ref{nadif})). We shall use boldface
symbols in order to emphasize that a geometric space/object/ construction is
adapted to a N--connection structure. Similar symbols will be written in
"non-boldface" form for any geometric object/equation if the adapting to a
N--connection splitting is not considered. For convenience, we shall use
primed, underlined indices etc ., and the Einstein summation convention on
repeated up--low indices will be assumed if not explicitly stated.

A manifold $(\mathbf{V})$ endowed with a nontrivial N--connection structure $%
\mathbf{N,}$ with nonzero $W_{\alpha \beta }^{\gamma },$ is called
nonholonomic. In a similar form, we can introduce the concept of
nonholonomic supermanifold which depends on the type of nonholonomic
distributions used for the definition of the supersymmetric/ supergravity
structures \cite{vtamsuper,vnpfinsl}. We shall omit such considerations in
this article and work only with the so-called bosonic part of certain
modified supergravity/ superstring theory.

We can elaborate any geometric construction, geometric physical theory with
a N--adapted differential and integral calculus and a corresponding
variational formalism for (modified) gravity theories using the N--elongated
operators (\ref{nader}) and (\ref{nadif}). Using frame transformations, such
constructions can be re-defined with respect to arbitrary frame of
reference. We say that the geometric constructions are performed with
distinguished objects (in brief, d--objects) whenever the coefficients are
determined with respect to N--adapted (co) frames and their tensor products.
For instance, a vector $Y(u)\in T\mathbf{V}$ can be parameterized as a
d--vector, $\mathbf{Y}=$ $\mathbf{Y}^{\alpha }\mathbf{e}_{\alpha }=\mathbf{Y}%
^{i}\mathbf{e}_{i}+\mathbf{Y}^{a}e_{a},$ or $\mathbf{Y}=(hY,vY),$ with $hY=\{%
\mathbf{Y}^{i}\}$ and $vY=\{\mathbf{Y}^{a}\}.$ Equivalently, we can write
this as $\mathbf{Y}=$ $Y^{\alpha }e_{\alpha
}=Y^{i}e_{i}+Y^{a}e_{a}=Y^{i}\partial _{i}+Y^{a}\partial _{a}.$ Similarly,
using the same technique we can determine and compute the coefficients of
d--tensors, N--adapted differential forms, d--connections, d--spinors etc.
If we do not adapt the N--adapted form, the conventional h- and v--splitting
of formulas is not preserved under general frame/coordinate transforms.

In this article, we shall work with a special class of linear connections: A
distinguished connection, d--connection, $\mathbf{D}=(\ ^{h}\mathbf{D},\ ^{v}%
\mathbf{D})$ is a linear connection, one preserving, under parallel
transport, the N--connection splitting (\ref{ncon}). A general linear
connection $D$ is not adapted to a chosen $h$-$v$--decomposition, i.e. it is
not a d--connection. A well known example is the Levi--Civita, LC,
connection in GR which is not a d--connection even if it can be written with
respect to N--adapted frames. Any d--connection $\mathbf{D}$ defines an
operator of covariant derivative, $\mathbf{D}_{\mathbf{X}}\mathbf{Y}$, for a
d--vector $\mathbf{Y}$ in the direction of a d--vector $\mathbf{X}.$ With
respect to N--adapted frames (\ref{nader}) and (\ref{nadif}), we can compute
N--adapted coefficients for $\mathbf{D}=\{\mathbf{\Gamma }_{\ \alpha \beta
}^{\gamma }=(L_{jk}^{i},L_{bk}^{a},C_{jc}^{i},C_{bc}^{a})\}$ or of any
linear connection $D$ defined by certain geometric/ physical principles, see
details and explicit formulas in Refs. \cite{sv2001,vsingl,svvvey,tgovsv,bubv}. The
N-adapted coefficients $\mathbf{\Gamma }_{\ \alpha \beta }^{\gamma }$ are
defined and computed geometrically for the h--v--components of $\mathbf{D}_{%
\mathbf{e}_{\alpha }}\mathbf{e}_{\beta }:=$ $\mathbf{D}_{\alpha }\mathbf{e}%
_{\beta }$ using $\mathbf{X}=\mathbf{e}_{\alpha }$ and $\mathbf{Y}=\mathbf{e}%
_{\beta }.$\footnote{%
Any d--connection $\mathbf{D}$ is characterized by the corresponding
d--torsion, $\mathbf{T,}$ nonmetricity, $\mathbf{Q},$ and d--curvature, $%
\mathbf{R},$ tensors (for N--adapted constructions, it is used the term
d--tensor). Such values are defined in standard form, when for any
d--connection $\mathbf{D}$ and d--vectors $\mathbf{X,Y\in }T\mathbf{V,}$
\begin{equation*}
\mathbf{T}(\mathbf{X,Y}):=\mathbf{D}_{\mathbf{X}}\mathbf{Y}-\mathbf{D}_{%
\mathbf{Y}}\mathbf{X}-[\mathbf{X,Y}],\ \mathbf{Q}(\mathbf{X}):=\mathbf{D}_{%
\mathbf{X}}\mathbf{g}\ \mathbf{R}(\mathbf{X,Y}):=\mathbf{D}_{\mathbf{X}}%
\mathbf{D}_{\mathbf{Y}}-\mathbf{D}_{\mathbf{Y}}\mathbf{D}_{\mathbf{X}}-%
\mathbf{D}_{\mathbf{[X,Y]}}.
\end{equation*}%
The N--adapted coefficients are correspondingly labeled using $h$- and $v$%
--indices,
\begin{equation*}
\mathbf{T}=\{\mathbf{T}_{\ \alpha \beta }^{\gamma }=\left( T_{\ jk}^{i},T_{\
ja}^{i},T_{\ ji}^{a},T_{\ bi}^{a},T_{\ bc}^{a}\right) \},\mathbf{Q}=\mathbf{%
\{Q}_{\ \alpha \beta }^{\gamma }\},\ \mathbf{R}=\mathbf{\{R}_{\ \beta \gamma
\delta }^{\alpha }=\left( R_{\ hjk}^{i}\mathbf{,}R_{\ bjk}^{a},R_{\
hja}^{i},R_{\ bja}^{c},R_{\ hba}^{i},R_{\ bea}^{c}\right) \},
\end{equation*}%
see explicit formulas in \cite{sv2001,vsingl,svvvey,tgovsv,bubv}.}

Any metric tensor $\mathbf{g}$ on a nonholonomic pseudo--Riemannian manifold
$\mathbf{V}$ can be parameterized in off--diagonal form,
\begin{equation}
\mathbf{g}=\underline{g}_{\alpha \beta }du^{\alpha }\otimes du^{\beta },%
\mbox{\ where \ }\underline{g}_{\alpha \beta }=\left[
\begin{array}{cc}
g_{ij}+N_{i}^{a}N_{j}^{b}g_{ab} & N_{j}^{e}g_{ae} \\
N_{i}^{e}g_{be} & g_{ab}%
\end{array}%
\right] ,  \label{ofdans}
\end{equation}%
with respect to a dual local coordinate basis $du^{\alpha }.$ Equivalently,
we can write a metric as a d--tensor (d--metric)
\begin{equation}
\widehat{\mathbf{g}}=g_{\alpha }(u)\mathbf{e}^{\alpha }\otimes \mathbf{e}%
^{\beta }=g_{i}(x)dx^{i}\otimes dx^{i}+g_{a}(x,y)\mathbf{e}^{a}\otimes
\mathbf{e}^{a},  \label{dm1}
\end{equation}%
in brief, $\mathbf{g}=(h\mathbf{g},v\mathbf{g})=\widehat{\mathbf{g}},$ with
respect to a tensor product of N--adapted dual frame (\ref{nadif}). For a $%
2+2$ splitting, any d-metric can be written in two $2\times 2$ block
diagonal form. We emphasize that a metric $\mathbf{g}$ (\ref{ofdans}) with
N--coefficients $N_{j}^{e}$ is generic off--diagonal if the anholonomy
coefficients $W_{\alpha \beta }^{\gamma }$ (\ref{anhcoef}) are not zero. It
is more convenient to work with d--metrics, canonical d--connections and
N--adapted frames if we want to decouple and find general solutions for
certain fundamental geometric/physical equations.

Following such geometric conditions, there are two very important linear
connection structures determined by the same metric structure : {%
\begin{equation}
\mathbf{g}\rightarrow \left\{
\begin{array}{ccccc}
\nabla : &  & \nabla \mathbf{g}=0;\ \mathbf{T[\nabla ]}=0, &  &
\mbox{ the
Levi--Civita connection;} \\
\widehat{\mathbf{D}}: &  & \widehat{\mathbf{D}}\ \mathbf{g}=0;\ h\widehat{%
\mathbf{T}}=0,\ v\widehat{\mathbf{T}}=0. &  &
\mbox{ the canonical
d--connection.}%
\end{array}%
\right.  \label{lcconcdcon}
\end{equation}%
The LC--connection }$\nabla $ can be introduced without any N--connection
structure but can be always canonically distorted to a necessary type of
d--connection. The canonical d--connection $\widehat{\mathbf{D}}$ depends
generically on a prescribed nonholonomic $h$- and $v$-splitting. In formulas
(\ref{lcconcdcon}), $h\widehat{\mathbf{T}}$ and $\ v\widehat{\mathbf{T}}$
are respective torsion components which vanish on conventional h- and
v--subspaces. There are also nonzero torsion components, $hv\widehat{\mathbf{%
T}}$, with nonzero mixed indices with respect to a N-adapted basis (\ref%
{nader}) and/or (\ref{nadif}).

On any (pseudo) Riemannian manifold $\mathbf{V,}$ all geometric
constructions can be performed equivalently with $\nabla $ and/or $\widehat{%
\mathbf{D}}$ and related via a canonical distortion relation
\begin{equation}
\widehat{\mathbf{D}}\mathbf{[g,N]}=\nabla \mathbf{[g,N]}+\widehat{\mathbf{Z}}%
\mathbf{[g,N]}.  \label{distr}
\end{equation}%
In this formula both the linear connections and the distorting tensor $%
\widehat{\mathbf{Z}}$ are uniquely determined by some data $(\mathbf{g,N)}$.
The d--tensor $\widehat{\mathbf{Z}}$ is an algebraic combination of
coefficients $\widehat{\mathbf{T}}_{\ \alpha \beta }^{\gamma }$. The
N--adapted coefficients for $\widehat{\mathbf{D}}$ and the corresponding
torsion, $\widehat{\mathbf{T}}_{\ \alpha \beta }^{\gamma }$, Ricci
d--tensor, $\widehat{\mathbf{R}}_{\ \beta \gamma }$, and Einstein d--tensor,
$\widehat{\mathbf{E}}_{\ \beta \gamma }$, can be computed in a simple way in
N--adapted form, see details in \ \cite{sv2001,vsingl,svvvey,tgovsv,bubv}. The canonical
distortion relation (\ref{distr}) can be used for computing respective
distortion tensors of the Riemiann, Ricci and Einstein tensors and
corresponding curvature scalars [such values are also uniquely determined by
data $(\mathbf{g,N)].}$ Thus, any (pseudo) Riemannian geometry and gravity
theory based on such a (pseudo) geometry can be equivalently formulated
using $(\mathbf{g,\nabla )}$ and/or $(\widehat{\mathbf{g}},\widehat{\mathbf{D%
}}).$

Explicitly, the N--adapted coefficients of $\widehat{\mathbf{D}}=\{$ $%
\widehat{\mathbf{\Gamma }}_{\ \alpha \beta }^{\gamma }=(\widehat{L}_{jk}^{i},%
\widehat{L}_{bk}^{a},\widehat{C}_{jc}^{i},\widehat{C}_{bc}^{a})\}$ in (\ref%
{lcconcdcon}) and (\ref{distr}) are
\begin{eqnarray}
\widehat{L}_{jk}^{i} &=&\frac{1}{2}g^{ir}\left( \mathbf{e}_{k}g_{jr}+\mathbf{%
e}_{j}g_{kr}-\mathbf{e}_{r}g_{jk}\right) ,\ \widehat{L}%
_{bk}^{a}=e_{b}(N_{k}^{a})+\frac{1}{2}h^{ac}\left( e_{k}h_{bc}-h_{dc}\
e_{b}N_{k}^{d}-h_{db}\ e_{c}N_{k}^{d}\right) ,  \notag \\
\widehat{C}_{jc}^{i} &=&\frac{1}{2}g^{ik}e_{c}g_{jk},\ \widehat{C}_{bc}^{a}=%
\frac{1}{2}h^{ad}\left( e_{c}h_{bd}+e_{c}h_{cd}-e_{d}h_{bc}\right) .
\label{candcon}
\end{eqnarray}%
The N--adapted coefficients of nonholonomically induced torsion $\widehat{%
\mathcal{T}}$ $=\{\widehat{\mathbf{T}}_{\ \alpha \beta }^{\gamma }\}$ are
computed using $\ $(\ref{candcon}) for the d--metric (\ref{dm1}). Such
coefficients satisfy the conditions $\widehat{T}_{\ jk}^{i}=0$ and $\widehat{%
T}_{\ bc}^{a}=0,$ but with nontrivial h--v-- coefficients
\begin{equation}
\widehat{T}_{\ jk}^{i}=\widehat{L}_{jk}^{i}-\widehat{L}_{kj}^{i},\widehat{T}%
_{\ ja}^{i}=\widehat{C}_{jb}^{i},\widehat{T}_{\ ji}^{a}=-\Omega _{\
ji}^{a},\ \widehat{T}_{aj}^{c}=\widehat{L}_{aj}^{c}-e_{a}(N_{j}^{c}),%
\widehat{T}_{\ bc}^{a}=\ \widehat{C}_{bc}^{a}-\ \widehat{C}_{cb}^{a}.
\label{dtors}
\end{equation}

We can consider N--splitting with zero noholonomically induced d--torsion,
when $\widehat{\mathbf{T}}_{\ \alpha \beta }^{\gamma }=0,$ i.e.
\begin{equation}
\widehat{C}_{jb}^{i}=0,\Omega _{\ ji}^{a}=0\mbox{ and }\widehat{L}%
_{aj}^{c}=e_{a}(N_{j}^{c}).  \label{lccond}
\end{equation}%
These conditions follow from formulas (\ref{candcon}) and (\ref{dtors}), see
details in \cite{svvvey,tgovsv}. If the Levi--Civita conditions,
LC--conditions, (\ref{lccond}) are satisfied, we obtain in N--adapted frames
(\ref{nader}) and (\ref{nadif}) $\widehat{\mathbf{Z}}_{\ \alpha \beta
}^{\gamma }=0$ and $\widehat{\mathbf{\Gamma }}_{\ \alpha \beta }^{\gamma
}=\Gamma _{\ \alpha \beta }^{\gamma }.$\footnote{%
The definition and the frame/coordinate transformation laws of a
d--connection are different from that of a "usual" linear connection. In
general, $\widehat{\mathbf{D}}\neq \nabla )$ but we can impose additional
conditions on coefficients $(\mathbf{g}_{\alpha \beta },N_{j}^{c})$ which
allows us to generate LC--configurations.}

Using nonholonomic variables $(\widehat{\mathbf{g}},\widehat{\mathbf{D}}),$
we can introduce the Lagrange density $\ ^{g}\mathcal{L}=$ $\widehat{\mathbf{%
R}},$ where the scalar curvature $\widehat{\mathbf{R}}:=\widehat{\mathbf{g}}%
^{\alpha \beta }$ $\widehat{\mathbf{R}}_{\ \alpha \beta }$ is computed in
standard form. For simplicity, The Lagrange density for matter,$~^{m}%
\widehat{\mathcal{L}},$ takes a simple form as it depends only on the
coefficients of a metric field but not on their derivatives. The
energy--momentum d--tensor can be computed via N--adapted variational
calculus,
\begin{equation}
\ ^{m}\widehat{\mathbf{T}}_{\alpha \beta }:=-\frac{2}{\sqrt{|\widehat{%
\mathbf{g}}|}}\frac{\delta (\sqrt{|\widehat{\mathbf{g}}|}\ \ ^{m}\widehat{%
\mathcal{L}})}{\delta \widehat{\mathbf{g}}^{\alpha \beta }}=\ ^{m}\widehat{%
\mathcal{L}}\widehat{\mathbf{g}}^{\alpha \beta }+2\frac{\delta (\ ^{m}%
\widehat{\mathcal{L}})}{\delta \widehat{\mathbf{g}}_{\alpha \beta }},
\label{ematter}
\end{equation}%
for $|\widehat{\mathbf{g}}|=\det |\widehat{\mathbf{g}}_{\mu \nu }|=\det |%
\mathbf{g}_{\mu \nu }|=|\mathbf{g}|.$ Following an N--adapted variational
calculus with action
\begin{equation*}
\ ^{g}\mathcal{S+\ }^{m}\mathcal{S}=\int d^{4}u\sqrt{|\widehat{\mathbf{g}}|}%
(\ ^{g}\widehat{\mathcal{L}}+~^{m}\widehat{\mathcal{L}}),
\end{equation*}
we obtain the nonholonomically modified gravitational field equations
\begin{equation}
\widehat{\mathbf{R}}_{\mu \nu }=\ \widehat{\mathbf{\Upsilon }}_{\mu \nu },
\label{mfeq}
\end{equation}%
where $\ \widehat{\mathbf{\Upsilon }}_{\mu \nu }=\varkappa (\ ^{m}\widehat{%
\mathbf{T}}_{\alpha \beta }-\frac{1}{2}\widehat{\mathbf{g}}_{\alpha \beta }\
^{m}\widehat{\mathbf{T}}),$ for $\ ^{m}\widehat{\mathbf{T}}:=\widehat{%
\mathbf{g}}^{\mu \nu }\ ^{m}\widehat{\mathbf{T}}_{\mu \nu }$ and $\varkappa $
is the gravitational coupling constant.

The canonical d--connection $\widehat{\mathbf{D}}$ was used for elaborating
the AFDM as a geometric method of constructing exact solutions in geometric
flows and MGTs. Such a connection allows to decouple the gravitational and
matter field equations with respect to N--adapted frames of reference. In
Finsler geometry, similar decoupling properties also occur for the Cartan
d--connection, see \cite{vnhrf,vdq1}. The AFDM can not be applied completely
if we work from the very beginning only with $\nabla .$ After constructing
certain general classes of solutions for $\widehat{\mathbf{D}}$, or any
d--connection uniquely distorted and determined by $(\mathbf{g,\nabla )}$ or
$(\mathbf{g},\widehat{\mathbf{D}}),$ we can impose at the end the condition $%
\widehat{\mathbf{T}}=0$ and extract LC--configurations $\widehat{\mathbf{D}}%
_{\mid \widehat{\mathbf{T}}=0}=\nabla .$ In brief, the main principle of the
AFDM\ is to deform the LC connection to an auxiliary one defined in such a
way defined that it allows with ease the integration of certain important
geometric/ physical equations. At the end of the procedure, we can consider
imposing additional constraints if it is necessary to extract solutions with
zero, or any prescribed value, of d-connection.

\subsection{Modified Perelman's functionals in nonholonomic variables}

Let us consider a family of d--metrics $\mathbf{g}(\tau )=\mathbf{g}(\tau
,u) $ of signature $(+++-)$ and N--connections $\mathbf{N}(\tau )$
parameterized by a positive parameter $\tau ,0\leq \tau \leq \tau _{0}.$ Any
nonholonomic manifold $\mathbf{V}$ $\subset \mathbf{V}(\tau )$ can be
enabled with a double nonholonomic 2+2 and 3+1 splitting \cite{vvrfthbh}.
Respectively on the manifold are defined families of Lagrange densities $\
^{g}\mathcal{L}(\tau )$ and $\ ^{m}\widehat{\mathcal{L}}(\tau ).$ We can
consider local coordinates labeled as $u^{\alpha }=(x^{i},y^{a})=(x^{\grave{%
\imath}},u^{4}=t)$ for $i,j,k,...=1,2; a,b,c,...=3,4;$ and $\grave{\imath},%
\grave{j},\grave{k}=1,2,3.$ For 3+1 splitting, we can choose distributions
such that any open region $U\subset $ $\mathbf{V}$ is covered by a family of
3-d spacelike hypersurfaces $\widehat{\Xi }_{t}$ parameterized by a time
like parameter $t. $

For arbitrary frame transformations on 4-d nonholonomic Lorentz manifolds
with variables $(\mathbf{g}(\tau )$, $\widehat{\mathbf{D}}(\tau )),$ we
modify/ generalize the Perelman's functionals in the form
\begin{eqnarray}
\widehat{\mathcal{F}}(\tau ) &=&\int_{t_{1}}^{t_{2}}\int_{\widehat{\Xi }%
_{t}}e^{-\widehat{f}}\sqrt{|\mathbf{g}|}d^{4}u(\frac{1}{2}\widehat{R}+\frac{1%
}{2}\ ^{m}\widehat{\mathcal{L}}+|\widehat{\mathbf{D}}\widehat{f}|^{2}),
\label{fperelm4matter} \\
&&\mbox{ and }  \notag \\
\widehat{\mathcal{W}}(\tau ) &=&\int_{t_{1}}^{t_{2}}\int_{\widehat{\Xi }%
_{t}}\left( 4\pi \tau \right) ^{-3}e^{-\widehat{f}}\sqrt{|\mathbf{g}|}%
d^{4}u[\tau (\frac{1}{2}\widehat{R}\ +\frac{1}{2}\ ^{m}\widehat{\mathcal{L}}%
+|\ ^{h}\widehat{D}\widehat{f}|+|\ ^{v}\widehat{D}\widehat{f}|)^{2}+\widehat{%
f}-8],  \label{wfperelm4matt}
\end{eqnarray}%
where the normalizing function $\widehat{f}(\tau ,u)$ satisfies $%
\int_{t_{1}}^{t_{2}}\int_{\widehat{\Xi }_{t}}\left( 4\pi \tau \right)
^{-3}e^{-\widehat{f}}\sqrt{|\mathbf{g}|}d^{4}u=1.$ It should be noted that $%
\widehat{\mathcal{W}}$ does not have a character of entropy for
pseudo--Riemannian metrics as in the original Perelman's work \cite%
{perelman1}. In this paper, we consider $\widehat{R}+\ ^{m}\widehat{\mathcal{%
L}}$ instead of $R$ used in the former mathematical works and in our
previous articles. We go a step further and consider flows of the Lagrange
density for matter fields, $\ ^{m}\widehat{\mathcal{L}}$, and other
possibilities like modifying the term $\widehat{R}$ into a quadratic one, or
more general type of corrections and generalizations. In our approach, above
F- and W--functionals may characterize relativistic nonlinear hydrodynamic
flows of families of metrics and generalized connections \cite{vvrfthbh}. We
can compute entropy like values of type (\ref{fperelm4matter}) and (\ref%
{wfperelm4matt}) for any 3+1 splitting with 3-d closed hypersurface
fibrations $\widehat{\Xi }_{t}.$ In general, it is possible to work with any
class of functions $\widehat{f}(\tau ,u)$ which can be fixed for certain
constant values. In many cases, such a function is chosen in a non--explicit
form which allows us to study non--normalized geometric flows and decouple
certain systems of nonlinear PDEs.

If a family of nonholonomic Lorentz manifolds $\mathbf{V}(\tau )$ is
determined by a 3+1 splitting with a family of d--metrics $\mathbf{g}(\tau
)=[\mathbf{q}(\tau ),N(\tau )],$ for a 3-d hypersurface metrics $\mathbf{q}_{%
\grave{\imath}\grave{j}}(\tau )$ and lapse function $N(\tau )=-g_{4},$ we
can define and compute analogs of the above F-- and W--functionals on 3-d
space like hypersurfaces. Considering $\ _{\shortmid }\widehat{\mathbf{D}}=%
\widehat{\mathbf{D}}_{\mid \widehat{\Xi }_{t}}$ for the canonical
d--connection $\widehat{\mathbf{D}}$ defined on a 3-d hypersurface $\widehat{%
\Xi }_{t},$ when all values depend on temperature like parameter $\tau $, we
define also $\ _{\shortmid }\widehat{R}:=$ $\widehat{R}_{\mid \widehat{\Xi }%
_{t}}$ for a corresponding N--adapted fixing of the local systems of
references and necessary types of normalization functions. Using $(q_{\grave{%
\imath}})=(q_{i},q_{3}),$ the Perelman's functionals for 3-d spacelike
hypersurfaces parameterized in N--adapted form are defined
\begin{eqnarray}
\ _{\shortmid }\widehat{\mathcal{F}}(\tau ) &=&\int_{\widehat{\Xi }%
_{t}}e^{-f}\sqrt{|q_{\grave{\imath}\grave{j}}|}d\grave{x}^{3}(\frac{1}{2}\
_{\shortmid }\widehat{R}+\frac{1}{2}\ ^{m}{}_{\shortmid }\widehat{\mathcal{L}%
}+|\ _{\shortmid }\widehat{\mathbf{D}}f|^{2}),  \label{perelm3f} \\
&&\mbox{ and }  \notag \\
\ _{\shortmid }\widehat{\mathcal{W}}(\tau ) &=&\int_{\widehat{\Xi }%
_{t}}\left( 4\pi \tau \right) ^{-3}e^{-f}\sqrt{|q_{\grave{\imath}\grave{j}}|}%
d\grave{x}^{3}[\tau (\frac{1}{2}\ _{\shortmid }\widehat{R}+\frac{1}{2}\
^{m}{}_{\shortmid }\widehat{\mathcal{L}}+|\ \ _{\shortmid }^{h}\widehat{%
\mathbf{D}}f|+|\ \ _{\shortmid }^{v}\widehat{\mathbf{D}}f|)^{2}+f-6],
\label{perelm3w}
\end{eqnarray}%
where we have chosen a necessary type scaling function $f$ which satisfies $%
\int_{\widehat{\Xi }_{t}}\left( 4\pi \tau \right) ^{-3}e^{-f}\sqrt{|q_{%
\grave{\imath}\grave{j}}|}d\grave{x}^{3}=1.$ These functionals \ transform
into standard Perelman functionals for 3-d Riemannian metrics on $\ \widehat{%
\Xi }_{t}$ if $\ _{\shortmid }\widehat{\mathbf{D}}\rightarrow \ _{\shortmid
}\nabla .$ To consider only such functionals is enough for 4-d stationary
configurations. We can compute their evolution on a time interval by
introducing additional integration along a timelike curve if certain
coefficients of the metric and connections are modified to depend on the
timelike coordinate.

Let us discuss possible physical implications of above F- and W-functionals,
normalizing (scaling) functions, and their relativistic generalizations
and/or nonholonomic deformations. G. Perelman wrote (see section 1,
paragraphs 1.3 and 1.4$^{\ast }$, in \cite{perelman1}) that such a
"F-functional and its first variation formula can be found in the literature
on the string theory, where it describes the low energy effective action;
the function $f$ is called diaton field;...". He also remarked that exists a
natural interpretation of $F$ in terms of Bochner-Lichnerovicz formulas
(respectively, for differential forms and spinors). For 3-d Riemannian
configurations, we can chose $f$ and respective normalization conditions in
a form when variations of such functionals with the measure $e^{-f}\sqrt{|q_{%
\grave{\imath}\grave{j}}|}d\grave{x}^{3}$ result in appropriate
diffeomorphisms turning the evolution equations from weakly parabolic into
strongly parabolic. This considerably simplify the proof of important
theorems on short time existence and uniqueness of solutions. Such results
can be used for various models of relativistic evolution with 3+1 splitting.

The choice and interpretation of $F$- and $W$-functionals depend also on the
class of physical models where such values are introduced. Such values can
be associated to some critical order parameters and respective
thermodynamical values in nonlinear sigma models \cite%
{polyakov,friedan1,friedan2,friedan3}, for nonlinear sigma models \cite%
{nitta,tsey,oliy,carfora}, with generalizations for models of (non)
holonomic commutative and noncommutative gravity theories \cite%
{bakaslust,vnhrf,vnrflnc,streets}. Fixing certain values of $f$ and/or
relating such a normalization function to a scalar field $\phi $ (see Ref.
\cite{muen01} and developments in next sections of the present work), the
relativistically generalized Perelman's functionals (\ref{fperelm4matter}) and
(\ref{wfperelm4matt}) transforms into effective actions for $R^{2}$ gravity
and/or equivalent Einstein equations with scalar fields. In a general sense,
such functionals describe both relativistic evolution scenarios and, for
respective self-similar fixed configurations, actions for various classes of
non-Riemannian gravitational theories. Using variables involving the
canonical d--connection $\widehat{\mathbf{D}},$ we can prove that the
gravitational field equations in such models can be integrated in general
off-diagonal forms depending, in principle, on all spacetime coordinates
\cite{sv2001,vsingl,svvvey,tgovsv,vtamsuper,bubv}.

Cosmological models with locally anisotropic and/or inhomogeneous generic
off-diagonal metrics in MGTs (with possible mass terms, non-minimal
couplings, noncommutative and/or supersymmetric parameters etc.) have been
studied in our works \cite{vcosmsol1,vcosmsol2,vcosmsol3,vcosmsol4,vcosmsol4a,vcosmsol5}%
. Those classes of solutions were found with effective and/or locally
anisotropic polarization and variation of constants, deformed horizons,
nontrivial vacuum structure etc. In this work, we study explicit examples
when such cosmological models can be derived from generalized Perelman
functionals for certain normalized self-similar configurations resulting in
relativistic Ricci solitons being equivalent to $R^{2}$ gravity (reproducing
modified Starobinsky cosmological scenarios). Here we also note that such
constructions are motivated both by fundamental results in modern
mathematics and by the fact that $\widehat{\mathcal{F}}$-- and $\widehat{%
\mathcal{W}}$--functionals allows to elaborate an unified approach to
geometric evolution theories, string theory, cosmology of MGTs and geometric
thermodynamics. The last statement follows from the fact that $\ _{\shortmid
}\widehat{\mathcal{W}}$ \ (\ref{perelm3w}) transforms into the standard
Perelman's entropy if $\widehat{\mathbf{D}}$ is constrained to the
Levi-Civita configurations. Various generalized Hamiltonians for geometric
evolution and MGTs are generated by respective self-similar configurations
determined by a corresponding choice, or fixing, of normalized f-function
with possible interpretation as a dilaton/ effective scalar field.

\subsection{Nonholonomic geometric evolution of $R^{2}$ gravity with
conformally coupled matter}

In the standard geometric analysis applied to Ricci flow theory,
mathematicians are usually interested in studying the flow evolution of
geometric objects determined by the metric structure (usually, of Riemannian
signature or metrics related to K\"{a}hler configurations). For applications
to the real world and generate realistic physical models and in modern
cosmology, physicists have to create geometric relativistic models/ theories
with Lorentz signature and relate to MGTs and GR and investigate possible
geometric evolution scenarios (more complex, and less determined, than
nonlinear diffusion processes) using exact solutions and nonholonomic
deformations.

\subsubsection{Geometric evolution functionals and modified flow equations}

Let us conformally rescale the metric in (\ref{fperelm4matter}) and (\ref%
{wfperelm4matt}) in the following way, $\ g_{\mu \nu }\rightarrow \widetilde{%
g}_{\mu \nu }=g_{\mu \nu }e^{-\ln |1+2\widetilde{t}|}$, for $\sqrt{2/3}\phi
=\ln |1+2\widetilde{t}|$, and introduce a specific Lagrange density for
matter
\begin{equation}
\ ^{m}\widehat{\mathcal{L}}=-\frac{1}{2}\mathbf{e}_{\mu }\phi \ \mathbf{e}%
^{\mu }\phi -V(\phi ).  \label{sclagr}
\end{equation}%
In the above formula, we consider a nonlinear potential for scalar field $%
\phi $
\begin{eqnarray}
V(\phi ) &=&\mu ^{2}(1-e^{-\sqrt{2/3}\phi })^{2},\mu =const,
\label{starobpot} \\
\mbox{ when }\ V(\phi &\gg &0)\rightarrow \mu ^{2},\ V(\phi =0)=0,\ V(\phi
\ll 0)\sim \mu ^{2}e^{-2\sqrt{2/3}\phi },  \notag
\end{eqnarray}%
and employ the N--elongated partial derivative and differential operators,
respectively, (\ref{nader}) and (\ref{nadif}). From (\ref{perelm3f}) and (%
\ref{perelm3w}), respectively, we get the functionals {\small
\begin{eqnarray}
\widehat{\mathcal{F}}(\tau ) &=&\int_{t_{1}}^{t_{2}}\int_{\widehat{\Xi }%
_{t}}e^{-\widehat{f}}\sqrt{|\mathbf{g}|}d^{4}u(\frac{1}{2}\widehat{R}-\frac{1%
}{2}\mathbf{e}_{\mu }\phi \ \mathbf{e}^{\mu }\phi -\ V(\phi )+|\widehat{%
\mathbf{D}}\widehat{f}|^{2}),  \label{ffsc} \\
&&\mbox{ and }  \notag \\
\widehat{\mathcal{W}}(\tau ) &=&\int_{t_{1}}^{t_{2}}\int_{\widehat{\Xi }%
_{t}}\left( 4\pi \tau \right) ^{-3}e^{-\widehat{f}}\sqrt{|\mathbf{g}|}%
d^{4}u[\tau (\frac{1}{2}\widehat{R}\ -\frac{1}{2}\mathbf{e}_{\mu }\phi \
\mathbf{e}^{\mu }\phi -\ V(\phi )+|\ ^{h}\widehat{D}\widehat{f}|+|\ ^{v}%
\widehat{D}\widehat{f}|)^{2}+\widehat{f}-8],  \label{wfsc}
\end{eqnarray}%
} These values present a relativistic generalization of Perelman's
functionals (describing geometric evolution of 3-d Riemannian metrics) to
the case of normalized evolution of a class of MGTs with nonholonomically
deformed connection $\widehat{\mathbf{D}}$ and nonlinear potential for
scalar field $\phi $ (\ref{starobpot}). Fixing explicit values of $\widehat{f%
},$ we can generate effective actions for nonholonomic Ricci soliton
configurations which are equivalent to actions of respective
gravitational-scalar models related to $R^{2}$ gravity. For applications of
geometric flows methods in modern gravity and cosmology theories, it present
a special interest to study, and find physical important solutions, of the
systems of generalized Hamilton equations which can be derived from (\ref%
{ffsc}) and/or (\ref{wfsc}):

We can follow an N--adapted variational calculus with the above functionals.
The procedure is similar to that presented in N--adapted form in \cite%
{vnhrf,vnrflnc}.\footnote{%
In abstract geometric form, such proofs can be performed by changing $%
\partial _{\mu }\rightarrow \mathbf{e}_{\mu }$ and $\nabla \rightarrow
\widehat{\mathbf{D}}$ in the method presented in Perelman's preprint \cite%
{perelman1}, or any detailed proof in \cite{monogrrf1,monogrrf2,monogrrf3}}
As a result, we derive a system of nonlinear PDEs that generalize the R.
Hamilton equations to nonholonomic geometric evolution with canonical
deformation of the linear connection structure adapted to $h$-- and $v$%
--splitting of relativistic systems $(g_{\mu \nu },N_{i}^{a},\widehat{%
\mathbf{D}},\phi ),$
\begin{eqnarray}
\partial _{\tau }g_{ij} &=&-2(\widehat{R}_{ij}-\ \Upsilon _{ij}(\phi )),\
\label{ricciflowr2} \\
\partial _{\tau }g_{ab} &=&-2(\widehat{R}_{ab}-\Upsilon _{ab}(\phi )),
\notag \\
\widehat{R}_{ia} &=&\widehat{R}_{ai}=0;\widehat{R}_{ij}=\widehat{R}_{ji};%
\widehat{R}_{ab}=\widehat{R}_{ba};  \notag \\
\partial _{\tau }\phi &=&-2(\widehat{\square }\phi +\frac{d\ V(\phi )}{d\phi
});  \notag \\
\partial _{\tau }f &=&-\widehat{\square }f+\left\vert \widehat{\mathbf{D}}%
f\right\vert ^{2}-\ ^{h}\widehat{R}-\ ^{v}\widehat{R}+\widehat{\square }\phi
+V(\phi ).  \notag
\end{eqnarray}%
The conditions $\widehat{R}_{ia}=0$ and $\widehat{R}_{ai}=0$ are necessary
if we want to keep the total metric to be symmetric under Ricci flow
evolution, see \cite{vnonsymmet} for theories of geometric evolution with
nonsymmetric metrics. The general relativistic character of 4-d geometric
flow evolution is encoded in operators like $\widehat{\square }=\widehat{%
\mathbf{D}}^{\alpha }\widehat{\mathbf{D}}_{\alpha },$ d-tensor components $%
\widehat{R}_{ij}$ and $\widehat{R}_{ab},$ theirs scalars $\ ^{h}\widehat{R}%
=g^{ij}\widehat{R}_{ij}$ and $\ ^{v}\widehat{R}=g^{ab}\widehat{R}_{ab}$ with
data $(g_{ij},g_{ab},\widehat{\mathbf{D}}_{\alpha }).$ The matter source
d--tensor $\ \Upsilon _{\alpha \beta }(\phi )=(\Upsilon _{ij}(\phi
),\Upsilon _{ab}(\phi ))$ in the above formulae is computed as in (\ref{mfeq}%
) for the energy momentum tensor (\ref{ematter}) determined by (\ref{sclagr}%
), i.e.
\begin{equation*}
\ ^{\phi }\mathbf{T}_{\mu \nu }(\phi )=\widehat{\mathbf{D}}_{\mu }\phi
\widehat{\mathbf{D}}_{\nu }\phi -\mathbf{g}_{\mu \nu }\left[ \frac{1}{2}%
\widehat{\mathbf{D}}_{\alpha }\phi \widehat{\mathbf{D}}^{\alpha }\phi +\
V(\phi )\right] ,
\end{equation*}%
when the potential $\ V(\phi )$ is chosen in a form corresponding to
evolution of a MGT with quadratic gravity and applications in modern
cosmology.

\subsubsection{Modified Ricci solitons and $R^{2}$ gravity}

For self--similar fixed configurations with $\tau =\tau _{0},$ when $%
\partial _{\tau }g_{\alpha \beta }=0,$ $\partial _{\tau }\phi =0,$ the
modified Hamilton equations (\ref{ricciflowr2}) transform into canonically
nonholonomically deformed gravitational and matter field equations for $%
R^{2} $ gravity supplemented with scalar field (studied for $\widehat{%
\mathbf{D}}\rightarrow \nabla $ in \cite{kounnas3} and references therein),%
\begin{eqnarray}
\widehat{\mathbf{R}}_{\alpha \beta } &=&\Upsilon _{\alpha \beta }
\label{nhesceq} \\
\widehat{\square }\phi +\frac{d\ V(\phi )}{d\phi } &=&0.  \notag
\end{eqnarray}%
into a relativistic nonholonomic variant of Ricci soliton equations (for
simplicity, we omit the last formula working with non-normalized geometric
flow evolution). These equations can be derived alternatively (equivalently)
from the action
\begin{eqnarray}
S &=&\int d^{4}u\sqrt{|g|}\left( \frac{1}{2}\widehat{\mathbf{R}}-3\frac{%
\mathbf{e}_{\mu }\widetilde{t}\mathbf{e}^{\mu }\widetilde{t}}{(1+2\widetilde{%
t})^{2}}-\mu ^{2}\frac{(2\widetilde{t})^{2}}{(1+2\widetilde{t})^{2}}\right)
\notag \\
&=&\int d^{4}u\sqrt{|g|}\left( \frac{1}{2}\widehat{\mathbf{R}}-\frac{1}{2}%
\mathbf{e}_{\mu }\phi \mathbf{e}^{\mu }\phi -V(\phi )\right) ,
\label{auxact2}
\end{eqnarray}%
which for any $\tau =\tau _{0}$ and corresponding parametrization of
normalization function can be considered as the corresponding limits of
functionals (\ref{ffsc}) \ and (\ref{wfsc}).

Let us prove that the equations (\ref{nhesceq}) really define a relativistic
version of Ricci soliton equations modeling $\mathcal{R}^{2}$ gravity with
the action
\begin{equation}
S=\int d^{4}u\sqrt{|g|}\left( \frac{1}{2}\mathcal{R}+\frac{1}{16\mu ^{2}}%
\mathcal{R}^{2}\right) .  \label{starobact}
\end{equation}%
Considering $\widetilde{t}$ as a Lagrange multiplier, we can replace in the
above equation $\mathcal{R}^{2}\rightarrow 2\widetilde{t}\widehat{\mathbf{R}}%
-8\mu ^{2}\widetilde{t}^{2}$ (working in nonholonomic variables with $%
\mathcal{R}\rightarrow \widehat{\mathbf{R}}$). Using the equation of motion
for $\widetilde{t}$ as an effective scalar field,%
\begin{equation*}
\widetilde{t}=\widehat{\mathbf{R}}/8\mu ^{2},
\end{equation*}%
we can reproduce a theory with $\widehat{\mathbf{R}}^{2}$ term determined by
a standard action for the $R^{2}$ gravity if $\widehat{\mathbf{D}}%
\rightarrow \nabla .$ Thus the action (\ref{starobact}) can be written in an
equivalent form (in the Jordan frame)%
\begin{equation}
S=\int d^{4}u\sqrt{|g|}\left[ \frac{1}{2}\left( 1+2\widetilde{t}\right)
\widehat{\mathbf{R}}-8\mu ^{2}\widetilde{t}^{2}\right] .  \label{actjordan}
\end{equation}%
This action is equivalent to actions (\ref{auxact2}) and (\ref{starobact}).

It is well known that the pure $\mathcal{R}^{2}$ gravity constitutes an
example of a minimal version of a scale invariant theory without ghosts when
instead of (\ref{actjordan}) we take
\begin{equation*}
S=\int d^{4}u\sqrt{|g|}\frac{1}{16\mu ^{2}}\widehat{\mathbf{R}}%
^{2}\rightarrow S=\int d^{4}u\sqrt{|g|}\left[ \frac{1}{2}(2\widetilde{t}%
\widehat{\mathbf{R}}-8\mu ^{2}\widetilde{t}^{2})\right] .
\end{equation*}%
This theory is invariant (both for $\nabla $ and $\widehat{\mathbf{D}}$)
under global dilatation symmetry with a constant $\sigma ,$ $g_{\mu \nu
}\rightarrow e^{-2\sigma }g_{\mu \nu },\widetilde{t}\rightarrow e^{2\sigma }%
\widetilde{t}. $ Passing from the Jordan to the Einstein frame with a
redefinition $\phi =\sqrt{3/2}\ln |2\widetilde{t}|,$ we obtain
\begin{equation*}
\ ^{E}S=\int d^{4}u\sqrt{|g|}\left( \frac{1}{2}\widehat{\mathbf{R}}-\frac{1}{%
2}\mathbf{e}_{\mu }\phi \ \mathbf{e}^{\mu }\phi -2\mu ^{2}\right) ,
\end{equation*}%
where the scalar potential $\ ^{\phi }V(\phi )$ in (\ref{auxact2}) is
transformed into the cosmological constant term $\mu ^{2}$ which can be
positive / negative / zero, respectively for de Sitter / anti de Sitter /
flat space. The field equations derived from $\ ^{E}S$ are
\begin{eqnarray}
\widehat{\mathbf{R}}_{\mu \nu }-\mathbf{e}_{\mu }\phi \ \mathbf{e}_{\nu
}\phi -2\mu ^{2}\widehat{\mathbf{g}}_{\mu \nu } &=&0,  \label{riccisol4d} \\
\widehat{\mathbf{D}}^{2}\phi &=&0.  \label{aux2}
\end{eqnarray}%
Such equations constitute an example of relativistic nonholonomically
deformed Ricci soliton equation considered in \cite{muen01}.

We conclude that the $\mathcal{R}^{2}$ gravity for any linear connection
determined in a metric compatible form by the metric structure can be
modelled as a relativistic Ricci soliton configuration with effective scalar
fields. In a more general context, we can preserve such an interpretation
but keep in mind that additional nonholonomic deformations can be determined
by a nonlinear scalar potential $\ ^{\phi }V(\phi ),$ or other type
modifications with contributions of matter fields and (we shall study in
next sections) from supergravity theories. Homogeneous relativistic Ricci
solitons constitute explicit examples of physically important cosmological
spaces. \ For large positive values of $\widetilde{t}$ and integrable
configurations, the relativistic Ricci soliton is described by an
approximate de Sitter space, which grows exponentially with Hubble parameter
proportional to $\varsigma ,$%
\begin{equation*}
3H^{2}=\mu ^{2},H=\frac{\dot{a}}{a}
\end{equation*}%
for $a(t)$ denoting the scale factor of the metric $%
ds^{2}=-dt^{2}+a^{2}(dx^{i})^{2}.$ Starobinsky \cite{star1}\ proposed to
describe a realistic inflationary cosmology using contributions from $R^{2}$
gravity. \ In order to create a successful density perturbation, $\delta
\rho /\rho \sim 10^{-5},$ the mass scale must be as low as $\mu \sim
10^{-5}M\sim 10^{13}GeV.$ Cosmological scenarios can be modified by generic
off--diagonal interactions and geometric flows determining polarizations and
running of physical constants (see, for instance, \cite{flambaum}).

\subsection{The $SO(1,1+k)$ generalizations of geometric flow and pure $%
\mathcal{R}^{2}$ gravity models}

\label{ssmscal}In order to realize realistic physical models, we generalize
the pure $\mathcal{R}^{2}$ geometric flow, Ricci solitons and gravity models
by adding extra matter fields.\footnote{%
We shall use a "caligraphic" symbol $\mathcal{R}$ in order to emphasize that
such scalar curvature can be for any necessary type of linear connection.}
For conformally coupled scalars, fermions and gauge bosons, one attributes
corresponding scale weights in order to promote the global scale symmetry of
the matter sector (at the classical level of theory) to a local conformal
symmetry. It is considered that the interactions are of gauge or Yukawa type
and that the scalar potential is quartic. For instance, the scalar potential
from (\ref{sclagr}) is completed by a quartic one, $\ ^{c}V(\Phi _{\overline{%
i}}),$ with corresponding modifications of $\ ^{\phi }V(\phi )\rightarrow \
^{\mu }V(\phi ,\Phi _{\overline{i}}),$ when (respectively)
\begin{eqnarray*}
\ ^{c}V(\Phi _{\overline{i}}) &=&\lambda _{\overline{i}\overline{j}\overline{%
k}\overline{l}}\Phi _{\overline{i}}\Phi _{\overline{j}}\Phi _{\overline{k}%
}\Phi _{\overline{l}},\mbox{ for certain constants }\lambda _{\overline{i}%
\overline{j}\overline{k}\overline{l}},\mbox{ and }\  \\
\ ^{\mu }V(\phi ,\Phi _{\overline{i}}) &=&\mu ^{2}(1+e^{-\sqrt{2/3}\phi
}\Phi _{\overline{i}}^{2}/6)^{2},\mbox{ or } \\
&=&\mu ^{2}(1-e^{-\sqrt{2/3}\phi }+e^{-\sqrt{2/3}\phi }\Phi _{\overline{i}%
}^{2}/6)^{2},\mbox{ for scale non--invarint models}.
\end{eqnarray*}%
We define
\begin{equation}
\ \overline{V}(\phi ,\Phi _{\overline{i}}):=-e^{-2\sqrt{2/3}\phi }\ ^{c}V-\
^{\mu }V,  \label{escalarso}
\end{equation}%
for some scalar fields $\Phi _{\overline{l}}$ with values in $SO(1,1+k),$
for $k=1,2,3,...,$ where $\overline{i},\overline{j},\overline{k}%
...=2,3,...,k+1.$ Our notations differ from those in \cite{kounnas3} (see
this paper on additional speculations about classical and quantum $SO(1,1+k)$
theories and their generalizations for supergravity and superstring ones)
and is motivated by constructions to be used for constructing exact
solutions.
\begin{eqnarray}
\widehat{\mathcal{F}}(\tau ) &=&\int_{t_{1}}^{t_{2}}\int_{\widehat{\Xi }%
_{t}}e^{-\widehat{f}}\sqrt{|\mathbf{g}|}d^{4}u\frac{1}{2}(\widehat{R}-%
\mathbf{e}_{\mu }\phi \ \mathbf{e}^{\mu }\phi -e^{-\sqrt{2/3}\phi }\mathbf{e}%
_{\mu }\Phi _{\overline{i}}\ \mathbf{e}^{\mu }\Phi _{\overline{i}}-\
\overline{V}(\phi ,\Phi _{\overline{i}})+|\widehat{\mathbf{D}}\widehat{f}%
|^{2})  \label{ffso} \\
&&\mbox{ and }  \notag \\
\widehat{\mathcal{W}}(\tau ) &=&\int_{t_{1}}^{t_{2}}\int_{\widehat{\Xi }%
_{t}}\left( 4\pi \tau \right) ^{-3}e^{-\widehat{f}}\sqrt{|\mathbf{g}|}%
d^{4}u[\tau (\frac{1}{2}\widehat{R}\ -\frac{1}{2}\mathbf{e}_{\mu }\phi \
\mathbf{e}^{\mu }\phi -e^{-\sqrt{2/3}\phi }\mathbf{e}_{\mu }\Phi _{\overline{%
i}}\ \mathbf{e}^{\mu }\Phi _{\overline{i}}-\ \overline{V}(\phi ,\Phi _{%
\overline{i}})  \notag \\
&&+|\ ^{h}\widehat{D}\widehat{f}|+|\ ^{v}\widehat{D}\widehat{f}|)^{2}+%
\widehat{f}-8].  \label{wfso}
\end{eqnarray}%
We can compensate contributions of terms like $\mathbf{e}_{\mu }\phi \
\mathbf{e}^{\mu }\phi $ and/or $e^{-\sqrt{2/3}\phi }\mathbf{e}_{\mu }\Phi _{%
\overline{i}}\ \mathbf{e}^{\mu }\Phi _{\overline{i}}$ by choosing a
corresponding set of normalization functions $\widehat{f}$ but the nonlinear
scalar potential $\ ^{\Phi }V$ will contribute always to geometric flow
evolution. These generalizations of Perelman's functions are defined in such
forms which include $SO(1,1+k)$ generalizations of $\mathcal{R}^{2}$ gravity
models, which admit further extensions to supergravity theories (see next
section) and new applications in elaborating cosmological evolution
scenarios.

For evolution of symmetric d--metrics, the functionals result in modified
Hamilton equations
\begin{eqnarray}
\partial _{\tau }g_{ij} &=&-2(\widehat{R}_{ij}-\ \Upsilon _{ij}(\phi ,\Phi _{%
\overline{i}})),\   \label{ghameqso} \\
\partial _{\tau }g_{ab} &=&-2(\widehat{R}_{ab}-\Upsilon _{ab}(\phi ,\Phi _{%
\overline{i}})),  \notag \\
\widehat{R}_{ia} &=&\widehat{R}_{ai}=0;\widehat{R}_{ij}=\widehat{R}_{ji};%
\widehat{R}_{ab}=\widehat{R}_{ba};  \notag \\
\partial _{\tau }\phi &=&-2(\widehat{\square }\phi +\frac{\partial \
\overline{V}}{\partial \phi });  \notag \\
\partial _{\tau }\Phi _{\overline{i}} &=&-2(\widehat{\square }\Phi _{%
\overline{i}}+\frac{\partial \ \overline{V}}{\partial \Phi _{\overline{i}}});
\notag \\
\partial _{\tau }f &=&-\widehat{\square }f+\left\vert \widehat{\mathbf{D}}%
f\right\vert ^{2}-\ ^{h}\widehat{R}-\ ^{v}\widehat{R}+\widehat{\square }\phi
+\ \ V(\phi ).  \notag
\end{eqnarray}%
The matter source d--tensor $\ \Upsilon _{\alpha \beta }(\phi ,\Phi _{%
\overline{i}})=(\Upsilon _{ij}(\phi ,\Phi _{\overline{i}}),\Upsilon
_{ab}(\phi ,\Phi _{\overline{i}}))$ in the above formulae is computed as in (%
\ref{mfeq}) for the energy momentum tensor (\ref{ematter}) determined by (%
\ref{escalarso}), with
\begin{equation}
\ ^{m}\widehat{\mathcal{L}}\rightarrow \ \widehat{\mathcal{L}}(\phi ,\Phi _{%
\overline{i}})=-\frac{1}{2}\mathbf{e}_{\mu }\phi \ \mathbf{e}^{\mu }\phi -%
\frac{1}{2}e^{-\sqrt{2/3}\phi }\mathbf{e}_{\mu }\Phi _{\overline{i}}\
\mathbf{e}^{\mu }\Phi _{\overline{i}}-\ ^{\Phi }V(\phi ,\Phi _{\overline{i}%
}).  \label{matterfields}
\end{equation}%
One gets,
\begin{equation*}
\ \ \widehat{\mathbf{T}}_{\alpha \beta }(\phi ,\Phi _{\overline{i}}):=-\frac{%
2}{\sqrt{|\widehat{\mathbf{g}}|}}\frac{\delta (\sqrt{|\widehat{\mathbf{g}}|}%
\ \ \widehat{\mathcal{L}}(\phi ,\Phi _{\overline{i}}))}{\delta \widehat{%
\mathbf{g}}^{\alpha \beta }}=\widehat{\mathcal{L}}(\phi ,\Phi _{\overline{i}%
})\widehat{\mathbf{g}}^{\alpha \beta }+2\frac{\delta (\ \widehat{\mathcal{L}}%
(\phi ,\Phi _{\overline{i}}))}{\delta \widehat{\mathbf{g}}_{\alpha \beta }},
\end{equation*}%
and
\begin{equation*}
\ \widehat{\mathbf{\Upsilon }}_{\mu \nu }(\phi ,\Phi _{\overline{i}%
})=\varkappa (\widehat{\mathbf{T}}_{\alpha \beta }(\phi ,\Phi _{\overline{i}%
})-\frac{1}{2}\widehat{\mathbf{g}}_{\alpha \beta }\ \widehat{\mathbf{T}}%
(\phi ,\Phi _{\overline{i}})),
\end{equation*}%
when the Lagrange density $\ \ \widehat{\mathcal{L}}(\phi ,\Phi _{\overline{i%
}})$ is chosen in such a form which correspond to $SO(1,k+1)$ models of $%
\mathcal{R}^{2}$ gravity theory studied in \ \cite{kounnas3}.

For self-similar configurations and a fixed value of parameter $\tau _{0}$
(when terms with $\partial _{\tau }$ are fixed to zero), we obtain the
equations of nonholonomic relativistic Ricci solitons with values in a
scalar manifold
\begin{equation*}
\mathcal{M}(\phi ,\Phi _{\overline{i}})=\mathcal{H}^{k+1}\equiv
SO(1,1+k)/SO(1+k)
\end{equation*}%
which is isomorphic to a maximally symmetric spaces, ie. the hyperbolic
space $\mathcal{H}^{k+1}$ which is an Euclidean $AdS$ space with
non-negative cosmological constant $\mu ^{2}.$ The nonholonomic structure is
given by the base spacetime $\mathbf{V}.$ Such equations can be obtained
equivalently by an N--adapted variational calculus. The action in the
Einstein frame is
\begin{equation*}
S=\int d^{4}u\sqrt{|g|}\frac{1}{2}\left[ \widehat{R}-\mathbf{e}_{\mu }\phi \
\mathbf{e}^{\mu }\phi \mathcal{-}e^{-\sqrt{2/3}\phi }\mathbf{e}_{\mu }\Phi _{%
\overline{i}}\ \mathbf{e}^{\mu }\Phi _{\overline{i}}-\overline{V}(\phi ,\Phi
_{\overline{i}})\right] .
\end{equation*}

Let us prove that the nonholonomic $SO(1,k+1)$ Ricci solitons model a $%
\mathcal{R}^{2}$ theory of gravity. Write the corresponding action in the
Jordan N--adapted frame with a Lagrange multiplier field $\widetilde{t}$ of
type%
\begin{equation*}
S=\int d^{4}u\sqrt{|g|}\frac{1}{2}\left( (2\widetilde{t}-\frac{1}{6}\Phi _{%
\overline{i}}^{2})\mathcal{R-}\mathbf{e}_{\mu }\Phi _{\overline{i}}\ \mathbf{%
e}^{\mu }\Phi _{\overline{i}}-2\ ^{c}V(\Phi _{\overline{i}})-8\mu ^{2}%
\widetilde{\phi }^{2}\right) ,
\end{equation*}%
where the linear term $\mathcal{R}\rightarrow \widehat{R}$ is dressed by the
fields $\Phi _{\overline{i}}^{2}$ which has the conformal weight $w_{\Phi
}=1.$ Performing a conformal rescaling of the metric%
\begin{equation*}
g_{\mu \nu }\rightarrow g_{\mu \nu }e^{-\ln |2\widetilde{t}-\frac{1}{6}\Phi
_{\overline{i}}^{2}|},
\end{equation*}%
with $\sqrt{2/3}\phi =\ln |2\widetilde{t}-\frac{1}{6}\Phi _{\overline{i}%
}^{2}|,$ \ we obtain a nonholonomic $\widehat{R}^{2}$ theory with
conformally coupled matter is given by action%
\begin{equation*}
S=\int d^{4}u\sqrt{|g|}\frac{1}{2}\left( \frac{1}{8\mu ^{2}}\widehat{R}^{2}-%
\frac{1}{6}\Phi _{\overline{i}}^{2}\ \widehat{R}\mathcal{-}\mathbf{e}_{\mu
}\Phi _{\overline{i}}\ \mathbf{e}^{\mu }\Phi _{\overline{i}}-2\ ^{c}V(\Phi _{%
\overline{i}})+...\right) .
\end{equation*}%
In this formula, ellipses denote the fermionic and gauge boson matter parts
which do not affect the vacuum structure of the model (minimal extension
with fermionic and gauge boson parts of the action are invariant under
conformal transformations). We note that the canonical Einstein term $R$ is
absent since it is not permitted by the scale symmetry. Such a term in the
action breaks the classical invariance since it shifts the field $2%
\widetilde{t}\rightarrow 2\underline{t}=2\widetilde{t}+1.$ \ Both last
actions present explicit examples of two measure theories studied, for
instance, in \cite{guend1,guend2,rajpootvacaru} and references therein.

The nonholonomic Ricci soliton configurations are of three types which
correspond to the following three phases of the equivalent $SO(1,1+k)$
modified gravity of the type $R^{2}:$

\begin{enumerate}
\item The scale invariant de Sitter era when the induced square mass of the
canonically normalized fields $\Phi _{\overline{i}}$ satisfy the condition $%
m_{\overline{i}}^{2}=\frac{2\mu ^{2}}{3}(1-e^{-\sqrt{2/3}\phi })>0$ and $\
^{\mu }V\rightarrow \mu ^{2}.$

\item The flat space era when for any vacuum the total potential $\ ^{t}V=\
^{c}V+\ ^{\mu }V=0.$ When there are flat directions in $\ ^{c}V,$ there is a
degeneracy vacuum in the flat direction of $\ ^{\mu }V,$ which allows to
compute the mean values $<\Phi _{\overline{i}}^{2}=e^{-\sqrt{2/3}\phi }-1>,$
with $\ ^{c}V=\ ^{\mu }V=0.$

\item The scale non--invariant era which is characterized by a more complex
structure of the vacuum, see section 2.3.3 in \cite{kounnas3}.
\end{enumerate}

We note that under geometric flow evolution the vacuum structure of the MGTs
can be changed by relating different types of nonholonomic Ricci soliton
configurations. Such solutions of generalized Hamilton equations (including
self-similar configurations) will be presented in following sections.

\section{Supersymmetric Extensions of Perelman's Functionals and Ricci
Solitons for $\mathcal{R}^{2}$ Su\-per\-gravity}

\label{sec3}Our main goal is to investigate possible connections with gauged
supergravity of scale invariant geometric flows in a de Sitter background
extending the $R^{2}$ gravity plus conformally invariant matter theories.
The constructions will be performed in nonholonomic variables which will
allow one to construct exact solutions in following sections. We shall work
with the canonical d--connection which for $SU(1,1+k)$ extensions posses a
quadratic scalar curvature $\widehat{\mathcal{R}}^{2}$ considering that via
nonholonomic constraints $\widehat{\mathbf{D}}_{\mid \widehat{\mathbf{T}}%
}=\nabla $ we can always extract standard $R^{2}$ and related GR
configurations.

\subsection{Nonholonomic variables and the minimal scale invariant $%
SU(1,1+k) $ supersymmetric extension of Perelman's functionals}

The minimal N--adapted supersymmetric extension of the $SO(1,1+k)$ $\mathcal{%
R}^{2}$ - model to $SU(1,1+k)$ configurations is realized by introducing the
supersymmetric scalar partners of $\widetilde{t}$ and $\Phi _{\overline{i}%
}^{2}$ considered on a nonholonomic manifold $\mathbf{V}.$ We shall use
complex fields%
\begin{equation*}
\psi ^{I}=\{T=\widetilde{t}+ib,z^{\overline{i}}=|z^{\overline{i}}|e^{i\theta
^{\overline{i}}}\},\mbox{ for }|z^{\overline{i}}|=\Phi ^{\overline{i}}/\sqrt{%
6}.
\end{equation*}%
Introducing the complex function $Y$ and it partial complex derivatives,
\begin{equation*}
Y=T+\overline{T}-|z^{\overline{i}}|^{2}, \,\,\, Y_{I}=\frac{\partial Y}{%
\partial \psi ^{I}}, \,\,\, Y_{\overline{I}}=\frac{\partial Y}{\partial \psi
^{\overline{I}}}, \,\,\, Y_{I\overline{I}}=\frac{\partial ^{2}Y}{\partial
\psi ^{I}\partial \overline{\psi }^{\overline{I}}},
\end{equation*}%
we can formulate scale invariant $SU(1,1+k)$ supergeometric flow and
supergravity models. In such terms, the supersymmetric extension of
generalized Perelman's functionals (\ref{ffso}) and (\ref{wfso}) can be
performed in certain equivalent forms following respective formulae: {\small
\begin{eqnarray}
\widehat{\mathcal{F}}(\tau ) &=&\int_{t_{1}}^{t_{2}}\int_{\widehat{\Xi }%
_{t}}e^{-\widehat{f}}\sqrt{|\mathbf{g}|}d^{4}u\frac{1}{2}[\widehat{R}-K_{I%
\overline{I}}\mathbf{e}_{\mu }\psi ^{I}\mathbf{e}^{\mu }\psi ^{\overline{I}%
}-\ ^{E}V+|\widehat{\mathbf{D}}\widehat{f}|^{2}]  \label{ffssm} \\
&=&\int_{t_{1}}^{t_{2}}\int_{\widehat{\Xi }_{t}}e^{-\widehat{f}}\sqrt{|%
\mathbf{g}|}d^{4}u\frac{1}{2}[\widehat{R}-\frac{3}{4Y^{2}}(\mathcal{D}_{\mu }%
\mathcal{D}^{\mu }+\mathbf{J}_{\mu }\mathbf{J}^{\mu })+\frac{3}{Y}Y_{I%
\overline{I}}\mathbf{e}_{\mu }\psi ^{I}\mathbf{e}^{\mu }\overline{\psi }^{%
\overline{I}}-\ ^{E}V+|\widehat{\mathbf{D}}\widehat{f}|^{2}]  \notag \\
&=&\int_{t_{1}}^{t_{2}}\int_{\widehat{\Xi }_{t}}e^{-\widehat{f}}\sqrt{|%
\mathbf{g}|}d^{4}u\frac{1}{2}[\widehat{R}-\frac{1}{2}\mathbf{e}_{\mu }\phi
\mathbf{e}^{\mu }\phi -\frac{3}{4}e^{-2\sqrt{2/3}\phi }\mathbf{J}_{\mu }%
\mathbf{J}^{\mu }-3e^{-\sqrt{2/3}\phi }\mathbf{e}_{\mu }z^{\overline{i}}%
\mathbf{e}^{\mu }\overline{z}^{\overline{i}}-\ ^{E}V+|\widehat{\mathbf{D}}%
\widehat{f}|^{2}]  \notag
\end{eqnarray}%
and
\begin{eqnarray}
\widehat{\mathcal{W}}(\tau ) &=&\int_{t_{1}}^{t_{2}}\int_{\widehat{\Xi }%
_{t}}\left( 4\pi \tau \right) ^{-3}e^{-\widehat{f}}\sqrt{|\mathbf{g}|}%
d^{4}u[\tau (\widehat{R}-K_{I\overline{I}}\mathbf{e}_{\mu }\psi ^{I}\mathbf{e%
}^{\mu }\psi ^{\overline{I}}-\ ^{E}V+|\ ^{h}\widehat{D}\widehat{f}|+|\ ^{v}%
\widehat{D}\widehat{f}|)^{2}+\widehat{f}-8]  \notag \\
&=&\int_{t_{1}}^{t_{2}}\int_{\widehat{\Xi }_{t}}\left( 4\pi \tau \right)
^{-3}e^{-\widehat{f}}\sqrt{|\mathbf{g}|}d^{4}u[\tau (\widehat{R}-\frac{3}{%
4Y^{2}}(\mathcal{D}_{\mu }\mathcal{D}^{\mu }+\mathbf{J}_{\mu }\mathbf{J}%
^{\mu })+\frac{3}{Y}Y_{I\overline{I}}\mathbf{e}_{\mu }\psi ^{I}\mathbf{e}%
^{\mu }\overline{\psi }^{\overline{I}}-\ ^{E}V  \notag \\
&&+|\ ^{h}\widehat{D}\widehat{f}|+|\ ^{v}\widehat{D}\widehat{f}|)^{2}+%
\widehat{f}-8]  \label{wfssm} \\
&=&\int_{t_{1}}^{t_{2}}\int_{\widehat{\Xi }_{t}}\left( 4\pi \tau \right)
^{-3}e^{-\widehat{f}}\sqrt{|\mathbf{g}|}d^{4}u[\tau (\widehat{R}-\frac{1}{2}%
\mathbf{e}_{\mu }\phi \mathbf{e}^{\mu }\phi -\frac{3}{4}e^{-2\sqrt{2/3}\phi }%
\mathbf{J}_{\mu }\mathbf{J}^{\mu }-3e^{-\sqrt{2/3}\phi }\mathbf{e}_{\mu }z^{%
\overline{i}}\mathbf{e}^{\mu }\overline{z}^{\overline{i}}  \notag \\
&&-\ ^{E}V+|\ ^{h}\widehat{D}\widehat{f}|+|\ ^{v}\widehat{D}\widehat{f}%
|)^{2}+\widehat{f}-8].  \notag
\end{eqnarray}%
} The nonholonomic and supersymmetric variables in the above formulae are
defined in the following forms:

\begin{itemize}
\item The auxiliary vector field $\mathbf{A}_{\mu }$ and the axial current $%
\mathbf{J}_{\mu }$ (neglecting fermionic contributions, see details in \cite%
{51.0} for holonomic configurations) are given in terms of the scalar fields
in the following expression
\begin{equation*}
\mathbf{A}_{\mu }=\frac{3\mathbf{J}_{\mu }}{2Y}=\frac{3i}{2Y}(Y_{\overline{I}%
}\mathbf{e}_{\mu }\overline{\psi }^{\overline{I}}-Y_{I}\mathbf{e}_{\mu }\psi
^{I}),
\end{equation*}%
for
\begin{equation*}
\mathbf{J}_{\mu }=i(Y_{\overline{I}}\mathbf{e}_{\mu }\overline{\psi }^{%
\overline{I}}-Y_{I}\mathbf{e}_{\mu }\psi ^{I}),\mathcal{D}_{\mu }=Y_{%
\overline{I}}\mathbf{e}_{\mu }\overline{\psi }^{\overline{I}}+Y_{I}\mathbf{e}%
_{\mu }\psi ^{I}\equiv \mathbf{e}_{\mu }Y.
\end{equation*}%
These d-vector fields are necessary for the supersymmetric extension of the
model and are a member of the $\mathcal{R}$ supermultiplet. They appear
naturally together with the Einstein term $R.$

\item In supergravity, the potential $\ ^{c}V$ can be written in terms of
the $Y$ function and superpotential $\widetilde{W}(\psi ^{I})$ (see details
in \cite{kounnas3} on different supersymmetric generalizations of $R^{2}$
theory with corresponding choices of $\widetilde{W};$ in this work, we use "$%
\widetilde{symbol}$" in order to avoid confusions with $W$ from anholonomy
relations (\ref{anhcoef})),%
\begin{eqnarray}
\ ^{c}V &=&Y^{2}\ ^{E}V,  \label{sssourc} \\
\ ^{E}V &=&e^{K}\{(\widetilde{W}_{I}+K_{I}\widetilde{W})K^{I\overline{I}}(%
\overline{\widetilde{W}}_{\overline{I}}+K_{\overline{I}}\overline{\widetilde{%
W}})-3|\widetilde{W}|^{2}\}+D-\mbox{ terms }.  \notag
\end{eqnarray}

\item $K=-3\ln Y$ is the K\"{a}hler potential (a real function of scalars)
which defines the symplectic metric%
\begin{equation*}
K_{I\overline{I}}=\frac{\partial ^{2}K}{\partial \psi ^{I}\partial \psi ^{%
\overline{I}}}=\frac{3}{Y^{2}}(Y_{I}Y_{\overline{I}}-YY_{I\overline{I}}).
\end{equation*}

\item Like in the $SO(1,1+k)$ case, we use the no scale field $\phi $
determined by conditions $\sqrt{\frac{2}{3}}\phi =\ln Y.$
\end{itemize}

We use different supersymmetric N--adapted variables in order to be \ able
to model for self--similar fixed configurations different models of modified
supergravity in the so--called "old" and "new" formalisms corresponding to
constructions in Refs. \cite%
{14.1,14.2,51.0,52.1,52.2,53.1,53.2,53.3,53.4,53.5}. The nonholonomic
versions of Perelman's like superfunctionals (\ref{ffssm}) and (\ref{wfssm})
are postulated in such forms which result (for self-similar Ricci
supersoliton configurations) in mentioned type modified supergravity
theories. The normalizing function and superfields, in respective
nonholonomic variables, are parameterized in such a form which allows to
describe alternatively the Ricci supersoliton configurations as certain $%
\mathcal{R}^{2}$ supergravity theory (described by an effective
supersymmetric action, see below (\ref{effsupact}), and exact cosmological
type solutions). These way, such supersymmetric generalizations of the
Hamilton-Perelman theory (with respective evolution supersymmetric Hamilton
like equations, to be derived in the next subsection) have explicit
motivations from modern supergravity and superstring theory. The
cosmological solutions for such effective $\mathcal{R}^{2}$ supergeometric
flows and supergravity will be constructed in explicit form providing new
types of supersymmetric accelerating cosmological scenarios.

\subsection{Geometric flow equations with supersymmetric $SU(1,1+k)$
modifications}

The explicit form of modified relativistic Hamilton equations depend on the
type of nonholonomic variables, parameterizations and normalization
functions we chose for functionals (\ref{ffssm}) and (\ref{wfssm}). For
instance, the normalization function can be redefined $\widehat{f}%
\rightarrow f$ and constrained to satisfy the conditions%
\begin{equation}
|\widehat{\mathbf{D}}f|^{2}=\frac{3}{4}e^{-2\sqrt{2/3}\phi }\mathbf{J}_{\mu }%
\mathbf{J}^{\mu }+3e^{-\sqrt{2/3}\phi }\mathbf{e}_{\mu }z^{\overline{i}}%
\mathbf{e}^{\mu }\overline{z}^{\overline{i}}.  \label{normalcond}
\end{equation}%
Applying these variables and an N-adapted variational formalism (for
instance, to the first above functional) we obtain the following
supersymmetrically modified Hamilton equations
\begin{eqnarray}
\partial _{\tau }g_{ij} &=&-2(\widehat{R}_{ij}-\ ^{E}\Upsilon _{ij}),\
\label{ssymeveq} \\
\partial _{\tau }g_{ab} &=&-2(\widehat{R}_{ab}-\ ^{E}\Upsilon _{ab}),  \notag
\\
\widehat{R}_{ia} &=&\widehat{R}_{ai}=0; \,\,\, \widehat{R}_{ij}=\widehat{R}%
_{ji}; \,\,\, \widehat{R}_{ab}=\widehat{R}_{ba};  \notag \\
\partial _{\tau }\phi &=&-2(\widehat{\square }\phi +\frac{\partial \ ^{E}V}{%
\partial \phi });  \notag \\
\partial _{\tau }f &=&-\widehat{\square }f+\left\vert \widehat{\mathbf{D}}%
f\right\vert ^{2}-\ ^{h}\widehat{R}-\ ^{v}\widehat{R}+\widehat{\square }\phi
+\ ^{E}V.  \label{normalizeq}
\end{eqnarray}%
The matter source d--tensor $\ ^{E}\Upsilon _{\alpha \beta }=(\ ^{E}\Upsilon
_{ij},\ ^{E}\Upsilon _{ab})$ in the gravitational part of the above formulae
is computed as in (\ref{mfeq}) for the energy momentum tensor (\ref{ematter}%
) determined by (\ref{sssourc}), with
\begin{equation*}
\ ^{m}\widehat{\mathcal{L}}\rightarrow \ ^{E}\widehat{\mathcal{L}}=-\frac{1}{%
2}\mathbf{e}_{\mu }\phi \ \mathbf{e}^{\mu }\phi -\ ^{E}V.
\end{equation*}%
This allows to define
\begin{eqnarray}
\ ^{E}\widehat{\mathbf{\Upsilon }}_{\alpha \beta } &=&\varkappa (\ \ ^{E}%
\widehat{\mathbf{T}}_{\alpha \beta }-\frac{1}{2}\widehat{\mathbf{g}}_{\alpha
\beta }\ \ ^{E}\widehat{\mathbf{T}})  \label{ssource} \\
&=&\varkappa \left[ -\mathbf{e}_{\alpha }\phi \ \mathbf{e}_{\beta }\phi +%
\widehat{\mathbf{g}}_{\alpha \beta }\mathbf{e}_{\mu }\phi \ \mathbf{e}^{\mu
}\phi +\widehat{\mathbf{g}}_{\alpha \beta }\ ^{E}V\right] \mbox{ for }\
\notag \\
\ ^{E}\widehat{\mathbf{T}}_{\alpha \beta } &=&\ ^{E}\widehat{\mathcal{L}}%
\widehat{\mathbf{g}}_{\alpha \beta }+2\frac{\delta (\ \ ^{E}\widehat{%
\mathcal{L}})}{\delta \widehat{\mathbf{g}}^{\alpha \beta }}=\left( -\frac{1}{%
2}\mathbf{e}_{\mu }\phi \ \mathbf{e}^{\mu }\phi -\ ^{E}V\right) \widehat{%
\mathbf{g}}_{\alpha \beta }-\mathbf{e}_{\alpha }\phi \ \mathbf{e}_{\beta
}\phi ,  \notag \\
\ ^{E}\widehat{\mathbf{T}} &=&-3\mathbf{e}_{\mu }\phi \ \mathbf{e}^{\mu
}\phi -4\ ^{E}V,  \notag
\end{eqnarray}%
where the Lagrange density $\ \ ^{E}\widehat{\mathcal{L}}$ is chosen in such
a form which correspond to $SU(1,1+k)$ models of $\mathcal{R}^{2}$ gravity
theory studied in \ \cite{kounnas3} and referenced therein.

The form of the system of nonlinear PDEs (\ref{ssymeveq}) depend on the type
of normalizing conditions we impose on geometric flows. The nonholonomic
constraint (\ref{normalcond}) encode the evolution dynamics of fields $%
\mathbf{J}_{\mu },$ and/or $\mathbf{A}_{\mu },$ and $z^{\overline{i}}$ into
normalizations functions and geometric pseudo--Riemannian and K\"{a}hler
structure. For other types of nonholonomic conditions, the evolution
equations for such fields appear in the modified Hamilton equations. For
MGTs, we can compute such contributions considering actions for
gravitational and matter fields associated with functionals (\ref{ffssm})
and (\ref{wfssm}) for certain self--similar configurations.

\subsection{Nonholonomic Ricci solitons with supersymmetric $SU(1,1+k)$
modifications and the $\mathcal{R}^{2}$ gravity theory}

Fixing to zero the terms with $\partial _{\tau }$ for fixed value of
parameter $\tau _{0},$ we obtain the equations nonholonomic Ricci solitons
with supersymmetric modifications determined by $\ ^{E}V.$ Depending on the
type of nonholonomic variables, such equations can be alternatively derived
from actions which present nonholonomic deformations in $\mathcal{R}^{2}$
gravity. Writing such action with standard $\mathcal{R}\rightarrow \widehat{%
\mathbf{R}}$ term in two sets of variables, we obtain%
\begin{eqnarray}
\ ^{E}S &=&\int d^{4}u\sqrt{|g|}\left[ \frac{1}{2}Y\left( \widehat{\mathbf{R}%
}\mathcal{+}\frac{2}{3}\mathbf{A}_{\mu }\mathbf{A}^{\mu })-\mathbf{J}_{\mu }%
\mathbf{A}^{\mu }+3Y_{I\overline{I}}\mathbf{e}_{\mu }\psi ^{I}\mathbf{e}%
^{\mu }\overline{\psi }^{\overline{I}}-\ ^{c}V\right) \right]  \notag \\
&=&\int d^{4}u\sqrt{|g|}\left[ \frac{1}{2}\widehat{\mathbf{R}}-K_{I\overline{%
I}}\mathbf{e}_{\mu }\psi ^{I}\mathbf{e}^{\mu }\psi ^{\overline{I}}-\ ^{E}V%
\right]  \notag \\
&=&\int d^{4}u\sqrt{|g|}\left[ \frac{1}{2}\widehat{\mathbf{R}}-\frac{3}{%
4Y^{2}}(\mathcal{D}_{\mu }\mathcal{D}^{\mu }+\mathbf{J}_{\mu }\mathbf{J}%
^{\mu })+\frac{3}{Y}Y_{I\overline{I}}\mathbf{e}_{\mu }\psi ^{I}\mathbf{e}%
^{\mu }\overline{\psi }^{\overline{I}}-\ ^{E}V\right]  \notag \\
&=&\int d^{4}u\sqrt{|g|}\left[ \frac{1}{2}\widehat{\mathbf{R}}-\frac{1}{2}%
\mathbf{e}_{\mu }\phi \mathbf{e}^{\mu }\phi -\frac{3}{4}e^{-2\sqrt{2/3}\phi }%
\mathbf{J}_{\mu }\mathbf{J}^{\mu }-3e^{-\sqrt{2/3}\phi }\mathbf{e}_{\mu }z^{%
\overline{i}}\mathbf{e}^{\mu }\overline{z}^{\overline{\overline{i}}}-\ ^{E}V%
\right] .  \label{effsupact}
\end{eqnarray}

Let us consider the main geometric and physical properties of such
supersymmetric Ricci soliton configurations. The kinetic part of the $%
SU(1,1+k)$ geometric evolution and supergravity theory is manifestly scale
invariant in a form which is very similar to the $SO(1,1+k)$
non-supersymmetric model. In the Einstein frame the scale symmetry acts as
follows:
\begin{equation*}
T\rightarrow e^{2\sigma }T, \,\,\, z^{\overline{i}}\rightarrow e^{\sigma }z^{%
\overline{i}}, \,\,\, Y\rightarrow e^{2\sigma }Y, \,\,\, e^{\alpha \phi
}\rightarrow e^{2\sigma }e^{\alpha \phi }, \,\,\, b\rightarrow e^{2\sigma }b.
\end{equation*}%
The scale symmetry constraint implies that $\widetilde{W}$ has scaling
weight 3: $\widetilde{W}\rightarrow e^{3\sigma }\widetilde{W},$ see \cite%
{kounnas3}. This requirement leads to the following three possibilities (or
linear combinations of these):

\begin{enumerate}
\item The anti de Sitter (AdS) realization of the $\mathcal{R}^{2}$ theory,
when $\widetilde{W}=cT^{3/2},c=const,$ give rise to a scale invariant model
with negative cosmological term, i.e. the parameter $\mu ^{2}\,\ $is
negative. The theory is conformally equivalent to a pure $\mathcal{R}^{2}$
supersymmetric theory. The potential turns out to be negative%
\begin{equation*}
\ ^{E}V=e^{K}\left( \frac{|K_{T}\widetilde{W}+\widetilde{W}_{T}|^{2}}{K_{TT}}%
-3|\widetilde{W}|^{2}\right) =\frac{3|c|^{2}}{8}\left( \frac{|T|b^{2}}{%
\widetilde{t}^{3}}-\frac{|T|(\widetilde{t}^{2}+b^{2})}{\widetilde{t}^{3}}%
\right) =-\frac{3|c|^{2}}{8}\sqrt{1+\frac{b^{2}}{\widetilde{t}^{2}}}.
\end{equation*}%
There is also a non-trivial gravitino mass term, $\ m_{3/2}^{2}=e^{K}|%
\widetilde{W}|^{2}=\frac{|c|^{2}}{8}\left( 1+\frac{b^{2}}{\widetilde{t}^{2}}%
\right) ^{3/2}.$ The stationary point occurs at $b=0.$ The classical vacuum
corresponds to an $AdS$ space with non-vanishing gravitino mass term $%
m_{3/2}^{2}=\frac{|c|^{2}}{8}$ and $V=-\frac{3|c|^{2}}{8}.$

\item Flat space realizations and the connection to no-scale models are
possible for $\widetilde{W}=c_{\overline{i}\overline{j}\overline{k}}z^{%
\overline{i}}z^{\overline{j}}z^{\overline{k}},$ with nontrivial constants $%
c_{\overline{i}\overline{j}\overline{k}}.$ For instance, under the so called
$U(1)_{R}$ gauging, $\ U(1)_{R}:z^{\overline{i}}\rightarrow e^{iw}z^{%
\overline{i}}\mbox{ and }W\rightarrow e^{3iw}\widetilde{W}.$ The classical
vacua of the model are characterized by $\widetilde{W}_{i}=0$ and $\mathcal{D%
}^{\mu }=0$ (F--flatness and D-flatness) with $\ ^{E}V=0.$ The no scale
modulus $\phi $ and the gravitino mass remain undetermined at the classical
level. In this class of models the F-part of the potential can never
generate a non-zero cosmological term. The only remaining way is via a
contribution from non trivial Fayet-Iliopoulos D--terms (FI-term, with
nontrivial antisymmetric field $B_{\mu \nu }).$ We cite section 4 in \cite%
{kounnas3} on emergence of inflationary potentials for certain anomaly free
consistency conditions, which must be valid at the quantum level of the
theory.

\item De Sitter realizations of the $\mathcal{R}^{2}$ theory are possible
for $\widetilde{W}=c_{\overline{k}}z^{\overline{k}}T,$ with nontrivial
vacuum constants.The classical vacuum of the theory is de Sitter space if
and only if there is non-trivial contribution from a non trivial $U(1)_{R}$
symmetry D--term necessary for the stabilization of the $z^{\overline{k}}$
fields. Under this particular case with $U(1)_{R},$ the superpotential
transforms with a non-trivial phase giving rise to a non-zero FI term. It
was shown that $N=1$ superstring constructions provide metastable de Sitter
vacua supported by FI $D_{R}$--terms associated to the several anomalous $%
U(1)_{R}^{A}$ gauge symmetries. The superstring resolution of these
anomalies is achieved via local axion shifts promoting the several $U(1)_{R}$
symmetries to $U(1)_{t}$ or $U(1)_{d}$ symmetries.
\end{enumerate}

Ricci soliton configurations with $SU(1,1+k)$ modifications can be generated
by different types of generating and integration functions and effective
sources following the AFDM. The off--diagonal nonlinear interactions may
preserve the realizations 1-3 above or change substantially the vacuum
structure of certain classes of solutions.

%%%%%%%%444444444444444444444444444444444444444444444%%%

\section{Cosmological Solutions for Supersymmetric Modifications of Ricci
Flows \& $R^{2}$ Gravity}

\label{sec4} We develop the AFDM \cite{sv2001,vsingl,svvvey,tgovsv,bubv} in order to
construct in general forms certain classes of locally anisotropic and
inhomogeneous cosmological solutions for geometric flows of Ricci soliton
configurations modelling $R^{2}$ gravity. Metrics of such solutions are not
stationary (like we considered for black ellipsoid solutions in $R^{2}$
gravity \cite{muen01}) but depend in general form on all spacetime
coordinates and on evolution parameter $\tau $, i.e. $\mathbf{g}_{\alpha
\beta}=\mathbf{g}_{\alpha \beta} (\tau ,x^{i},y^{3},y^{4}=t).$ Such
geometric flows and Ricci soliton configurations are with nontrivial
evolution of nonholonomically induced torsion. Having constructed certain
general classes of solutions, we can always consider additional constraints
for geometric flows with zero torsion and/or cosmological metrics of type $%
\mathbf{g}_{\alpha \beta }(\tau ,t)$ encoding variations of constants,
off--diagonal deformations of standard cosmological solutions to
nonholonomic cosmological Ricci solitons and other type cosmological
solutions in MGTs.

\subsection{PDEs for time like dependence of off--diagonal geometric flows
and Ricci solitons}

Using N--adapted 2+2 frame and coordinate transformations of the metric and
source $\ ^{E}\Upsilon _{\alpha \beta },$
\begin{eqnarray*}
\mathbf{g}_{\alpha \beta }(\tau ,x^{i},t) &=&e_{\ \alpha }^{\alpha ^{\prime
}}(\tau ,x^{i},y^{a})e_{\ \beta }^{\beta ^{\prime }}(\tau ,x^{i},y^{a})%
\widehat{\mathbf{g}}_{\alpha ^{\prime }\beta ^{\prime }}(\tau ,x^{i},y^{a})%
\mbox{ \,\,\,\,\,\, and } \\
\ \ ^{E}\Upsilon _{\alpha \beta }(\tau ,x^{i},t) &=&e_{\ \alpha }^{\alpha
^{\prime }}(\tau ,x^{i},y^{a})e_{\ \beta }^{\beta ^{\prime }}(\tau
,x^{i},y^{a})\widehat{\Upsilon }_{\alpha ^{\prime }\beta ^{\prime }}(\tau
,x^{i},y^{a}),
\end{eqnarray*}%
for a time like coordinate $y^{4}=t$ ($i^{\prime },i,k,k^{\prime },...=1,2$
and $a,a^{\prime },b,b^{\prime },...=3,4),$ we introduce another type of
ansatz for metrics (comparing to those in \cite{muen01}, when $\partial _{t}$
was considered as a Killing vector for various classes of prime and target
metrics). The new canonical parameterizations of d--metric and N--connection
coefficients and (effective) sources are with generic dependence on time
like coordinate $t$, applicable even for certain cases when there will be
considered solutions with Killing like symmetry on $\partial _3$ where $y^3$
is a spacelike coordinate. This will allow us to decouple and solve
physically important systems of PDEs in order to generate, in general,
inhomogeneous and locally anisotropic solutions for relativistic geometric
flows and MGTs. The generic off--diagonal metric ansatz is taken in the form
\begin{eqnarray}
\mathbf{g} &=&\mathbf{g}_{\alpha ^{\prime }\beta ^{\prime }}\mathbf{e}%
^{\alpha ^{\prime }}\otimes \mathbf{e}^{\beta ^{\prime }}=g_{i}(\tau
,x^{k})dx^{i}\otimes dx^{j}+\omega ^{2}(\tau ,x^{k},y^{3},t)h_{a}(\tau
,x^{k},t)\mathbf{e}^{a}\otimes \mathbf{e}^{a}  \label{offdans} \\
&=&q_{i}(\tau ,x^{k})dx^{i}\otimes dx^{i}+q_{3}(\tau ,x^{k},y^{3},t)\mathbf{e%
}^{3}\otimes \mathbf{e}^{3}-\breve{N}^{2}(\tau ,x^{k},y^{3},t)\mathbf{e}%
^{4}\otimes \mathbf{e}^{4},  \label{lapsnonh} \\
\mathbf{e}^{3} &=&dy^{3}+n_{i}(\tau ,x^{k},t)dx^{i}, \,\,\,\, \mathbf{e}%
^{4}=dt+w_{i}(\tau ,x^{k},t)dx^{i}.  \notag
\end{eqnarray}%
The parametrization (\ref{lapsnonh}) is written as a 4--d metric with
extension of a 3--d metric $q_{ij}=diag(q_{\grave{\imath}})=(q_{i},q_{3})$
on a hypersurface $\widehat{\Xi }_{t}$ \ where
\begin{equation}
q_{3}=g_{3}=\omega ^{2}h_{3}\mbox{ and }\breve{N}^{2}(\tau ,u)=-\omega
^{2}h_{4}=-g_{4},  \label{shift1}
\end{equation}%
are related to the lapse function $\breve{N}(\tau ,u).$

The ansatz (\ref{offdans}) are characterized by geometric evolution on
parameter $\tau $ of such N--connection and d-metric coefficients and are
denoted by
\begin{eqnarray}
\ N_{i}^{3}(\tau ) &=&n_{i}(\tau ,x^{k},t); \,\,\,\, N_{i}^{4}=w_{i}(\tau
,x^{k},t)\mbox{ and }  \label{coefft} \\
g_{i^{\prime }j^{\prime }}(\tau ) &=&diag[g_{i}], \,\,\,\,
g_{1}=g_{2}=q_{1}=q_{2}=e^{\psi (\tau ,x^{k})};\   \notag \\
\,\,\,\, g_{a^{\prime }b^{\prime }}(\tau ) &=&diag[\omega ^{2}h_{a}],
\,\,\,\, h_{a}=h_{a}(\tau , \,\,\,\, x^{k},y^{3}), \,\,\,\, q_{3}=\omega
^{2}h_{3}, \,\,\,\, \breve{N}^{2}=\breve{N}^{2}(\tau ,x^{k},y^{3},t).  \notag
\end{eqnarray}%
To be able to construct exact solutions in explicit form we must
parameterize sources with respect to N--adapted frames in certain diagonal
form and with dependencies on $\tau $ and $(x^{i},t),$
\begin{equation}
\widehat{\Upsilon }_{\alpha \beta }=diag[\Upsilon _{i}(\tau );\Upsilon
_{a}(\tau )],\mbox{ for }\Upsilon _{1}(\tau )=\Upsilon _{2}(\tau )=%
\widetilde{\Upsilon }(\tau ,x^{k}), \,\,\, \Upsilon _{3}(\tau )=\Upsilon
_{4}(\tau )=\Upsilon (\tau ,x^{k},t).  \label{sourc2}
\end{equation}%
We suppose also that the scalar and other matter fields can be described
with respect to N--adapted frames when the exact solutions for $\omega =1$
are with Killing symmetry on $\partial _{3}$ which is a space like vector.
Geometric methods of constructing exact solutions elaborated in \cite%
{sv2001,vsingl,svvvey,tgovsv,bubv} for "non--Killing" configurations allow us to
construct very general classes of generic off--diagonal solutions depending
on all spacetime variables.

In N--adapted frames, we consider certain special conditions for the
effective scalar field $\phi $ when $\mathbf{e}_{\alpha }\phi =\phi _{\alpha
}^{[0]}=const.$ This results in $\widehat{\mathbf{D}}^{2}\phi =0.$ We
restrict our models to configurations of $\phi ,$ which can be encoded into
N--connection coefficients
\begin{eqnarray*}
\mathbf{e}_{i}\phi &=&\partial _{i}\phi -n_{i}\phi ^{\ast }-w_{i}\phi
^{\diamond }=\ \phi _{i}^{[0]};\,\,\,\,\phi ^{\ast }=\phi
_{3}^{[0]};\,\,\,\,\phi ^{\diamond }=\phi _{4}^{[0]};\,\,\,\, \\
\mbox{ for }\ \phi _{1}^{[0]} &=&\phi _{2}^{[0]}\mbox{ and }\phi
_{3}^{[0]}=\phi _{4}^{[0]}.
\end{eqnarray*}%
In this part of our work, there are considered brief denotations for partial
derivatives $a^{\bullet }=\partial _{1}a,\,\,\,\,b^{\prime }=\partial
_{2}b,\,\,\,\,$and$\,\,\,\,h^{\diamond }=\partial _{4}h=\partial _{t}h.$ We
used $\phi ^{\ast }=\partial _{3}\phi $ in \cite{muen01}. This way we encode
the scalar field configurations into additional source $~\ \widetilde{%
\Upsilon }(\phi )=~\widetilde{\Lambda }_{0}(\phi )=const$ \ and $\ \Upsilon
(\phi )=\Lambda _{0}(\phi )=const,$ where notations $\widetilde{\Lambda }%
_{0}(\phi )$ and $\Lambda _{0}(\phi )$ emphasize that such constants are
fixed for respective functionals $\widetilde{\Upsilon }(\phi )$ and $%
\Upsilon (\phi ).$

Self--consistent running of N--connection coefficients and scalar fields
under geometric flows can be modeled for
\begin{equation}
\mathbf{e}_{\alpha }\phi (\tau ,x^{k},y^{a})=\phi _{\alpha }^{[0]}+\ \ \phi
_{\bot \alpha }^{[0]}(\tau )  \label{scfl}
\end{equation}%
which modifies the effective h- and v--sources as%
\begin{equation}
~\widetilde{\Upsilon }(\phi )=~\widetilde{\Lambda }_{0}(\phi )+\widetilde{%
\check{\Lambda}}(\phi ,\tau )\mbox{ and }\ \Upsilon =\ \Lambda _{0}(\phi )+\
\check{\Lambda}(\phi ,\tau ).  \label{scfsourc}
\end{equation}%
We can use such effective sources with small values of $\ \widetilde{\check{%
\Lambda}}(\phi ,\tau )$ and $\ \check{\Lambda}(\phi ,\tau )$ and formula (%
\ref{ssource}) in order to compute additional deformations of the evolution
and modified gravitational field equations (see similar discussion in \cite%
{muen01}).

In order to study important physical effects of geometric flows, we use the
possibility to encode matter fields and various supersymmetric contributions
into effective sources by imposing additional conditions on the normalizing
functions $f,$ or any nonholonomic modification to $\widehat{f}.$ Thus for
scalar fields we get,
\begin{equation*}
\ ^{E}\widehat{\mathbf{\Upsilon }}_{\alpha \beta } = \varkappa (\ \ ^{E}%
\widehat{\mathbf{T}}_{\alpha \beta }-\frac{1}{2}\widehat{\mathbf{g}}_{\alpha
\beta }\ \ ^{E}\widehat{\mathbf{T}}) = \varkappa \left[ -\mathbf{e}_{\alpha
}\phi \ \mathbf{e}_{\beta }\phi +\widehat{\mathbf{g}}_{\alpha \beta }(%
\mathbf{e}_{\mu }\phi \ \mathbf{e}^{\mu }\phi +\ ^{E}V)\right] .
\end{equation*}%
We write
\begin{equation}
\ ^{E}\widehat{\mathbf{\Upsilon }}_{\alpha \beta }=\ \varkappa \widehat{%
\mathbf{g}}_{\alpha \beta }\ ^{E}V  \label{effssourc}
\end{equation}%
for two classes of nonholonomic configurations: a) if $\ \mathbf{e}_{\beta
}\phi \approx 0$ is very small (this can be considered for gravitational
field equations) and/or b) we choose such a normalization function in (\ref%
{ffssm}) and (\ref{wfssm}), i.e. for geometric flows, that $|\widehat{%
\mathbf{D}}\widehat{f}|^{2}=\frac{1}{2}\mathbf{e}_{\mu }\phi \mathbf{e}^{\mu
}\phi $. For such non--normalized geometric flows, the effective source are
determined by $\ ^{E}V$ encoding supergravity modifications of geometric
flows and Ricci soliton configurations.

There are also other types of normalization conditions with N--elongated
partial derivatives of the scalar field introduced in (\ref{normalizeq}).
Such conditions can be expressed in the following three equivalent forms
\begin{eqnarray}
K_{I\overline{I}}\mathbf{e}_{\mu }\psi ^{I}\mathbf{e}^{\mu }\psi ^{\overline{%
I}}+|\widehat{\mathbf{D}}\widehat{f}|^{2} &=&0;  \label{threenormaliz} \\
-\frac{3}{4Y^{2}}(\mathcal{D}_{\mu }\mathcal{D}^{\mu }+\mathbf{J}_{\mu }%
\mathbf{J}^{\mu })+\frac{3}{Y}Y_{I\overline{I}}\mathbf{e}_{\mu }\psi ^{I}%
\mathbf{e}^{\mu }\psi ^{\overline{I}}+|\widehat{\mathbf{D}}\widehat{f}|^{2}]
&=&0;  \notag \\
\frac{1}{2}\mathbf{e}_{\mu }\phi \mathbf{e}^{\mu }\phi -\frac{3}{4}e^{-2%
\sqrt{2/3}\phi }\mathbf{J}_{\mu }\mathbf{J}^{\mu }-3e^{-\sqrt{2/3}\phi }%
\mathbf{e}_{\mu }z^{\overline{i}}\mathbf{e}^{\mu }\overline{z}^{\overline{i}%
}+|\widehat{\mathbf{D}}\widehat{f}|^{2}] &=&0.  \notag
\end{eqnarray}%
Supersymmetric contributions change the nonholonomic structure. The effect
of this is that the normalizing function $\widehat{f}$ is different for
different types of nonholonomic variables. Nevertheless, this is not a
problem for finding exact solutions even in the non-normalized form. The
main idea is to construct certain classes of physically important solutions
for certain special normalisation when the decoupling of PDEs is possible.
Fixing certain evolution points for $\tau$, the metrics and connections will
determine corresponding cosmological Ricci soliton configurations and/or
MGTs.

We conclude that supersymmetric $SU(1,1+k)$ models are characterized by
modified geometric evolution equations (\ref{ssymeveq}) when all terms
depending on $\phi $ can be encoded into normalizing function $f$ using the
equation (\ref{normalizeq}) keeping a nontrivial value of potential $\
^{E}V. $ This results in , $\ ^{m}\widehat{\mathcal{L}}\rightarrow \ ^{E}%
\widehat{\mathcal{L}}\rightarrow -\ ^{E}V$ for a N--adapted parametrization $%
\ ^{E}\widehat{\mathbf{\Upsilon }}_{\mu \nu }\rightarrow diag[~\ ^{E}%
\widetilde{\Upsilon },\ ^{E}\Upsilon ].$ In such cases, sources of type (\ref%
{scfsourc}) will be encoded into certain configurations $[~\ ^{E}\widetilde{%
\Upsilon },\ ^{E}\Upsilon ],$ when $~\ ^{E}\widetilde{\Upsilon }\neq \
^{E}\Upsilon .$ For any of normalizations (\ref{threenormaliz}), or source (%
\ref{effssourc}), we shall be able to integrate in explicit form the
geometric flow equations for any $\ ^{E}\widetilde{\Upsilon }(\tau ,x^{k})$
and $\ ^{E}\Upsilon =\ \varkappa \ ^{E}V(\tau ,x^{k},t).$ In particular, we
can consider stationary distributions of matter with $\ ^{E}\widetilde{%
\Upsilon }(\tau ,x^{k})=$ $\ ^{E}\Upsilon =\ \varkappa \ ^{E}V(\tau ,x^{k})$
which under geometric flows and via generic off--diagonal interactions
result in inhomogeneous and locally anisotropic cosmological metrics $%
\widehat{\mathbf{g}}_{\alpha \beta }(\tau ,x^{k},y^{3},t).$

\subsubsection{Geometric flows and time evolution of d--metric coefficients}

In this work, the solutions will be generated with respect to two generating
functions $\psi (\tau ,x^{i})$ $\ $and $\Psi (\tau ,x^{i},t).$ The first one
will be determined to generate solutions for certain 2-d Poisson type
equations with (effective) source modified by geometric flows. To understand
the properties of the second one, we consider a set of coefficients $\alpha
_{\beta }=(\alpha _{i},\alpha _{3}=0,\alpha _{4})$ determined by a
generating function $\Psi $ when
\begin{eqnarray}
\alpha _{i} &=&h_{3}^{\diamond }\partial _{i}\Psi /\Psi , \,\,\,\, \alpha
_{4}=h_{3}^{\diamond }\ \Psi ^{\diamond }/\Psi , \,\,\,\, \gamma =\left( \ln
|h_{3}|^{3/2}/|h_{4}|\right) ^{\diamond }  \label{coefgenf} \\
\mbox{ for }\Psi &:=&h_{3}^{\diamond }/\sqrt{|h_{3}h_{4}|}.  \label{genfunct}
\end{eqnarray}

A tedious computation (see details of such computation in \cite%
{svvvey,tgovsv}) of the N--adapted components of the Ricci tensors for $%
\widehat{\mathbf{D}},$ ansatz for d--metric (\ref{offdans}) and
supersymmetric sources (\ref{sourc2}) prove that we can transform the
nonholonomic Ricci evolution equations (\ref{ssymeveq}) into the followng
system of nonlinear PDEs,
\begin{eqnarray}
\psi ^{\bullet \bullet }+\psi ^{\prime \prime } &=&2(\ ^{E}\widetilde{%
\Upsilon }-\frac{1}{2}\partial _{\tau }\psi ),  \label{rf1} \\
\ \Psi ^{\diamond }h_{3}^{\diamond } &=&2h_{3}h_{4}(\ ^{E}\Upsilon -\partial
_{\tau }\ln |\omega ^{2}h_{3}|)\Psi ,  \label{rf2} \\
\partial _{\tau }\ln |\omega ^{2}h_{3}| &=&\partial _{\tau }\ln |\omega
^{2}h_{4}|=\partial _{\tau }\ln |\breve{N}^{2}|\ ,  \label{rf2a} \\
n_{i}^{\diamond \diamond }+\gamma n_{i}^{\diamond } &=&0,  \label{rf3a} \\
\alpha _{4}w_{i}-\alpha _{i} &=&0,\   \label{rf4a} \\
\mathbf{e}_{k}\omega &=&\partial _{k}\omega +n_{k}\omega ^{\ast
}+w_{k}\omega ^{\diamond }=0,  \label{rfc}
\end{eqnarray}%
The unknown functions are $\psi (\tau ,x^{i}),\omega (\tau
,x^{k},y^{3},t),h_{a}(\tau ,x^{k},t),n_{i}(\tau ,x^{k},t)$ and $n_{i}(\tau
,x^{k},t).$ The first two equations contain possible additional sources
determined by other effective polarized cosmological constants or matter
fields written as $\ ^{E}\widetilde{\Upsilon }(\tau ,x^{k})$ and $\
^{E}\Upsilon (\tau ,x^{k},t).$ The system (\ref{rf1})--(\ref{rfc}) has an
important decoupling property when we can find $\psi$ from the first
equation, $h_3$ and $h_4$ from the second and third equation and so on for
any unknown function.

\subsubsection{Inhomogeneous cosmological Ricci soliton equations}

For such models, we consider fixed evolution parameter configurations with $%
\partial _{\tau }g_{\alpha \beta }=0$ and $\tau =\tau _{0}.$ The equations (%
\ref{rf1}), (\ref{rf2}) and (\ref{rf2a}) transform into self-similar Ricci
soliton equations which for the off--diagonal ansatz can be written in the
form
\begin{eqnarray}
\ \psi ^{\bullet \bullet }( \backepsilon +\ \psi ^{\prime \prime
}(\backepsilon )&=&2\ \ ^{E}\widetilde{\Upsilon }(\backepsilon )\mbox{ and }
\label{rs1} \\
\ \ \ \Psi ^{\diamond }( \backepsilon )\ \ h_{3}^{\diamond }(\backepsilon
)&=&2\ \ h_{3}(\backepsilon )\ h_{4}(\backepsilon )\ \ ^{E}\Upsilon
(\backepsilon )\ \ \Psi (\backepsilon ).  \label{rs2}
\end{eqnarray}%
In this paper, we write $\ \psi (\backepsilon ,\tau ,x^{i})$ and $\ \Psi
(\backepsilon ,\tau ,x^{i},y^{4}=t)$ instead of, respectively, $\ \ _{\flat
}\psi (\tau ,x^{i})$ and $\ _{\flat }\Psi (\tau ,x^{i},y^{3})$ used in \cite%
{muen01} for generating stationary $R^{2}$ Ricci solitons (for a fixed
evolution parameter those solutions depend generically only on space like
coordinates $(x^{i},y^{3})$). For nonhomogeneous and locally anisotropic
Ricci solitonic configurations, the time like evolution is possible being
determined by nonholonomically deformed Einstein equations.

The equation (\ref{rs1}) is just the 2-d Poisson equation which can be
solved in general form for any prescribed source $\ ^{E}\widetilde{\Upsilon }%
(\backepsilon ,x^{k}).$ We note that such an equation is obtained with
respect to N--adapted frames and that the right hand side encode "static"
h--distributions of matter and possible supersymmetric deformations.

The system of nonlinear PDEs (\ref{genfunct}) and (\ref{rs2}) can be
integrated for any source $\ ^{E}\Upsilon (\backepsilon ,x^{k},t)$ encoding
contributions from supergravity modified theories and matter fields. We
point some key differences in constructing such solutions compared to
various classes of solutions studied in \cite%
{svvvey,tgovsv,bubv,vcosmsol1,vcosmsol2,vcosmsol3,vcosmsol4,vcosmsol4a}. The generic
off-diagonal cosmological solutions posses a nonlinear symmetry for
re--definition of generating function, $\Psi \longleftrightarrow \widetilde{%
\Psi },$ and (effective source), $\ ^{E}\Upsilon (\backepsilon
)\longleftrightarrow \Lambda _{0}\neq 0,$ when
\begin{equation*}
\Lambda _{0}(\ \Psi ^{2}(\backepsilon ))^{\diamond }=|\ ^{E}\Upsilon |(%
\widetilde{\Psi }^{2}(\backepsilon ))^{\diamond },\mbox{
or  }\Lambda _{0}\ \Psi ^{2}(\backepsilon )=\ \widetilde{\Psi }%
^{2}(\backepsilon )|\ ^{E}\Upsilon (\backepsilon )|-\int dt\ \widetilde{\Psi
}^{2}(\backepsilon )|\ ^{E}\Upsilon (\backepsilon )|^{\diamond }.
\end{equation*}%
This property can be used for re--definition of generation and source
functions in order to simplify the method of generating exact solutions and
in order to find new classes of solutions by nonholonomic deformations.

For generating off--diagonal locally anisotropic nonsingular cosmological
solutions depending on $y^{4}=t,$ we have to consider generating functions
involving $\ \Psi ^{\diamond }(\backepsilon ).$ The system (\ref{rf2})--(\ref%
{rf4a}) with Killing symmetry on $\partial _{3}$ leads to the following
system of nonlinear PDEs%
\begin{eqnarray}
\ \ \Psi ^{\diamond }( \backepsilon )h_{3}^{\diamond }(\backepsilon ) &=& 2\
h_{3}(\backepsilon )h_{4}(\backepsilon )\ \ ^{E}\Upsilon (\backepsilon )\
\Psi (\backepsilon ),  \label{rsa1} \\
\sqrt{|\ h_{3}(\backepsilon )h_{4}(\backepsilon )|}\ \Psi ( \backepsilon )
&=& \ h_{3}^{\diamond }(\backepsilon ),  \label{rsa1a} \\
\ \ \Psi ^{\diamond }( \backepsilon )\ w_{i}(\backepsilon )-\partial _{i}\
\Psi (\backepsilon ) &=& 0,\   \label{rsa2} \\
\ n_{i}^{\diamond \diamond }( \backepsilon )+\left( \ln |\
h_{3}(\backepsilon )|^{3/2}/|\ h_{4}(\backepsilon )|\right) ^{\diamond }\
n_{i}^{\diamond }(\backepsilon ) &=& 0.\   \label{rsa3}
\end{eqnarray}%
This system for nonholonomic Ricci solitons with explicit dependence on time
like coordinate $t$ and for a fixed parameter $\tau _{0}$ can be integrated
in certain general forms by prescribing $\ ^{E}\Upsilon (\backepsilon )$ and
$\Psi (\backepsilon )$ and finding solutions "step by step".

In order to integrate the above system of equations, we apply the AFDM
following the key steps described in details in \cite{svvvey,tgovsv,bubv,muen01}.
Introducing the function
\begin{equation}
q^{2}:=\epsilon _{3}\epsilon _{4}\ _{\backepsilon }h_{3}\ _{\backepsilon
}h_{4},  \label{qf2}
\end{equation}%
for $\epsilon _{3}=1$ and $\epsilon _{4}=-1$ (which is determined by the
chosen signature of the metric). The equations (\ref{rsa1}) and (\ref{rsa1a}%
) can be expressed respectively as
\begin{equation}
\ \ \Psi ^{\diamond }(\backepsilon )\ h_{3}^{\diamond }(\backepsilon
)=-2q^{2}\ \ \ ^{E}\Upsilon (\backepsilon )\ \Psi (\backepsilon )%
\mbox{
and }\ \ h_{3}^{\diamond }(\backepsilon )=q\ \Psi (\backepsilon ).
\label{rsa1b}
\end{equation}%
Introducing $\ h_{3}^{\diamond }(\backepsilon )$ from the second equation
into the first one, we find%
\begin{equation}
q=-\frac{1}{2}\frac{\ \Psi ^{\diamond }(\backepsilon )}{\ ^{E}\Upsilon
(\backepsilon )}.  \label{qf1}
\end{equation}%
Substituting this value in the second equation of (\ref{rsa1b}) and
integrating on $t,$ we find
\begin{equation}
\ \ h_{3}(\backepsilon )=h_{3}^{[0]}(x^{k})-\frac{1}{4}\int dt\frac{(\Psi
(\backepsilon ))^{\diamond }}{\ ^{E}\Upsilon (\backepsilon )},  \label{h3n}
\end{equation}%
where $h_{3}^{[0]}(x^{k})$ is an integration function. The coefficient $%
h_{4} $ follows from considering (\ref{qf1}), (\ref{qf2}) and formula (\ref%
{h3n}),%
\begin{equation}
\ h_{4}(\backepsilon )=-\frac{1}{4h_{3}}\left[ \frac{\ \ \Psi ^{\diamond
}(\backepsilon )}{\ \ ^{E}\Upsilon (\backepsilon )}\right] ^{2}=\frac{1}{4}%
\frac{\left[ \ \Psi ^{\diamond }(\backepsilon )\right] ^{2}}{(\ ^{E}\Upsilon
(\backepsilon ))^{2}}\left( h_{3}^{[0]}-\frac{1}{4}\int dt\frac{(\Psi
^{2}(\backepsilon ))^{\diamond }}{\ ^{E}\Upsilon (\backepsilon )}\right)
^{-1}.  \label{h4n}
\end{equation}

The N--connection coefficients $n_{i}(x^{k},t)$ are found by integrating two
times on $t$ in (\ref{rsa3}) using the value of coefficient $\gamma $ (\ref%
{coefgenf}) found from (\ref{h3n}) and (\ref{h4n}). The first integration
results in $n_{i}^{\diamond }(\backepsilon )=\ _{2}n_{i}(x^{k})|\
h_{4}(\backepsilon )|/|\ h_{3}(\backepsilon )|^{3/2}$ with an integration
functions $\ _{2}n_{i}(x^{k}).$ Integrating again on $t$ and considering
other set of integration functions $\ _{1}n_{k}(x^{k}),$ we find
\begin{eqnarray*}
\ _{\backepsilon }n_{k}( \backepsilon ,\tau _{0}) &=& \ _{1}n_{k}+\
_{2}n_{k}\int dt\ \frac{\ h_{4}(\backepsilon )}{|h_{3}(\backepsilon )|^{3/2}}%
=\ _{1}n_{k}+\ _{2}\widetilde{n}_{k}\int dt\ \frac{\left[ \ \Psi ^{\diamond
}(\backepsilon )\right] ^{2}}{|\ h_{3}(\backepsilon )|^{5/2}\ (\
^{E}\Upsilon (\backepsilon ))^{2}} \\
&=&\ _{1}n_{k}+\ _{2}n_{k}\int dt\frac{(\Psi (\backepsilon ))^{2}}{(\
^{E}\Upsilon (\backepsilon ))^{2}}\left\vert h_{3}^{[0]}-\frac{1}{4}\int dt%
\frac{(\ \Psi ^{2}(\backepsilon ))^{\diamond }}{\ ^{E}\Upsilon (\backepsilon
)}\right\vert ^{-5/2},
\end{eqnarray*}%
with redefined $\ _{2}n_{i}\rightarrow \ _{2}\widetilde{n}_{k}(x^{i})$
including certain constant coefficients before the source and generating
function. For any generating function $\Psi (\backepsilon ),$ we can solve
the linear algebraic equations (\ref{rsa2}) and find the second sub-set of
N--connection coefficients, $\ w_{i}(\backepsilon )=\partial _{i}\ \Psi
(\backepsilon )/\ \Psi ^{\diamond }(\backepsilon ).$

Summarizing in this subsection, we obtain the set of formulae for computing
the coefficients of a d--metric and a N--connection which determine, in
general form and dynamical in time, Ricci solitons with Killing symmetry on $%
\partial _{3}$ as solutions for the system (\ref{rf1})--(\ref{rf4a}),%
\begin{eqnarray}
\ g_{i}( \backepsilon ,\tau _{0}) &=&e^{\ \psi (\backepsilon ,\tau
_{0},x^{k})}\mbox{ as
a solution of 2-d Poisson equations (\ref{rs1})};  \notag \\
\ h_{3}( \backepsilon ,\tau _{0}) &=& -\frac{1}{4}\frac{(\Psi
^{2}(\backepsilon ))^{\diamond }}{(\ ^{E}\Upsilon (\backepsilon ))^{2}}%
\left( h_{3}^{[0]}-\frac{1}{4}\int dt\frac{(\ \Psi ^{2}(\backepsilon
))^{\diamond }}{\ ^{E}\Upsilon (\backepsilon )}\right) ^{-1};
\label{solut1t} \\
\ h_{4}( \backepsilon ,\tau _{0}) &=& h_{4}^{[0]}(x^{k})-\frac{1}{4}\int dt%
\frac{(\Psi ^{2}(\backepsilon ))^{\diamond }}{\ ^{E}\Upsilon (\backepsilon )}%
;  \notag \\
\ \ n_{k}( \backepsilon ,\tau _{0})&=&\ _{1}n_{k}+\ _{2}n_{k}\int dt\frac{%
(\Psi ^{\diamond }(\backepsilon ))^{2}}{(\ ^{E}\Upsilon (\backepsilon ))^{2}}%
\left\vert h_{3}^{[0]}-\frac{1}{4}\int dt\frac{(\Psi ^{2}(\backepsilon
))^{\diamond }}{\ ^{E}\Upsilon (\backepsilon )}\right\vert ^{-5/2};  \notag
\\
\ w_{i}( \backepsilon ,\tau _{0}) &=& \partial _{i}\Psi (\backepsilon )\ /\
\Psi ^{\diamond }(\backepsilon );  \notag \\
\ \ \omega (\, \backepsilon ,\tau _{0}) &=& \omega \lbrack \ \Psi
(\backepsilon ),\ \ ^{E}\Upsilon (\backepsilon )]%
\mbox{ is any solution of the 1st order
system (\ref{rfc})}.  \notag
\end{eqnarray}%
The d--metric coefficients are determined functionally, respectively, by
generating functions and sources, $\ g_{i}[\psi (\backepsilon ),~\ ^{E}%
\widetilde{\Upsilon }(\backepsilon )]$ and $\ h_{a}[\Psi (\backepsilon ),\ \
^{E}\Upsilon (\backepsilon )].$ We can solve the equations (\ref{rfc}) for a
nontrivial $\ \ \omega ^{2}(\backepsilon )=|\ \ h_{3}(\backepsilon )|^{-1}.$
The quadratic elements for such solutions with nonholonomically induced
torsion are parameterized as
\begin{eqnarray}
ds^{2} &=&\ g_{\alpha \beta }(\backepsilon ,x^{k},t)du^{\alpha }du^{\beta
}=e^{\ \psi (\backepsilon )}[(dx^{1})^{2}+(dx^{2})^{2}]+  \notag \\
&&\ \ \omega ^{2}(\backepsilon )\ h_{3}[\Psi (\backepsilon ),\ \
^{E}\Upsilon (\backepsilon )][dy^{3}+(\ _{1}n_{k}+\ _{2}\widetilde{n}%
_{k}\int dt\frac{(\Psi ^{\diamond }(\backepsilon ))^{2}}{\ (\ ^{E}\Upsilon
(\backepsilon ))^{2}|h_{3}(\backepsilon )|^{5/2}})dx^{k}]^{2}-
\label{riccisolt} \\
&&\frac{\ \omega ^{2}(\backepsilon )}{4\ h_{3}(\backepsilon )}\left[ \frac{\
\Psi ^{\diamond }(\backepsilon )}{\ ^{E}\Upsilon (\backepsilon )}\right]
^{2}\ [dt+\frac{\partial _{i}\ \Psi (\backepsilon )}{\ \Psi ^{\diamond
}(\backepsilon )}dx^{i}]^{2}.  \notag
\end{eqnarray}%
This class of metrics define also exact cosmological inhomogeneous and
locally anisotropic solutions for the canonical d--connection $\widehat{%
\mathbf{D}}$ in $\mathcal{R}^{2}$ gravity with nonholonomically induced
torsion and effective scalar field encoded into a nonholonomically polarized
vacuum. We can impose additional constraints on generating functions and
sources in order to extract Levi--Civita configurations (LC--configurations)
as it is described in section \ref{lccons}. If in above formulas, we take
instead of $\ ^{E}\Upsilon (\backepsilon )=\ ^{E}V,$ for instance, $\
^{E}\Upsilon (\backepsilon )=\ \overline{V}(\phi ,\Phi _{\overline{i}})$ (%
\ref{matterfields}), we generate nonholonomic cosmological Ricci soliton
solutions generalizing the constructions for $SO(1,k+1)$ models of $\mathcal{%
R}^{2}$ gravity theory \cite{kounnas3,muen01}.

\subsubsection{Geometric evolution of cosmological Ricci solitons with
factorized dependence on flow parameter}

For simplicity, we construct solutions with Killing symmetry on $\partial
_{3}$ (when $\omega =1)$ and $\ ^{E}\Upsilon (\backepsilon )=\Upsilon
_{\lbrack 0]}=const.$ If $\ ^{E}\Upsilon (\backepsilon )$ is not constant,
it is a more difficult task to construct exact solutions in explicit form
(we shall consider such examples in section \ref{cgfdmncon}). Considered
here are generated functions, $\psi (\tau ,x^{k})$ and $\Psi (\tau
,x^{k},t), $ and effective sources, $\ \widetilde{\Upsilon }(\tau ,x^{k})$
and $\Upsilon (\tau ,x^{k},t),$ depending in factorized form on flow
parameter $\tau $,
\begin{eqnarray}
&&\psi (\tau ,x^{k})=\ \psi _{\bot }(\tau )+\ _{\backepsilon }\psi
(\backepsilon ,x^{k}),\Psi =\ \ \Psi _{\bot }(\tau )\ \ \Psi (\backepsilon
,x^{k},t),  \notag \\
&&\mbox{ for }h_{3}=\ \ h_{\bot 3}(\tau )\ \ h_{3}(\backepsilon
,x^{k},t),h_{4}=\ \ h_{\bot 4}(\tau )\ \ h_{4}(\backepsilon ,x^{k},t)
\label{factoriz} \\
&&\mbox{ and }~\widetilde{\Upsilon }(\tau ,x^{k})=\ \widetilde{\Upsilon }%
_{\bot }(\tau )+\ \ ^{E}\widetilde{\Upsilon }(\backepsilon ,x^{k}),\Upsilon
(\tau ,x^{k},t)=\ \Upsilon _{\bot }(\tau )+\ ^{E}\Upsilon (\backepsilon
,x^{k},t),  \notag
\end{eqnarray}%
see also (\ref{scfsourc}).

In N--adapted form (for parameterizations (\ref{factoriz})), the system of
geometric flow equations (\ref{rf1})--(\ref{rf4a}) transforms into
\begin{eqnarray}
 \psi ^{\bullet \bullet }( \backepsilon )+\ \psi ^{\prime \prime
}(\backepsilon ) &=& 2  \ ^{E}\widetilde{\Upsilon }(\backepsilon ),\
 \partial _{\tau }\ \ \psi _{\bot }(\tau ) =  2\ \widetilde{\Upsilon }_{\bot }(\tau ); \label{rff1} \\ \
 \Psi ^{\diamond }( \backepsilon )\  h_{3}^{\diamond
}(\backepsilon ) &=& 2\ h_{\bot 4}  h_{4}(\backepsilon ) h_{3}(\backepsilon )( \widetilde{\Upsilon }_{\bot }(\tau )+\Upsilon
_{\lbrack 0]}-\partial _{\tau }\ln |\ h_{\bot 3}|)\ \Psi (\backepsilon ),\
\label{rff2}  \\
\partial _{\tau }\ln |\ \ h_{\bot 3}| &=&\partial _{\tau }\ln |\ h_{\bot
4}|=\partial _{\tau }\ln |\breve{N}^{2}|\ ,  \label{rff2a} \\
\ n_{i}^{\diamond \diamond }( \backepsilon )+\gamma \ (\backepsilon
)n_{i}^{\diamond }(\backepsilon ) &=& 0,  \label{rff3a} \\
\ \alpha _{4}( \backepsilon )w_{i}(\backepsilon )-\alpha _{i}(\backepsilon )
&=& 0.  \label{rff4a}
\end{eqnarray}%
The coefficients $\ \gamma (\backepsilon ),\alpha _{i}(\backepsilon )$ and $%
\alpha _{4}(\backepsilon )$ are computed following formulas (\ref{coefgenf})
and (\ref{genfunct}) by taking corresponding values $\ h_{a}(\backepsilon )$
and $\ \ \Psi (\backepsilon )$ for the cosmological Ricci solitons. We can
choose such classes of integration functions when the N--connection
coefficients are completely determined by the data for a Ricci soliton but
the d--metric coefficients are with factorized $\tau $ evolution. The system
(\ref{rff1})--(\ref{rff4a}) can be integrated "step by step" as follows:

The first equation in (\ref{rff1}) is just the 2-d Poisson equation for $\ \
\psi (\backepsilon ,x^{k})$ corresponding to the solution in the first line
of (\ref{solut1t}). The second equation in that line, for $\psi _{\bot
}(\tau ),$ can be solved and expressed as
\begin{equation*}
e^{\ \psi _{\bot }}=A_{0}e^{2\int d\tau \ \widetilde{\Lambda }(\tau
)},A_{0}=const,
\end{equation*}%
where the integration constant can be set to $A_{0}=1.$ This corresponds to
possible variation of constants induced by effective scalar fields and
effective cosmological constant and other possible matter sources.

To model the evolution of certain Ricci soliton configurations described by (%
\ref{rff2}) it is necessary to satisfy the conditions $|\ h_{\bot 3}|=|\
h_{\bot 4}|$ and
\begin{equation*}
\ \ h_{\bot 4}[1+\frac{1}{\Upsilon _{\lbrack 0]}}\left( \ \Upsilon _{\bot
}(\tau )-\frac{\partial _{\tau }\ h_{\bot 4}}{\ \ h_{\bot 4}}\right) ]\ =1.
\end{equation*}%
The solution of this equation is
\begin{eqnarray}
\ h_{\bot 3}(\tau ) &=&1+\varepsilon _{\bot }(\tau ),\mbox{ for }
\label{tauh3} \\
\ \varepsilon _{\bot }(\tau ) &=&S_{0}e^{\lambda _{1}\tau }+S_{1}e^{\lambda
_{1}\tau }\int d\tau e^{-\lambda _{1}\tau }[\Upsilon _{\bot }(\tau )],
\label{tauh3s}
\end{eqnarray}%
with integration constants $S_{0}$ and $S_{1}$ and $\lambda _{1}:=(\Upsilon
_{\lbrack 0]}).$ \ Such configurations have physical importance if there is
an interval $0\leq \tau \leq \tau _{0}$ with$\ ^{0}h_{4}\rightarrow 1$ for
increasing $\tau _{0}.$ For certain deformations of stationary solutions in
MGTs, \ the function $\ \varepsilon _{\bot }(\tau ),|\ \varepsilon _{\bot
}(\tau )|\ll 1,$ has to be found from experimental data. We can express
\begin{equation*}
h_{a}=|\ \ h_{\bot a}(\tau )|\ \ h_{a}(\backepsilon ,x^{k},t)
\end{equation*}%
where $\ h_{a}(\backepsilon )$ are those from (\ref{solut1t}) but with
\begin{equation*}
\ \ h_{3}(\backepsilon )=h_{3}^{[0]}(x^{k})-\frac{1}{4\Upsilon _{\lbrack 0]}}%
(\ \Psi (\backepsilon ))^{2}\ \ \mbox{ and
}\ \ h_{4}(\backepsilon )=-\frac{1}{4\ \ h_{3}(\backepsilon )}(\frac{\ \Psi
^{\diamond }(\backepsilon )}{\Upsilon _{\lbrack 0]}})^{2}.
\end{equation*}

Summarizing the above solutions, we obtain the following sets of d--metric
coefficients,
\begin{eqnarray}
g_{1}(\tau ,x^{k}) &=&g_{2}=e^{\ \psi _{\bot }}e^{\ \ \ \psi (\backepsilon
,x^{k})}\mbox{ for }e^{\ \psi _{\bot }}=A_{0}e^{2\int d\tau \widetilde{%
\Lambda }(\tau )},A_{0}=const;  \notag \\
h_{3}(\tau ,x^{k},t) &=&|\ h_{\bot 3}(\tau )|\ \left[ h_{3}^{[0]}(x^{k})-%
\frac{1}{4\Upsilon _{\lbrack 0]}}(\ \ _{\backepsilon }\Psi )^{2}\right] ;\
\notag \\
h_{4}(\tau ,x^{k},t) &=&-|\ h_{\bot 3}(\tau )|\frac{1}{4\ h_{3}(\backepsilon
)}\left( \frac{\ \Psi ^{\diamond }(\backepsilon )}{\Upsilon _{\lbrack 0]}}%
\right) ^{2}\mbox{ for }\ h_{\bot 3}(\tau )\mbox{ taken as in
(\ref{tauh3})};  \label{geomflcoef}
\end{eqnarray}%
and N--connection coefficients,%
\begin{eqnarray*}
n_{k}( \backepsilon ,\tau ,x^{i},t) &=&\ _{1}n_{k}(\tau ,x^{i})+\
_{2}n_{k}(\tau ,x^{i})\int dt\ \frac{h_{4}}{|h_{3}|^{3/2}}=\ _{1}n_{k}(\tau
,x^{i})+\ _{2}\widetilde{n}_{k}(\tau ,x^{i})\int dt\ \frac{(\Psi
(\backepsilon ))^{2}}{|h_{4}(\backepsilon )|^{5/2}} \\
&=&\ _{1}n_{k}(\tau ,x^{i})+\ _{2}\widetilde{n}_{k}(\tau ,x^{i})\int dt(\Psi
(\backepsilon ))^{2}\left\vert h_{3}^{[0]}(x^{k})-\frac{1}{4\Upsilon
_{\lbrack 0]}}(\Psi (\backepsilon ))^{2}\right\vert ^{-5/2}; \\
w_{i}( \backepsilon ,x^{k},t) &=& \partial _{i}\ \ \ \Psi (\backepsilon )/\
\ \ \Psi ^{\diamond }(\backepsilon );
\end{eqnarray*}%
for certain re-defined integration and generation functions. In the above
formulae, the generation functions and sources, the integration functions
and the constants depend on the evolution parameter $\tau $. This determines
additional anisotropic polarizations of physical values and running of
physical constants. The off--diagonal terms $w_{i}(\backepsilon ,x^{k},t)$
do not depend on the evolution parameter. If we take $_{2}n_{k}=0$ and $\
_{1}n_{k}=\ _{1}n_{k}(x^{k}),$ not all N--connection coefficients now depend
on geometric evolution parameter being determined by a prescribed
cosmological Ricci soliton configuration. Such configurations can be
restricted to LC ones.

The generic off--diagonal quadratic elements with coefficients (\ref%
{geomflcoef}) are for solutions of relativistic geometric flows that induce
anisotropic polarizations and running constants of cosmological Ricci
solitons,
\begin{eqnarray}
ds^{2} &=&g_{\bot \alpha \beta }(\backepsilon ,\tau ,x^{k},t)du^{\alpha
}du^{\beta }=e^{2\int d\tau \widetilde{\Lambda }(\tau )}e^{\ ^{1}\psi
(x^{k})}[(dx^{1})^{2}+(dx^{2})^{2}]+  \label{runningconst} \\
&&+\{1+\ \varepsilon _{\bot }(\tau )\}\{\ h_{3}(\backepsilon ,x^{k},t)\left[
\ dy^{3}+(\ _{1}n_{k}(\tau ,x^{i})+\ _{2}n_{k}(\tau ,x^{i})\int dt\ \frac{%
h_{4}(\backepsilon )}{|h_{3}(\backepsilon )|^{3/2}})dx^{k}\right] ^{2}
\notag \\
&&\ -\frac{1}{4\ h_{4}(\backepsilon )}\left( \frac{\ \Psi ^{\diamond
}(\backepsilon )}{\Upsilon _{\lbrack 0]}}\right) ^{2}\ \left[ dt+\frac{%
\partial _{i}\ \Psi (\backepsilon )}{\ \ \Psi ^{\diamond }(\backepsilon )}%
dx^{i}\right] ^{2}\},  \notag
\end{eqnarray}

The nonholonomic geometric flow evolution described by the above
cosmological d--metrics is for the canonical d--connection $\widehat{\mathbf{%
D}}$ in $R^{2}$ gravity with effective scalar field encoded into a
nonholonomically polarized vacuum. For realistic cosmological models, we
shall consider, for instance, modifications of the FLRW metric with explicit
running/ polarized physical constants, see section \ref{sec5}. Variations of
constants should be taken from certain observational data and theoretical
arguments (see, for instance, \cite{flambaum}).

\subsubsection{Geometric cosmological flows of effective sources determined
by evolution of d--metric and N--connection coefficients}

\label{cgfdmncon} Applying the AFDM, we can find exact solutions of the
geometric flow equations when the d--metric and N--connection coefficients
and generating functions depend in a general form on evolution parameter $%
\tau .$ For instance, we impose certain physically motivated constraints on
the generating functions and then compute some well--defined horizontal and
vertical effective sources.

We consider an effective source $\widetilde{\Lambda }_{0}(\tau )$ which
constraints the source with supersymmetric corrections $~\ ^{E}\widetilde{%
\Upsilon }$ and the generating function $\partial _{\tau }\psi $ to satisfy
the condition
\begin{equation*}
\ ^{E}\widetilde{\Upsilon }-\frac{1}{2}\partial _{\tau }\psi =\widetilde{%
\Lambda }_{0}(\tau ).
\end{equation*}%
Let us chose $\psi (\tau ,x^{k})$ for (\ref{rf1}) as a solution of
parametric on $\tau $ 2-d Poisson equation,

\begin{equation*}
\psi ^{\bullet \bullet }+\psi ^{\prime \prime }=2\widetilde{\Lambda }%
_{0}(\tau ).
\end{equation*}%
In such an approach, the supersymmetric source (of any type 1-3 with
corresponding scale symmetry on $\widetilde{W}$ considered at the end of
section \ref{sec3}).

Another family of solutions can be generated if we use for the h--metric
factorizations (\ref{factoriz}) with $\psi (\tau ,x^{k})=\ \psi _{\bot
}(\tau )+\psi (\backepsilon ,x^{k})$ and $~\widetilde{\Upsilon }(\tau
,x^{k})=\ \widetilde{\Upsilon }_{\bot }(\tau )+\ \ ^{E}\widetilde{\Upsilon }%
(\backepsilon ,x^{k}).$

The next step is to integrate in certain general forms the equations for the
vertical components of d--metric. We can generate a class of solutions of
geometric flow equations (\ref{rf2})--(\ref{rfc}) for arbitrary $h_{3}(\tau
,x^{k},t),h_{3}^{\diamond }\neq 0$ (it can be considered also as a
generating function) if we introduce an effective cosmological constant $%
\Lambda _{0}$ and treat $\ ^{E}\Upsilon (\tau ,x^{k},t)$ as an effective
source determined by the condition
\begin{equation}
\ ^{E}\Upsilon -\partial _{\tau }\ln |\omega ^{2}h_{3}|=\Lambda _{0}\neq 0.
\label{effsourc}
\end{equation}%
As a result, the system of equations (\ref{genfunct}) and (\ref{rf2}) can be
written as
\begin{equation*}
\sqrt{|h_{4}|}=\frac{h_{3}^{\diamond }}{\Psi \sqrt{|h_{3}|}}\mbox{ and }%
h_{4}=\frac{\Psi ^{\diamond }}{\Psi }\frac{h_{3}^{\diamond }}{2h_{3}}\Lambda
_{0},
\end{equation*}%
for two unknown functions $h_{3}(\tau ,x^{k},t)$ and $\Psi (\tau ,x^{k},t). $
Using the square of the first equation with $h_{a}=\epsilon
_{a}|h_{a}|,\epsilon _{a}=\pm 1,$ we compute
\begin{equation}
\Psi ^{2}=B(\tau ,x^{k})-\frac{4}{\Lambda _{0}}h_{3}\mbox{ and } h_{4} =-%
\frac{(h_{3}^{\diamond })^{2}}{h_{3}[\Lambda _{0}B(\tau ,x^{k})-4h_{3}]}
\label{h3a}
\end{equation}%
for an integration function $B(\tau ,x^{k}).$

A class of solutions of (\ref{rf2a}) can be generated for any $h_{3}=h_{4}$
considering both such values determined by the same generating function $%
h_{4}(\tau ,x^{k},t).$ Using (\ref{h3a}), we can generate solutions with $%
h_{3}\neq h_{4}$ but it is difficult to solve in explicit form such
equations for arbitrary $\omega .$

We find the first set of N--connection coefficients by integrating two times
on $t$ in (\ref{rf3a}) using the condition $h_{3}=h_{4},$
\begin{equation}
n_{k}(\tau ,x^{i},t)=\ _{1}n_{k}(\tau ,x^{i})+\ _{2}\widetilde{n}_{k}(\tau
,x^{i})\int dt\ (\sqrt{|h_{3}|})^{-1}.  \label{na}
\end{equation}%
The algebraic equation (\ref{rf4a}) for the second set of N--connection
coefficients are solved in general form by considering the generating
function from (\ref{h3a}),%
\begin{equation}
w_{i}(\tau ,x^{k},t)=\frac{\partial _{i}\Psi }{\Psi ^{\diamond }}=\frac{%
\partial _{i}\Psi ^{2}}{\partial _{t}(\Psi ^{2})}=(h_{3}^{\diamond
})^{-1}\partial _{i}[h_{3}-\frac{\Lambda _{0}}{4}B(\tau ,x^{k})].  \label{wa}
\end{equation}

It is possible to generalize the class of solutions by introducing a
vertical conformal factor $\omega (\tau ,x^{k},t)$ as a solution of (\ref%
{rfc}),
\begin{equation*}
\partial _{k}\omega +n_{k}(\tau ,x^{i},t)\omega ^{\ast }+w_{k}(\tau
,x^{i},t)\omega ^{\diamond }=0,
\end{equation*}%
for N--connection coefficients determined by $h_{3}$ and respective
integration functions, see (\ref{wa}) an (\ref{na}). In particular, we can
take $\omega =1$ and generate solutions for geometric evolution of
stationary configurations. As solutions of the equations (\ref{rfc}), we can
consider also distributions of a scalar field subject to the conditions (\ref%
{scfl}) and (\ref{scfsourc}) resulting in modifications with an effective
cosmological constant.

Putting together the above formulae, we construct the following quadratic
line element
\begin{eqnarray}
&&ds^{2}=g_{\alpha \beta }(\tau ,x^{k},t)du^{\alpha }du^{\beta }=e^{\psi
(\tau ,x^{k})}[(dx^{1})^{2}+(dx^{2})^{2}]+\omega ^{2}(\tau
,x^{i},t)h_{3}(\tau ,x^{i},t)  \notag \\
&&\{\ [dy^{3}+(\ _{1}n_{k}(\tau ,x^{i})+\ _{2}\widetilde{n}_{k}(\tau
,x^{i})\int dt\sqrt{|h_{3}(\tau ,x^{i},t)|})^{-1})dx^{k}]^{2}+  \notag \\
&& [dt+\frac{\partial _{i}(4h_{3}(\tau ,x^{i},t)-\Lambda _{0}B(\tau ,x^{k}))%
}{4h_{3}^{\diamond }(\tau ,x^{i},t)}dx^{i}]^{2}\}.  \label{ggeomfl}
\end{eqnarray}%
This class of inhomogeneous cosmological solutions and their geometric flows
is general, with evolution of N--connection coefficients and respective
nonholonomically induced torsion. Such geometric flows may transform one
class of cosmological Ricci solitons into another one. In a more general
context, we can consider evolution of a (supersymmetric) MGT into another
MGT, or into certain classes of solutions in GR. We can always put at the
end certain constraints for zero torsion. Mutual transformations of classes
of (off-) diagonal solutions in GR can be described as some particular
examples of such geometric flow evolution models. Using as sources effective
potentials with supersymmetric corrections, $\ ^{E}\Upsilon \rightarrow \
^{E}V,$ we can encode such contributions into a nontrivial effective vacuum
structure of MGT, in particular, for $R^{2}$ or $R$ gravity models.

\subsection{Extracting geometric cosmological evolution of Levi--Civita
configurations}

\label{lccons}We are required to impose additional constraints and
parameterizations for the d--metric and N--connection coefficients in order
to satisfy the zero torsion conditions (\ref{lccond}) and extract
Levi--Civita, LC, configurations. For the generic off--diagonal ansatz (\ref%
{offdans}) with dependence on flow parameter $\tau ,$ such conditions are
equivalent to the system equations (see details in \cite{svvvey,tgovsv,bubv})
\begin{eqnarray}
w_{i}^{\diamond }(\tau ,x^{i},t) &=&\left[ \partial _{i}-w_{i}(\tau
,x^{i},t)\partial _{t}\right] \ln \sqrt{|h_{3}(\tau ,x^{i},t)|},[\partial
_{i}-w_{i}(\tau ,x^{i},t)\partial _{t}]\ln \sqrt{|h_{4}(\tau ,x^{i},t)|}=0,
\label{lccond1} \\
\partial _{i}w_{j}(\tau ,x^{i},t) &=&\partial _{j}w_{i}(\tau
,x^{i},t),n_{i}^{\diamond }(\tau ,x^{i},t)=0,\partial _{i}n_{j}(\tau
,x^{i},t)=\partial _{j}n_{i}(\tau ,x^{i},t),  \notag
\end{eqnarray}%
where we denote in brief $\partial _{4}h_{4}=\partial
_{t}h_{4}=h_{4}^{\diamond }.$ By imposing such constraints, we can always
make zero the d--torsion coefficients (\ref{dtors}).

Following the same procedure as in \cite{muen01}, for simplicity, for
classes of solutions with factorized parameterizations of d--metrics like (%
\ref{factoriz}), we prove that (\ref{lccond1}) can be satisfied if there are
considered constraints such as follows:\

\begin{enumerate}
\item We consider generating functions $\Psi =\check{\Psi}(\tau ,x^{k},t)$
for which
\begin{equation}
(\partial _{i}\check{\Psi}[\tau ])^{\diamond }=\partial _{i}(\check{\Psi}%
^{\diamond }[\tau ]),  \label{cond1}
\end{equation}

\item For sources, we take $\ ^{E}\Upsilon =const$ or express any effective
source as a functional $\ \ ^{E}\Upsilon (x^{i},t)=\ \ ^{E}\Upsilon \lbrack
\check{\Psi}(x^{i},t),\Lambda (\tau )]$ with parametric coordinate
dependencies.

\item The conditions that $\partial _{i}w_{j}=\partial _{j}w_{i}$ can be
expressed in conventional form via any function $\check{A}=\check{A}(\tau
,x^{k},t)$ for which
\begin{equation}
w_{i}=\check{w}_{i}=\partial _{i}\check{\Psi}/\check{\Psi}^{\diamond
}=\partial _{i}\check{A}.  \label{cond2}
\end{equation}%
If $\check{\Psi}$ is prescribed, we solve a system of first order PDEs which
allows to find a function $\check{A}[\check{\Psi}].$

\item We choose $\ _{1}n_{j}(\tau ,x^{k})=\partial _{j}n(\tau ,x^{k})$ for a
function $n(\tau ,x^{k}).$ It is also possible to consider, running on a
geometric flow parameter, like $n(\tau ,x^{k})$ a more generalized class of
integration functions.
\end{enumerate}

Variations of values $\ \ ^{E}\Upsilon (\tau ),~\widetilde{\Lambda }(\tau
),\ \Lambda (\tau ),\mu (\tau )$ etc. have to be taken from observational
data \cite{flambaum}. Dirac's idea on variation of physical constants is
considered in MGTs and geometric flows \cite{vacvis}. We conclude that
geometric flow solutions can explain possible locally anisotropic
polarizations and running of d--metric and N--connection coefficients and
running of fundamental physical constants.

\subsection{Small parametric deformations of off--diagonal cosmological
solutions}

\label{ssedef} We constructed and studied such off--diagonal deformations of
cosmological solutions in various MTGs, see \cite%
{vcosmsol2,vcosmsol3,vcosmsol4,vcosmsol4a,vcosmsol5} and references therein. For
geometric flows of stationary metrics in $R^{2},$ such constructions are
also possible \cite{muen01}. In general, it is not clear if various classes
of inhomogeneous and locally anisotropic cosmological solutions may have
physical importance. There were not constructed such off--diagonal
cosmological solutions for geometric flows and Ricci solitons considered in
other works. We can solve an number of theoretical, phenomenological and
observational problems if we consider sub--sets of solutions generated from
well known ones as deformations depending on a small parameter.
Mathematically, using the AFDM\ various cosmological solutions can be
constructed as exact ones for certain sets of prescribed parameters and
generating functions. In this section, we show how the so-called $%
\varepsilon $--deformation procedure on a small positive parameter $%
\varepsilon $ of an off--diagonal \ "prime" metric, $\mathbf{\mathring{g}}%
(\tau ,x^{i},y^{3},t,)$\textbf{\ }(for physically important known solutions
such a metric can be diagonalizable under coordinate tranfsorms\textbf{)}
into a \ "target" metric, $\mathbf{g}(\tau ,x^{i},y^{3},t),$ works for
cosmological geometric flows and Ricci solitons with generic dependence on $%
t.$

A "prime " pseudo--Riemannian cosmological metric $\mathbf{\mathring{g}}=[%
\mathring{g}_{i},\mathring{h}_{a},\mathring{N}_{b}^{j}]$ can be
parameterized in the form
\begin{eqnarray}
ds^{2} &=&\mathring{g}_{i}(\tau ,x^{k},t)(dx^{i})^{2}+\mathring{h}_{a}(\tau
,x^{k},t)(\mathbf{\mathring{e}}^{a})^{2},  \label{pm} \\
\mathbf{\mathring{e}}^{3} &=&dy^{3}+\mathring{n}_{i}(\tau ,x^{k},t)dx^{i},%
\mathbf{\mathring{e}}^{4}=dt+\mathring{w}_{i}(\tau ,x^{k},t)dx^{i}.  \notag
\end{eqnarray}%
We suppose that such a metric is diagonalizable via a coordinate transform $%
u^{\alpha ^{\prime }}=u^{\alpha ^{\prime }}(u^{\alpha })$%
\begin{equation*}
ds^{2}=\mathring{g}_{i^{\prime }}(\tau ,x^{k^{\prime }},t)(dx^{i^{\prime
}})^{2}+\mathring{h}_{a^{\prime }}(\tau ,x^{k^{\prime }},t)(dy^{a^{\prime
}})^{2},
\end{equation*}%
with $\mathring{w}_{i^{\prime }}=\mathring{n}_{i}^{\prime }=0.$ In arbitrary
systems of coordinates, $\mathring{W}_{\beta \gamma }^{\alpha }(u^{\mu })=0,$
see (\ref{anhcoef}). In order to avoid singular coordinate conditions it is
important to work with "formal" off--diagonal parameterizations when the
coefficients $\mathring{n}_{i}(\tau ,x^{k},t)$ and/or $\mathring{w}_{i}(\tau
,x^{k},t)$ are not zero but the anholonomy coefficients vanish for
diagonalizable solutions. We suppose that some data $(\mathring{g}_{i},%
\mathring{h}_{a})$ may define a cosmological solution in MGT or in GR (for
instance, a Friedman--Lama\^{\i}tre-Robertson-Walker, FLRW, metric). In
general, $\mathbf{\mathring{g}}$ (\ref{pm}) may not be a solution of any
geometric evolution/ gravitational field equations, but it will be
nonholonomically deformed into such solutions.

Let us prove that it is possible to realize a self--consistent procedure of $%
\varepsilon $--deforming prime cosmological metrics into certain target
generic off--diagonal cosmological metrics,
\begin{eqnarray}
ds^{2} &=&\eta _{i}(\tau ,x^{k},t)\mathring{g}_{i}(\tau
,x^{k},t)(dx^{i})^{2}+\eta _{a}(\tau ,x^{k},t)\mathring{g}_{a}(\tau
,x^{k},t)(\mathbf{e}^{a})^{2},  \label{targm} \\
\mathbf{e}^{3} &=&dy^{3}+\ ^{n}\eta _{i}(\tau ,x^{k},t)\mathring{n}_{i}(\tau
,x^{k},t)dx^{i},\mathbf{e}^{4}=dt+\ ^{w}\eta _{i}(\tau ,x^{k},t)\mathring{w}%
_{i}(\tau ,x^{k},t)dx^{i}.  \notag
\end{eqnarray}%
In this formula, the coefficients $(g_{\alpha }=\eta _{\alpha }\mathring{g}%
_{\alpha },\ ^{n}\eta _{i}n_{i},\ ^{w}\eta _{i}\mathring{w}_{i})$ define,
for instance, a cosmological Ricci soliton configuration determined by a
class of solutions (\ref{solut1t}). We can express
\begin{eqnarray}
\eta _{i} &=&\check{\eta}_{i}(\tau ,x^{k},t)[1+\varepsilon \chi _{i}(\tau
,x^{k},t)],\eta _{a}=1+\varepsilon \chi _{a}(\tau ,x^{k},t)\mbox{ and }
\label{smpolariz} \\
\ ^{n}\eta _{i} &=&1+\varepsilon \ \ ^{n}\eta _{i}(\tau ,x^{k},t),\ ^{w}\eta
_{i}=1+\varepsilon \ ^{w}\chi _{i}(\tau ,x^{k},t),  \notag
\end{eqnarray}%
for a small parameter $0\leq \varepsilon \ll 1,$ when $\mathbf{g=}\ \mathbf{g%
}(\varepsilon )\mathbf{=(}g_{i}(\varepsilon ),h_{a}(\varepsilon
),N_{b}^{j}(\varepsilon ))$ (\ref{targm}) $\rightarrow $ $\mathbf{\mathring{g%
}}$ (\ref{pm}) for $\varepsilon \rightarrow 0.$ We emphasize that in general
smooth limits are not possible for such nonholonomic deformations which can
be satisfied for arbitrary parameterizations of sources, generation and
integration functions, integration constants and general (effective)
sources. Nevertheless, it is possible to construct exact solutions for
certain types of well-defined conditions on generating functions resulting
in small deformations with clear physical interpretation of new classes of
solutions.

The goal of the next subsection is to analyze such conditions when $%
\varepsilon $-deformations with nontrivial N--connection coefficients can be
related to new classes of cosmological solutions of Ricci soliton and
gravitational field equations (with possible supergravity corrections) and
their factorized geometric flows.

\subsubsection{$\protect\varepsilon $--deformations for cosmological Ricci
solitons}

We provide a detailed proof for such deformations which can be applied for
similar $\varepsilon $--formulae considered, for instance in \cite%
{vcosmsol3,vcosmsol4,vcosmsol4a,vcosmsol5}. In this work, the formulae are provided in
certain forms which can be applied for generating cosmological Ricci
solitons.

Deformations of $h$-components of the cosmological d--metrics with $\tau
=\tau _{0}$ are characterized by
\begin{equation*}
g_{i}(\varepsilon ,\backepsilon ,x^{k})=\mathring{g}_{i}(x^{k},t)\check{\eta}%
_{i}(x^{k},t)[1+\varepsilon \chi _{i}(x^{k},t)]=e^{\psi (\backepsilon
,x^{k})}
\end{equation*}%
being a solution of the Poisson equation (\ref{rs1}). We do not consider
summation on repeating indices in this formula. For $\ \psi (\backepsilon
)=\ ^{0}\psi (\backepsilon ,x^{k})+\varepsilon \ \ ^{1}\psi (\backepsilon
,x^{k})$ and $\ ^{E}\widetilde{\Upsilon }(x^{k})=\ ^{E0}\widetilde{\Upsilon }%
(\backepsilon ,x^{k})+\varepsilon \ ^{E1}\widetilde{\Upsilon }(\backepsilon
,x^{k}),$ we compute the deformation polarization functions%
\begin{equation}
\chi _{i}=e^{\ \ ^{0}\psi (\backepsilon )}\ \ ^{1}\psi (\backepsilon )/%
\mathring{g}_{i}\check{\eta}_{i}\ \ ^{E0}\widetilde{\Upsilon }(\backepsilon
).  \label{aux2a}
\end{equation}%
In particular, we consider $\ ^{1}\psi (\backepsilon )=0$ and $\ ^{E1}%
\widetilde{\Upsilon }(\backepsilon )=0.$

Let us compute $\varepsilon $--deformations of $v$--components using
formulas for a general source $\ ^{E}\Upsilon (\backepsilon ,x^{i},t)$,
\begin{eqnarray}
\ h_{3}( \backepsilon ,\varepsilon ) &=&h_{3}^{[0]}(x^{k})-\frac{1}{4}\int dt%
\frac{(\Psi ^{2}(\backepsilon ))^{\diamond }}{\ \ ^{E}\Upsilon (\backepsilon
)}=(1+\varepsilon \chi _{3})\mathring{g}_{3}\ ;  \label{h3b} \\
h_{4}( \backepsilon ,\varepsilon ) &=& -\frac{1}{4}\frac{(\ \Psi ^{\diamond
}(\backepsilon ))^{2}}{(\ ^{E}\Upsilon (\backepsilon ))^{2}}\left(
h_{4}^{[0]}-\frac{1}{4}\int dt\frac{(\Psi ^{2}(\backepsilon ))^{\diamond }}{%
\ ^{E}\Upsilon (\backepsilon )}\right) ^{-1}=(1+\varepsilon \ \chi _{4})%
\mathring{g}_{4}.  \label{h4b}
\end{eqnarray}%
We parameterize the generation function in the form
\begin{equation}
\ \Psi (\backepsilon )=\Psi (\backepsilon ,\varepsilon )=\mathring{\Psi}%
(x^{k},t)[1+\varepsilon \chi (x^{k},t)]  \label{aux5}
\end{equation}%
and introduce this value in (\ref{h3b}). This allows us to compute
\begin{equation}
\chi _{3}=-\frac{1}{4\mathring{g}_{3}}\int dt\frac{(\mathring{\Psi}^{2}\chi
)^{\diamond }}{\ ^{E}\Upsilon (\backepsilon )}\mbox{ and }\int dt\frac{(%
\mathring{\Psi}^{2})^{\diamond }}{\ ^{E}\Upsilon (\backepsilon )}%
=4(h_{3}^{[0]}-\mathring{g}_{3}).  \label{cond2a}
\end{equation}%
As a result, we can find $\chi _{3}$ for any deformation $\chi $ from a time
like oriented family of 2-hypersurfaces $t=t(x^{k})$ defined in non-explicit
form from $\mathring{\Psi}=\mathring{\Psi}(x^{k},t)$ when the integration
function $h_{3}^{[0]}(x^{k}),$ the prime value $\mathring{g}_{3}(x^{k})$ and
the fraction $(\mathring{\Psi}^{2})^{\diamond }/\ \ ^{E}\Upsilon
(\backepsilon )$ satisfy the condition (\ref{cond2a}).

The formula for a hypersurface $\mathring{\Psi}(x^{k},t)$ can be written by
choosing an explicit value of $\ ^{E}\Upsilon (\backepsilon ).$ Introducing (%
\ref{aux5}) into (\ref{h3b}), we get%
\begin{equation*}
\chi _{4}=2(\chi +\frac{\mathring{\Psi}}{\mathring{\Psi}^{\diamond }}\chi
^{\diamond })-\chi _{3}=2(\chi +\frac{\mathring{\Psi}}{\mathring{\Psi}%
^{\diamond }}\chi ^{\diamond })+\frac{1}{4\mathring{g}_{3}}\int dt\frac{(%
\mathring{\Psi}^{2}\chi )^{\diamond }}{\ _{\backepsilon }^{E}\Upsilon }.
\end{equation*}%
Thus, we can compute $\chi _{3}$ for any data $\left( \mathring{\Psi},%
\mathring{g}_{3},\chi \right) $. Here we note that the formula for a
compatible source is
\begin{equation*}
\ \ ^{E}\Upsilon (\backepsilon )=\pm \mathring{\Psi}^{\diamond }/2\sqrt{|%
\mathring{g}_{4}h_{3}^{[0]}|},
\end{equation*}%
which transforms (\ref{cond2a}) into a time oriented family of 2-d
hypersurface formulas $t=t(x^{k})$ defined in non-explicit form from
\begin{equation}
\int dt\mathring{\Psi}=\pm (h_{3}^{[0]}-\mathring{g}_{3})/\sqrt{|\mathring{g}%
_{4}h_{3}^{[0]}|}.  \label{cond2b}
\end{equation}

The $\varepsilon $--deformations of d--metric and N--connection coefficients
$\ w_{i}(\backepsilon ,\tau _{0})=\partial _{i}\ \Psi (\backepsilon )/\Psi
^{\diamond }(\backepsilon )$ for nontrivial $\ \mathring{w}_{i}(\tau
_{0})=\partial _{i}\ \mathring{\Psi}/\ \mathring{\Psi}^{\diamond }$ are
found from formulas (\ref{aux5}) and (\ref{smpolariz}). We obtain%
\begin{equation*}
\ ^{w}\chi _{i}=\frac{\partial _{i}(\chi \ \mathring{\Psi})}{\partial _{i}\
\mathring{\Psi}}-\frac{(\chi \ \mathring{\Psi})^{\diamond }}{\mathring{\Psi}%
^{\diamond }}.
\end{equation*}%
In a similar way, we can compute the deformation on the $n$--coefficients
(we omit such details which are not important if we restrict our research
only to LC-configurations).

As a result of (\ref{aux2a})-(\ref{cond2b}), we obtain the following
coefficients for $\varepsilon $--deformations of a prime metric (\ref{pm})
into a target cosmological Ricci soliton metric:
\begin{eqnarray}
\ g_{i}(\varepsilon , \backepsilon ,\tau _{0}) &=& [1+\varepsilon \chi
_{i}(x^{k},t)]\mathring{g}_{i}\check{\eta}_{i}=[1+\varepsilon e^{\ \
^{0}\psi (\backepsilon )}\ \ ^{1}\psi (\backepsilon )/\mathring{g}_{i}\check{%
\eta}_{i}\ \ ^{E0}\widetilde{\Upsilon }(\backepsilon )]\mathring{g}_{i}%
\mbox{ is a solution of  (\ref{rs1})};  \notag \\
\ \ h_{3}(\varepsilon , \backepsilon ,\tau _{0}) &=& [1+\varepsilon \ \chi
_{3}]\mathring{g}_{3}=\left[ 1-\varepsilon \frac{1}{4\mathring{g}_{3}}\int dt%
\frac{(\mathring{\Psi}^{2}\chi )^{\diamond }}{\ \ ^{E}\Upsilon (\backepsilon
)}\right] \mathring{g}_{3};  \notag \\
\ h_{4}(\varepsilon , \backepsilon ,\tau _{0}) &=& [1+\varepsilon \ \chi
_{4}]\mathring{g}_{4}=\left[ 1+\varepsilon \ \left( 2(\chi +\frac{\mathring{%
\Psi}}{\mathring{\Psi}^{\diamond }}\chi ^{\diamond })+\frac{1}{4\mathring{g}%
_{3}}\int dt\frac{(\mathring{\Psi}^{2}\chi )^{\diamond }}{\ ^{E}\Upsilon
(\backepsilon )}\right) \right] \mathring{g}_{4};  \label{ersdef} \\
\ n_{i}(\varepsilon , \backepsilon ,\tau _{0}) &=& [1+\varepsilon \ ^{n}\chi
_{i}]\mathring{n}_{i}=\left[ 1+\varepsilon \ \widetilde{n}_{i}\int dt\ \frac{%
1}{(\ ^{E}\Upsilon (\backepsilon ))^{2}}\left( \chi +\frac{\mathring{\Psi}}{%
\mathring{\Psi}^{\diamond }}\chi ^{\diamond }+\frac{5}{8}\frac{1}{\mathring{g%
}_{3}}\frac{(\mathring{\Psi}^{2}\chi )^{\diamond }}{\ ^{E}\Upsilon
(\backepsilon )}\right) \right] \mathring{n}_{i};  \notag \\
\ \ w_{i}(\varepsilon , \backepsilon ,\tau _{0}) &= & [1+\varepsilon \
^{w}\chi _{i}]\mathring{w}_{i}=\left[ 1+\varepsilon (\frac{\partial
_{i}(\chi \ \mathring{\Psi})}{\partial _{i}\ \mathring{\Psi}}-\frac{(\chi \
\mathring{\Psi})^{\diamond }}{\mathring{\Psi}^{\diamond }})\right] \mathring{%
w}_{i}.  \notag
\end{eqnarray}%
The factor $\ \widetilde{n}_{i}(x^{k})$ is a re-defined integration function
including contributions from the prime metric.

The quadratic element for such inhomogeneous and locally anisotropic
cosmological spaces is parameterized in the form
\begin{eqnarray}
ds^{2} &=&\ \ g_{\alpha \beta }(\varepsilon ,\backepsilon
,x^{k},t)du^{\alpha }du^{\beta }=g_{i}\left( \varepsilon ,\backepsilon
,x^{k}\right) [(dx^{1})^{2}+(dx^{2})^{2}]+  \label{riccisoltdef} \\
&&\ \ h_{3}(\varepsilon ,\backepsilon ,x^{k},t)\ [dy^{3}+n_{i}(\varepsilon
,\backepsilon ,x^{k},t)dx^{i}]^{2}+h_{4}(\varepsilon ,\backepsilon
,x^{k},t)[dt+\ w_{k}\ (\varepsilon ,\backepsilon ,x^{k},t)dx^{k}]^{2},
\notag
\end{eqnarray}%
where the coefficients are taken from (\ref{ersdef}). We can subject
additional constraints in order to extract LC--configurations as we
considered in (\ref{cond1}) and (\ref{cond2}). Further approximations of
type $\ g_{\alpha \beta }(\varepsilon ,\backepsilon ,x^{k},t)\simeq
g_{\alpha \beta }(\varepsilon ,\backepsilon ,t)$ and/or for small
off--diagonal contributions are possible in order to make such solutions to
be compatible with experimental data. The source $\ ^{E}\Upsilon
(\backepsilon ,x^{k},t)$ can be approximated to a potential $\ ^{E}V(x^{k})$
for certain supergravity models as we considered at the end of section \ref%
{sec3}.

\subsubsection{Geometric flow evolution of $\protect\varepsilon $--deformed
cosmological Ricci solitons}

Employing (\ref{riccisoltdef}) into (\ref{runningconst}), we construct
quadratic elements with factorized geometric flow evolution of cosmological
Ricci solitons
\begin{eqnarray}
&&ds^{2} =g_{\bot \alpha \beta }(\varepsilon ,\backepsilon ,\tau
,x^{k},t)du^{\alpha }du^{\beta }=e^{2\int d\tau \widetilde{\Lambda }(\tau
)}\ \ g_{i}\left( \varepsilon ,\backepsilon ,x^{k}\right)
[(dx^{1})^{2}+(dx^{2})^{2}]+[1+\ \varepsilon _{\bot }(\tau )]
\label{runningconstdef} \\
&&\{h_{3}(\varepsilon ,\backepsilon ,x^{k},t)\ \left[ dy^{3}+n_{i}(%
\varepsilon ,\backepsilon ,x^{k},t)dx^{i}\right] ^{2}-\frac{1}{4
h_{3}(\varepsilon ,\backepsilon ,x^{k},t)}\left(\frac{\ \Psi ^{\diamond
}(\varepsilon ,\backepsilon )}{\ ^{E}\Upsilon (\backepsilon )}\right) ^{2}%
\left[dt+ w_{k}(\varepsilon ,\backepsilon ,x^{k},t)dx^{k}\right] ^{2}\}.
\notag
\end{eqnarray}%
For simplicity, we do not linearize in $\varepsilon $ in $(\Psi ^{\diamond
}(\varepsilon ,\backepsilon ))^{2}/\ h_{3}(\varepsilon ,\backepsilon ),$
which is determined by any generating function $\chi (x^{k},t)$ and
respective integration functions. In explicit form, we can consider $%
\varepsilon $--deformations of a FLRW metric with dependence on $\tau $
locally anisotropic polarizations of constants. Such effects result in
modified locally anisotropic cosmological scenarios. Having constructed an
explicit class of solutions, we can study limits of type $\ g_{\alpha \beta
}(\varepsilon ,\backepsilon ,\tau ,x^{k},t)\simeq \ g_{\alpha \beta
}(\varepsilon ,\backepsilon ,\tau ,t)$ for almost diagonal cosmological
metrics. %%%%%%%%55555555555555555555555555555555555555555555555555%

\section{Effective Mimetic $F(A,\mathcal{R}^{2})$ Theories for (Super)
Geometric Evolution and Modified Ricci Solitons}

\label{sec5}In this section, we elaborate on effective mimetic theories
associated with (super) geometric evolution and modified Ricci soliton
models with sources $\ ^{E}\widehat{\mathbf{\Upsilon }}_{\alpha \beta }=\
\varkappa \widehat{\mathbf{g}}_{\alpha \beta }\ ^{E}V$ (\ref{effssourc})
encoding supersymmetric contributions. It is supposed that running and
locally anisotropic polarizations of physical constants and geometric flows
of nontrivial vacuum configurations (described by generic off--diagonal
metrics and for generalized connections) can be modelled equivalently in the
framework of effective mimetic $F(A,\mathcal{R}^{2})$ theories. Cosmological
implications of generic off-diagonal solutions of type
\begin{equation*}
\mathbf{\check{g}}_{_{\alpha \beta }}=\left\{
\begin{array}{cc}
\ g_{\alpha \beta }(\backepsilon ,x^{k},t)\simeq \ g_{\alpha \beta
}(\backepsilon ,t), & \mbox{ see (\ref{riccisolt})}; \\
g_{\bot \alpha \beta }(\backepsilon ,\tau ,x^{k},t)\simeq g_{\bot \alpha
\beta }(\backepsilon ,\tau ,t), & \mbox{ see
(\ref{runningconst})}; \\
\ g_{\alpha \beta }(\tau ,x^{k},t)\simeq \ g_{\alpha \beta }(\tau ,t), & %
\mbox{ see (\ref{ggeomfl})}; \\
\ g_{\alpha \beta }(\varepsilon ,\backepsilon ,x^{k},t)\simeq g_{\alpha
\beta }(\varepsilon ,\backepsilon ,t), & \mbox{ see
(\ref{riccisoltdef})}; \\
g_{\bot \alpha \beta }(\varepsilon ,\backepsilon ,\tau ,x^{k},t)\simeq
g_{\bot \alpha \beta }(\varepsilon ,\backepsilon ,\tau ,t), &
\mbox{
see (\ref{runningconstdef})},%
\end{array}%
\right.
\end{equation*}%
will be studied in the Einstein and Jordan N--adapted and coordinate frames.
We shall also exemplify our findings that are equivalent to $\mathcal{R}^{2}$
and GR theory for well defined nonholonomic constraints extracting
LC--configurations and diagonalizations (after the solutions are constructed
in certain general forms) to metrics of\ type $g_{\alpha \beta }(\tau ,t).$

\subsection{Einstein N--adapted frames, associated mimetic geometric flows
and $F(A,\mathcal{R}^{2})$ gravity}

We use specific normalizatons for the nonholonomic geometric flows when the
resulting cosmological Ricci soliton configurations are described by certain
actions which define equivalently $F(A,\mathcal{R}^{2})$ gravity models. In
addition to working with supersymmetric modifications of $\mathcal{R}^{2}$
theories, we consider ceratin equivalent non-quadratic functional
dependencies in order to be able to study observational effects (different,
or equivalent for certain conditions) in modified gravity theories (see, for
instance, \cite%
{odintsrec,saridak,mavromat,capozzello,basstavr,vcosmsol1,vcosmsol2,vcosmsol3,vcosmsol4,vcosmsol4a,vcosmsol5}
and references therein) with generalized $F$--functional dependence on
certain quantities denoted by $A$ and $\mathcal{R}^{2}$ to be defined below.

\subsubsection{Mimetic transforms of Perelman's potentials and modified
Hamilton's equations}

Let us consider a cosmological geometric flow, or Ricci soliton solution $%
\mathbf{\check{g}}_{_{\alpha \beta }}.$ Redefining the normalization
function $\widehat{f}\rightarrow f$ in modified Perelman's functionals (\ref%
{threenormaliz}) and (\ref{threenormaliz}) for $\widehat{R}[\mathbf{\check{g}%
}],$ $\widehat{\mathbf{D}}[\mathbf{\check{g}}]$ and $\mathbf{\check{e}}_{\mu
}\phi =\widehat{\mathbf{D}}\phi ,$ we get
\begin{eqnarray}
&&-\frac{3}{4}e^{-2\sqrt{2/3}\phi }\mathbf{J}_{\mu }\mathbf{J}^{\mu }-3e^{-%
\sqrt{2/3}\phi }\mathbf{e}_{\mu }z^{\overline{i}}\mathbf{e}^{\mu }\overline{z%
}^{\overline{i}}-\ ^{E}V+|\widehat{\mathbf{D}}\widehat{f}|^{2}=  \notag \\
&& -\ell (\wp )e^{-\sqrt{2/3}\kappa \wp }\mathbf{\check{g}}^{\mu \nu }(%
\mathbf{\check{e}}_{\mu }\wp )(\mathbf{\check{e}}^{\mu }\wp )-\ ^{E}\check{V}%
\ (\tau ,\phi ,\wp )+|\widehat{\mathbf{D}}f|^{2},  \label{renorm2}
\end{eqnarray}%
where%
\begin{equation}
\ ^{E}\check{V}\ (\tau ,\phi ,\wp )=-U(\tau ,\phi )-\ ^{E}\widetilde{V}\
(\tau ,\wp )e^{-2\sqrt{2/3}\kappa \phi }+\ell (\tau ,\wp )e^{-2\sqrt{2/3}%
\kappa \phi }.  \label{suppot2}
\end{equation}%
We introduce new nonholonomic variables with two effective scalar fields, $%
\phi (u^{\alpha })$ and $\wp (u^{\alpha }),$ with explicit gravitational
coupling constant $\kappa $ and Lagrange multiplier $\ell (\wp )$ which in
the cosmological approximation is a function of type $\ell (t).$ The
potential $U(\phi )$ is considered as a generalization of the Starobinsky
potential \cite{star1}. In general, the relation between $\ ^{E}V(\tilde{W})$
and $\ ^{E}\check{V}(\phi ,\wp )$ is not explicit and depend on the type of
(anti) de Sitter / flat scenarios we consider to be determined by
supergravity corrections as it is studied in \cite{kounnas3}, see points 1-3
after formula (\ref{effsupact}). We analyze models with effective running of
constants of the type $b(\tau ),c_{\overline{i}\overline{j}\overline{k}%
}(\tau ),c_{\overline{k}}(\tau )$ etc. and fields $z^{\overline{i}}(\tau
,u^{\alpha })=z^{\overline{i}}(\wp )$ in sources for solutions $\mathbf{%
\check{g}}_{_{\alpha \beta }}$ encoding possible geometric evolution and
that can be equivalently modelled by $F(A,\mathcal{R}^{2})\simeq F(A,%
\mathcal{R})$ with very similar cosmological properties as it is discussed
in \cite{odintsrec,vcosmsol1,vcosmsol2,vcosmsol3,vcosmsol4,vcosmsol4a,vcosmsol5}. For
simplicity, we consider
\begin{eqnarray}
\ ^{E}\check{V}\ (\tau ,\phi ,\wp ) &=&\ ^{E}V(\tau ,\tilde{W})+\widetilde{%
\mathbf{Z}}  \label{supotmod} \\
\ ^{E}\mathbf{\check{\Upsilon}}_{\alpha \beta } &=&\ \varkappa \mathbf{%
\check{g}}_{\alpha \beta }\ ^{E}\check{V},  \notag
\end{eqnarray}%
in order to define below a corresponding class of functionals $F$ for a
fixed value $\tau _{0},$ i.e. for cosmological soliton configurations with
sources (\ref{effssourc}). In the above formulae, $\widetilde{\mathbf{Z}}$
include possible distortions and off--diagonal terms of the Ricci d--tensor
computed for $\widehat{\mathbf{D}}[\mathbf{\check{g}}_{_{\alpha \beta }}]$
and $\mathbf{\check{N}}=\{\check{N}_{i}^{a}\}$ computed for off--diagonal
cosmological solutions introduced in (\ref{distr}).

The system of modified Hamilton equations (\ref{ssymeveq}) corresponding to
normalization (\ref{renorm2}) and potential (\ref{supotmod}) which is
derived from modified Perelman's potentials functionals (\ref{threenormaliz}%
) and (\ref{threenormaliz}) can be written as
\begin{eqnarray}
\partial _{\tau }\check{g}_{ij} &=&-2[\widehat{R}_{ij}(\check{g}_{kl})-\ ^{E}%
\check{\Upsilon}_{ij}],\   \label{mhameqmim} \\
\partial _{\tau }\check{g}_{ab} &=&-2[\widehat{R}_{ab}(\check{g}_{cd})-\ ^{E}%
\check{\Upsilon}_{ab}],  \notag \\
\widehat{R}_{ia} &=&\widehat{R}_{ai}=0;\widehat{R}_{ij}=\widehat{R}_{ji};%
\widehat{R}_{ab}=\widehat{R}_{ba};  \notag \\
\partial _{\tau }\phi &=&-2(\widehat{\square }\phi +\frac{\partial \ \ ^{E}%
\check{V}}{\partial \phi });  \notag \\
\partial _{\tau }\wp &=&-2(\widehat{\square }\wp +\frac{\partial \ ^{E}%
\check{V}}{\partial \wp });  \notag \\
\partial _{\tau }f &=&-\widehat{\square }f+\left\vert \widehat{\mathbf{D}}%
f\right\vert ^{2}-\ ^{h}\widehat{R}-\ ^{v}\widehat{R}+\widehat{\square }\phi
+\widehat{\square }\wp +\ ^{E}\check{V},  \notag
\end{eqnarray}%
where the d 'Alambert operator $\widehat{\square }$ is defined by $\widehat{%
\mathbf{D}}[\mathbf{\check{g}}_{_{\alpha \beta }}].$ In the above notice the
effective nonholonomically modification under geometric flows Ricci tensors
and the two scalar fields involvement.

\subsubsection{Cosmological Ricci solitons and mimetic $F(A,\mathcal{R}^{2})$
gravity with supergravity modified potential}

For self--similar configurations of the system (\ref{mhameqmim}), we obtain
cosmological Ricci soliton spacetimes described by the action
\begin{equation}
\ ^{E}S=\int d^{4}u\sqrt{|\mathbf{\check{g}}|}\left[ \frac{1}{2\kappa ^{2}}%
\widehat{\mathbf{R}}[\mathbf{\check{g}}]-\frac{1}{2}\mathbf{\check{g}}^{\mu
\nu }(\mathbf{\check{e}}_{\mu }\phi )(\mathbf{\check{e}}_{\nu }\phi )-\ell
(\wp )e^{-\sqrt{2/3}\kappa \phi }\mathbf{\check{g}}^{\mu \nu }(\mathbf{%
\check{e}}_{\mu }\wp )(\mathbf{\check{e}}^{\mu }\wp )-\ ^{E}\check{V}\ (\tau
,\phi ,\wp )\right] ,  \label{cosmsoltwopact}
\end{equation}%
which is equivalent (up to nonholonomic deformations and re--definition of
normalization functions) to the action (\ref{effsupact}). This implies that
phenomenologically such models are similar to those from \cite%
{ferrara1,kallosh,ellis}. \ The action (\ref{effsupact}) describes a
nonholonomic supersymmetric and imaginary deformation and generalization of
the Starobinsky model \cite{star1} with potential
\begin{equation*}
U(\tau =\tau _{0},\phi )=\ ^{\phi }V(\phi )=\frac{3}{4}\kappa
^{2}M^{2}(1-e^{-2\sqrt{2/3}\kappa \phi })^{2},\mbox{ for }\mu =\frac{3}{4}%
\kappa ^{2}M^{2}=const,
\end{equation*}%
see (\ref{starobpot}). It is possible to calculate the observational induced
inflation for such two scalar models and $U(\phi )$ by N--adapting the
techniques developed in Refs. \cite{kaiser1,kaiser2,renaux}.

For any cosmological configuration determined by geometric and/or
off--diagonal flows, we can define an analogous MGT following such a
procedure:

Let us consider a functional $F(\tau ,A)$ and express
\begin{equation*}
U(\tau ,\phi )=\frac{A}{F^{\circ }(\tau ,A)}-\frac{F(\tau ,A)}{[F^{\circ
}(\tau ,A)]^{2}}
\end{equation*}%
for an auxiliary scalar field $A$ as in Ref. \cite{odintsrec}, but in our
case we may take at the end $A=\widehat{R}$ defined by $\widehat{\mathbf{D}}$%
, or (for certain additional assumptions) $A=R$ defined by $\nabla .$ In
this formula, $F^{\circ }:=dF/dA.$ For $\tau =\tau _{0},$ we consider that
such an auxiliary parametrization transforms (\ref{cosmsoltwopact}) into the
action%
\begin{eqnarray*}
\ ^{E}\widetilde{S} &=&\int d^{4}u\sqrt{|\mathbf{\check{g}}|}\{\frac{1}{%
2\kappa ^{2}}\widehat{\mathbf{R}}[\mathbf{\check{g}}]-\frac{1}{2}[\frac{%
F^{\circ \circ }(\tau ,A)}{F^{\circ }(\tau ,A)}]^{2}\mathbf{\check{g}}^{\mu
\nu }(\mathbf{\check{e}}_{\mu }A)(\mathbf{\check{e}}_{\nu }A)-\frac{1}{%
2\kappa ^{2}}[\frac{A}{F^{\circ }(\tau ,A)}-\frac{F(\tau ,A)}{[F^{\circ
}(\tau ,A)]^{2}}] \\
&&-\ ^{E}\widetilde{V}\ (\tau ,\wp )e^{-2\sqrt{2/3}\kappa \phi }+\ell (\tau
,\wp )e^{-\sqrt{2/3}\kappa \phi }\mathbf{\check{g}}^{\mu \nu }(\mathbf{%
\check{e}}_{\mu }\wp )(\mathbf{\check{e}}^{\mu }\wp )+\ell (\tau ,\wp )e^{-2%
\sqrt{2/3}\kappa \phi }\}.
\end{eqnarray*}%
In N--adapted frames $\mathbf{\check{e}}_{\mu },$ we imply the following
transforms and identifications:%
\begin{eqnarray*}
g_{\mu \nu } &=&e^{-2\sqrt{2/3}\kappa \phi }\mathbf{\check{g}}_{\mu \nu },%
\sqrt{|g|}=e^{-2\sqrt{2/3}\kappa \phi }\sqrt{|\mathbf{\check{g}}|} \\
\mbox{ and }\phi (\tau ,u^{\alpha }) &=&-\sqrt{\frac{3}{2\kappa ^{2}}}\ln
F^{\circ }(\tau ,A),\mbox{ i.e. }\phi (\tau ,t)=-\sqrt{\frac{3}{2\kappa ^{2}}%
}\ln F^{\circ }(\tau ,A(\tau ,t)).
\end{eqnarray*}%
The metric $g_{\mu \nu }$ is considered in the so--called Jordan metric if
we work (it is preferred for observational computations) with local
coordinates. The action $\ ^{E}\widetilde{S}$ transforms into
\begin{equation*}
\ ^{E,F}\widetilde{S}=\int d^{4}u\sqrt{|\mathbf{\check{g}}|}\{F^{\circ
}(\tau ,A)\left[ (\widehat{\mathbf{R}}[\mathbf{\check{g}}]-A\right] +F(A)-\
^{E}\widetilde{V}\ (\tau ,\wp )+\ell (\tau ,\wp )[g^{\mu \nu }(\partial
_{\mu }\wp )(\partial _{\nu }\wp )+1]\}.
\end{equation*}%
By varying this action with respect to $A,$ we obtain $A=\widehat{\mathbf{R}}%
,$ which means that this action is mathematically equivalent to the Jordan
frame action for a mimetic $F(\widehat{\mathbf{R}})$ theory, in our case,
with effective scalar potential $\ ^{E}\widetilde{V}\ (\tau ,\wp )$
(encoding contributions of geometric flows, off--diagonal terms and
supersymmetric sources) and a Lagrange multiplier $\ell (\tau ,\wp )$. In an
equivalent form, we consider that $\ ^{E}\widetilde{V}\ (\tau ,\wp )$ encode
all distortions of connections and work with a $F(R)$ theory with LC scalar
curvature $R$, when
\begin{equation}
\ ^{F}S=\int d^{4}u\sqrt{|g|}\{F[R(g_{\mu \nu })]-\ ^{E}\widetilde{V}\ (\tau
,\wp )+\ell (\tau ,\wp )[g^{\mu \nu }(\partial _{\mu }\wp )(\partial _{\nu
}\wp )+1]\}.  \label{actmimgr}
\end{equation}%
For small $\varepsilon $--deformations studied in section \ref{ssedef}, we
assume that $g_{\mu \nu }$ is a physical metric in the FLRW form with
possible additional dependence on $\tau $-parameter (to be determined in
next subsections)%
\begin{equation}
ds^{2}=a^{2}(\tau ,t)[(dx^{1})^{2}+(dx^{2})^{2}+(dy^{3})^{2}]-dt^{2},
\label{flat}
\end{equation}%
when $R=6(H^{\diamond }+2H^{2})$ is computed for the Hubble rate $H(\tau
,t)=a^{\diamond }/a.$ Mimetic gravity was introduced in \cite{mimet1,mimet2}%
. In this section, we derived the action (\ref{actmimgr}) which is similar
to the F(R) modification elaborated in \cite{odints2} with the difference
that in our case $\ ^{E}\widetilde{V}\ (\tau ,\wp )$ and $\ell (\tau ,\wp )$
encode possible contributions from (supersymmetric) geometric flows and
off--diagonal cosmological Ricci solitons. Different (supersymmetric)
geometric flow, Ricci soliton and MGT models are characterized by different $%
\ ^{E}\widetilde{V}\ (\tau ,\wp )$ and $\ell (\tau ,\wp );$ such values are
different also for different classes of the solutions of the same theory.

In non-explicit form, the physical metric is a functional on generic
off--diagonal auxiliary cosmological metric $\mathbf{\check{g}}_{\mu \nu }$
and auxiliary scalar field $\wp (\tau ,u^{\alpha }),$ i.e. $g_{\mu \nu
}=g_{\mu \nu }[\mathbf{\check{g}}_{\alpha \beta },\wp ],$ subject to
conditions%
\begin{equation}
g^{\mu \nu }(\mathbf{\check{g}}^{\alpha \beta },\wp )(\partial _{\mu }\wp
)(\partial _{\nu }\wp )=\mathbf{g}^{\mu \nu }(\mathbf{\check{g}}^{\alpha
\beta },\wp )(\mathbf{e}_{\mu }\wp )(\mathbf{e}_{\nu }\wp )=-1.
\label{scalcond}
\end{equation}%
Upon variation of $\ ^{F}S$ with respect to $g_{\mu \nu },$ we obtain the
effective gravitational field equations,%
\begin{eqnarray}
&&\frac{1}{2}g_{\mu \nu }F(R)-R_{\mu \nu }F^{\circ }(R)+[\nabla _{\mu
}\nabla _{\nu }-g_{\mu \nu }\square ]F^{\circ }(R)+  \label{fmodfeq} \\
&&\frac{1}{2}g_{\mu \nu }[-\ ^{E}\widetilde{V}\ (\tau ,\wp )+\ell (\tau ,\wp
)(g^{\alpha \beta }(\partial _{\alpha }\wp )(\partial _{\beta }\wp
)+1)]-\ell (\tau ,\wp )(\partial _{\mu }\wp )(\partial _{\nu }\wp )=0.
\notag
\end{eqnarray}%
We note that exact cosmological solutions in various MGTs were studied in
\cite{vcosmsol1,vcosmsol2,vcosmsol3,vcosmsol4,vcosmsol4a,vcosmsol5} for $\nabla _{\mu
}\rightarrow \widehat{\mathbf{D}}_{\mu }.$ In this work, it is more
convenient to work directly with LC-configurations of certain modified
theories and consider physical metrics for effective F--modifications. The
variation of action (\ref{actmimgr}) in terms of $\nabla _{\mu }$ with
respect to the effective scalar field $\wp $ results in the equation
\begin{equation}
\nabla ^{\mu }(\ell \partial _{\mu }\wp )=-\frac{1}{2}\frac{\partial \ ^{E}%
\widetilde{V}}{\partial \wp }.  \label{scfeq2}
\end{equation}
Finally, we emphasize that the equations (\ref{scalcond}) can be obtained by
varying the action (\ref{actmimgr}) on the Lagrange multiplier $\ell .$

\subsection{Reconstructing (super) geometric flow and MGTs cosmological
scenarios}

Resorting to observational cosmological data, we derive certain conditions
relating to the compatibility of effective MGTs encoding (super) geometric
nonholonomic flows with dependence on $\tau $--parameter.

\subsubsection{Compatible effective supersymmetric potentials, MGTs and
Lagrange multipliers}

For a flat FLRW background (\ref{flat}) and approximation that the auxiliary
metrics and scalar field depend only on flow parameter $\tau $ and
cosmological time $t,$ we write the equations (\ref{fmodfeq}), (\ref{scfeq2}%
) and (\ref{scalcond}) (see similar details in \cite{odintsrec}),
respectively, in the following form
\begin{eqnarray*}
6(H^{\diamond }+H^{2})F^{\circ }(R)-6H[F^{\circ }(R)]^{\diamond }-F(R)-\ell
\lbrack (\wp ^{\diamond })^{2}+1]+\ ^{E}\widetilde{V}(\tau ,\wp ) &=&0, \\
2[F^{\circ }(R)]^{\diamond \diamond }+4H[F^{\circ }(R)]^{\diamond
}+F(R)-2(H^{\diamond }+3H^{2})-\ell \lbrack (\wp ^{\diamond })^{2}-1]-\ ^{E}%
\widetilde{V}(\tau ,\wp ) &=&0, \\
2(\ell \wp ^{\diamond })^{\diamond }+6H\ell \wp ^{\diamond }-\frac{\partial
\ ^{E}\widetilde{V}}{\partial \wp } &=&0, \\
(\wp ^{\diamond })^{2}-1 &=&0.
\end{eqnarray*}%
The potential $\ ^{E}\widetilde{V}(\tau ,\wp )$ can be considered as an
effective source of all possible geometric flows, nonholonomic deformations,
running of constants depending on $\tau $ for certain admissible
modifications of mimetic type gravity theories.

In \cite{mimet1}, it was suggested the identification $t=\wp $ which is used
also in F(R) modified gravity \cite{odintsrec}. In our case, such
identifications in the first and second equations of the above system allow
to express the effective potential and Lagrange multiplier by corresponding
formulas depending on $F(R)$ and $H,$%
\begin{eqnarray*}
\ ^{E}\widetilde{V}(\tau ,\wp &=&t)=2[F^{\circ }(R)]^{\diamond \diamond
}+4H[F^{\circ }(R)]^{\diamond }+F(R)-2(H^{\diamond }+3H^{2})\mbox{ and } \\
\ell (\tau ,t) &=&-3H[F^{\circ }(R)]^{\diamond }+6(H^{\diamond }+H^{2})-%
\frac{1}{2}F(R).
\end{eqnarray*}%
Prescribing any value $\ ^{E}\widetilde{V}$ and possible modification $F(R),$
for instance, we can analyse if there is a viable $\ell (\tau ,t)$ for
certain $H$ taken from experimental data and compatible with the assumption
that $\tau =\tau _{0}.$ In different MGTs, one studies functionals of the
type
\begin{equation}
F(R)=p_{3}R+p_{1}R^{p_{2}}+c_{1}R^{c_{2}}  \label{classmgt}
\end{equation}%
with arbitrary parameters $p_{1},p_{2},p_{3},c_{1},c_{2}.$ In particular, we
can consider that $F(R)=R^{2}.$

The novelty of the reconstruction methods considered in \cite%
{odintsrec,vcosmsol1,vcosmsol2,vcosmsol3,vcosmsol4,vcosmsol4a,vcosmsol5} is that by
choosing any possible generic off--diagonal cosmological solution for a
modified (super) geometric flow and/or MGT we can produce any cosmological
model (in general, nonhomogeneous and local anisotropic and with nontrivial
torsion) and produce certain cosmological evolution scenarios which allows
us to verify their concordance with experimental data.

\subsubsection{Observational indices for double scalar models encoding
(super) geometric flows and Ricci solitons}

Our purpose is to find if a (super) symmetric mimetic effective scalar
potential $\ ^{E}\widetilde{V}(\tau ,\wp =t)$ and corresponding Lagrange
multiplier, both depending on geometric evolution parameter $\tau .$ Such a
formalism was elaborated for the calculation of the Jordan frame
observational indices \cite{bamba,much3,odintsrec} \ and developed for
generic off--diagonal cosmological solutions in MGTs in \cite%
{vcosmsol1,vcosmsol2,vcosmsol3,vcosmsol4,vcosmsol4a,vcosmsol5}.

For observational indices and all physical quantities, we can use functions
of the $e$-folding number $\mathcal{N}$ $\ $(we write a calligraphic symbol
in order to avoid confusions with $\mathbf{N}$ and $N$ used for the
N--connection and its coefficients) instead of cosmic time $t.$ Let us
denote
\begin{eqnarray*}
P^{\diamond } &=&\frac{\partial P}{\partial t}=H(\mathcal{N})\frac{\partial P%
}{\partial \mathcal{N}}=H(\mathcal{N})P^{\triangleleft }=\mathcal{H}%
P^{\triangleleft },\mbox{ for }P^{\triangleleft }:=\frac{\partial P}{%
\partial \mathcal{N}},\mathcal{H}:=H(\mathcal{N}); \\
P^{\diamond \diamond } &=&\frac{\partial ^{2}P}{\partial t^{2}}=H^{2}(%
\mathcal{N})P^{\triangleleft \triangleleft }+H(\mathcal{N})H^{\triangleleft
}P=\mathcal{H}^{2}P^{\triangleleft \triangleleft }+\mathcal{H}%
P^{\triangleleft },\mbox{ for }P^{\triangleleft \triangleleft }:=\frac{%
\partial ^{2}P}{\partial \mathcal{N}^{2}},\mathcal{H}^{\triangleleft
}:=H^{\triangleleft }(\mathcal{N}).
\end{eqnarray*}%
As a result of this, the slow-roll indices can be written in the following
form (such relations are valid for very small values)%
\begin{eqnarray*}
\epsilon (\tau ,\mathcal{N}) &=&-\frac{\mathcal{H}}{4\mathcal{H}%
^{\triangleleft }}\left( \frac{\mathcal{H}^{\triangleleft \triangleleft }/%
\mathcal{H}+6\mathcal{H}^{\triangleleft }/\mathcal{H+}(\mathcal{H}%
^{\triangleleft }/\mathcal{H)}^{2}}{3+\mathcal{H}^{\triangleleft }/\mathcal{H%
}}\right) ^{2}\ll 1, \\
\eta (\tau ,\mathcal{N}) &=&-\frac{18\mathcal{H}^{\triangleleft }/\mathcal{H}%
+6\mathcal{H}^{\triangleleft \triangleleft }/\mathcal{H}+(\mathcal{H}%
^{\triangleleft }/\mathcal{H})^{2}-(\mathcal{H}^{\triangleleft \triangleleft
}/\mathcal{H}^{\triangleleft })^{2}+6\mathcal{H}^{\triangleleft
\triangleleft }/\mathcal{H}^{\triangleleft }+2\mathcal{H}^{\triangleleft
\triangleleft \triangleleft }/\mathcal{H}^{\triangleleft }}{4(3+\mathcal{H}%
^{\triangleleft }/\mathcal{H})}\ll 1.
\end{eqnarray*}%
Considering the same formalism \cite{bamba,odintsrec} but with dependence on
$\tau ,$ we obtain respective formulae for the spectral index of primordial
curvature perturbations, $n_{s},$ and the scalar-to-tensor ratio, $r,$%
\begin{equation*}
n_{2}\simeq 1-6\epsilon (\tau ,\mathcal{N})+2\eta (\tau ,\mathcal{N})%
\mbox{
and }r=16\epsilon (\tau ,\mathcal{N}).
\end{equation*}

Let us approximate
\begin{equation*}
\mathcal{H}=[-q_{0}(\tau )e^{b_{0}(\tau )\mathcal{N}}+q_{1}(\tau
)]^{b_{1}(\tau )},
\end{equation*}%
where the values $q_{0}(\tau ),b_{0}(\tau ),q_{1}(\tau )$ and $b_{1}(\tau )$
may run with respect to geometric flow parameter and have to be chosen to
obtain concordance with observational data. For instance, two sets of values
\begin{equation*}
\begin{array}{ccc}
q_{0}= & 0.5,\mbox{ or } & 0.5; \\
b_{0}= & 0.024,\mbox{ or } & 0.0222; \\
q_{1}= & 12,\mbox{ or } & 10; \\
b_{1}= & 0.5,\mbox{ or } & 1;%
\end{array}%
\end{equation*}%
are compatible with the recent Plank observational data \cite{planck}: $%
n_{s}=0.9644\pm 0.0049$ and $r<0.10.$ The second column results in $%
n_{s}=0.96567\pm 0.0049$ and $r<0.0640848$ as it was computed in \cite%
{odintsrec}. There is certain flexibility in the $\tau $--running of
coefficients and type of MGTs. In that work, there were studied in details
the issues of existence and stability of de Sitter solution in the framework
of mimetic $F(R)$ gravity with potential and Lagrange multiplier for various
values of constants/parameters $p_{1},p_{2},p_{3},c_{1},c_{2}$ in functional
(\ref{classmgt}). If any of such models is not compatible with observations
for certain fixed data for $\tau =\tau _{0},$ we can obtain compatibility
(also exit from inflation, stability of dS vacua etc.) under evolution of
parameters and because of generic off--diagonal interactions which change
the effective functional $F(R).$

\subsection{Double scalar model calculations of observational indices for
cosmological Ricci (super) solitons and mimetic $F(A,\mathcal{R}^{2})$
gravity}

We can chose nonholonomic geometric flow variables and constraints for
cosmological Ricci solitons when the results of computation of physically
important values are related to observational indices in cosmology. For
well--defined conditions, such results can be related to and compared with
those in mimetic $F(A,\mathcal{R}^{2})$ gravity.

\subsubsection{Effective configuration space}

Let us provide an example of how the slow--roll indices for the two scalar
model of (\ref{cosmsoltwopact}) for the Starobinsky potential (\ref%
{starobpot}) work. For computations, we use the methods elaborated in \cite%
{kaiser2,odintsrec}. Consider a two--scalar formalism which is an example
with potentials considered in \ref{ssmscal} but for $\ ^{E}\check{V}\ (\tau
,\phi ,\wp )=\ ^{E}\check{V}\ (\tau ,\phi ^{I}),$ with labels $I,J=1,2$ and $%
(\phi ^{1}=\phi ,\phi ^{2}=\wp ).$ Introducing a conventional scalar field
configuration space endowed with metric
\begin{equation*}
G_{IJ}(\tau ,\phi ^{I})=\left(
\begin{array}{cc}
1 & 0 \\
0 & e^{-\sqrt{6}\kappa \phi ^{1}}\ell (\tau ,\phi ^{2})%
\end{array}%
\right) ,
\end{equation*}%
the action for our model can be written as
\begin{equation}
S=\int d^{4}u\sqrt{|\mathbf{\check{g}}|}\left[ \frac{1}{2\kappa ^{2}}%
\widehat{\mathbf{R}}[\mathbf{\check{g}}]-\frac{1}{2}G_{IJ}(\phi ^{K})\mathbf{%
\check{g}}^{\mu \nu }(\mathbf{\check{e}}_{\mu }\phi ^{I})(\mathbf{\check{e}}%
_{\nu }\phi ^{J})-\ ^{E}\check{V}\ (\tau ,\phi ^{I})\right] .
\label{configlagr}
\end{equation}

In the cosmological approximation with $g_{\alpha \beta }(\tau ,t)$ and $%
\phi ^{I}(\tau ,t),$ the equations of motion corresponding to (\ref%
{cosmsoltwopact}) in the form (\ref{configlagr}) are%
\begin{eqnarray}
H^{2} &=&\frac{\kappa ^{2}}{3}[\frac{1}{2}\sigma ^{\diamond }+\ ^{E}\check{V}%
\ ],  \label{configcosmeq} \\
H^{\diamond } &=&-\frac{\kappa ^{2}}{2}(\sigma ^{\diamond })^{2},  \notag \\
\sigma ^{\diamond \diamond }+3H\sigma ^{\diamond }+\widehat{\sigma }^{I}\
\partial (\ ^{E}\check{V})/\partial \phi ^{I} &=&0,  \notag
\end{eqnarray}%
for $\sigma ^{\diamond }:=\sqrt{G_{IJ}(\phi ^{I})^{\diamond }(\phi
^{J})^{\diamond }}$ and $\widehat{\sigma }^{I}\ :=(\phi ^{I})^{\diamond
}/\sigma ^{\diamond }.$ This system of equations is sensitive on the type of
potential $\ ^{E}\check{V}$. In configuration two-scalar space, we get a
very simplified picture encoding certain values running on geometric
evolution flow parameter $\tau$.

\subsubsection{Slow--roll parameters determined by supersymmetric potential
with geometric flow parametric dependence}

Using cosmological equations (\ref{configcosmeq}) in 2-d configuration
space, we compute in standard form the slow--roll parameters%
\begin{equation*}
\epsilon =-\frac{H^{\diamond }}{H^{2}}=\frac{3(\sigma ^{\diamond })^{2}}{%
(\sigma ^{\diamond })^{2}+\ ^{E}\check{V}\ }\mbox{ and }\eta =\frac{%
M_{\sigma \sigma }}{\kappa ^{2}\ ^{E}\check{V}},
\end{equation*}%
for
\begin{equation*}
M_{\sigma \sigma }=\phi ^{K}\phi ^{J}[\partial ^{2}(\ ^{E}\check{V}%
)/\partial \phi ^{K}\partial \phi ^{l}-\Gamma _{KJ}^{I}\partial (\ ^{E}%
\check{V})/\partial \phi ^{I}]/(\sigma ^{\diamond })^{2}
\end{equation*}
and $\Gamma _{KJ}^{I}$ are the Christoffel symbols of the configuration
metric $G_{IJ}. $ A tedious computation similar to that presented in Sec. V.
A and Appendix B of \cite{odintsrec} in the approximation of small parameter
$\epsilon ,$ using our values $^{E}\check{V}$ and \ $\ell $ result in the
formulae%
\begin{eqnarray*}
\epsilon (\tau ) &=&\frac{3[(\phi ^{\diamond })^{2}+e^{-\sqrt{6}\kappa \phi
}\ell (\tau ,\wp )(\wp ^{\diamond })^{2}]}{(\phi ^{\diamond })^{2}+e^{-\sqrt{%
6}\kappa \phi }\ell (\tau ,\wp )(\wp ^{\diamond })^{2}-U(\tau ,\phi )-\ ^{E}%
\widetilde{V}\ (\tau ,\wp )e^{-2\sqrt{2/3}\kappa \phi }+\ell (\tau ,\wp
)e^{-2\sqrt{2/3}\kappa \phi }} \\
\eta (\tau ) &\simeq &\epsilon -\sigma ^{\diamond \diamond }/H\sigma
^{\diamond }+\mathcal{O}(\epsilon ^{2}),
\end{eqnarray*}%
\begin{eqnarray}
\mbox{ for } \sigma ^{\diamond \diamond } &=&-3H\sqrt{1+e^{-\sqrt{6}\kappa
\phi }\ell (\tau ,\wp )}-\frac{e^{-\sqrt{6}\kappa \phi }}{1+e^{-\sqrt{6}%
\kappa \phi }\ell (\tau ,\wp )}  \notag \\
&&\{\phi ^{\diamond }[\sqrt{6}\kappa \ ^{E}\widetilde{V}\ (\tau ,\wp )-\ell
(\tau ,\wp )+\frac{\partial U}{\partial \phi }]+\wp ^{\diamond }\frac{%
\partial }{\partial \wp }[\ell -\ ^{E}\widetilde{V}\ (\tau ,\wp )]\};  \notag
\\
\sigma ^{\diamond } &=&\sqrt{(\phi ^{\diamond })^{2}+e^{-\sqrt{6}\kappa \phi
}\ell (\tau ,\wp )(\wp ^{\diamond })^{2}}.  \label{aux4}
\end{eqnarray}%
As a result, the respective spectral index of primordial curvature
perturbations, $n_{s},$ and the scalar-to-tensor ratio, $r,$ are determined
to be
\begin{equation*}
n_{s}=1-6\epsilon (\tau )+2\eta (\tau )\mbox{  and }r=16\epsilon (\tau ).
\end{equation*}%
These potential slow-roll parameters are related to the Hubble slow-roll
parameters which can be computed following formulas \cite{liddle}, where it
was proven that such parameters coincide at first order in the slow--roll
approximation. The second Hubble slow-roll parameter is
\begin{equation*}
\eta _{H}(\tau )=-H^{\diamond \diamond }/2H^{\diamond }H.
\end{equation*}%
Using the cosmological equations (\ref{configcosmeq}), we obtain the formula
$H^{\diamond \diamond }=-\kappa \sigma ^{\diamond }\sigma ^{\diamond
\diamond }.$ For $\eta _{H},$ one employs the formula%
\begin{equation*}
\eta _{H}(\tau )=\frac{3}{\kappa ^{2}}\frac{3H(\sigma ^{\diamond })^{2}+\phi
^{\diamond }\ \partial \ ^{E}\widetilde{V}/\partial \phi +\wp ^{\diamond }\
\partial \ ^{E}\widetilde{V}/\partial \wp }{(\sigma ^{\diamond
})^{2}[(\sigma ^{\diamond })^{2}+2\ ^{E}\widetilde{V}]},
\end{equation*}%
where values (\ref{aux4}) allow us to express $\eta _{H}$ in terms of $%
U,\ell $ and$\ ^{E}\widetilde{V}.$ We omit the resulting cumbersome formulae
in this work.

It should be noted that the Hubble and potential slow--roll parameters and
spectral indices are related,
\begin{equation*}
n_{s}=1-4\epsilon (\tau )+2\eta _{H}(\tau )\mbox{ and }n_{H}=-\epsilon (\tau
)+\eta (\tau ).
\end{equation*}

We reduced the \ (super) geometric evolution dynamics and cosmological Ricci
soliton models to mimetic $F(R)$ gravity seen as a dynamical system with
certain constants running with respect to $\tau $. It is possible to find
the fixed points of such dynamical systems and study the stability of such
points as in Ref. \cite{odintsrec}. Such a study for special classes of
solutions with running of cosmological constant will be undertaken in future
studies correlated other specific reconstruction methods.

\section{Discussions and Conclusions}

\label{sec6} In this article we presented a new theoretical framework that
allows us to study applications of geometric flow theory in modified (super)
gravity and cosmological theories in a consistent and relatively clear
manner, making it easier, we hope, for physicists to follow. The approach is
based on geometric techniques formulated for nonholonomic variables which
allow to decouple in general ways physically important geometric evolution
and modified gravitational field equations. This paper develops for modified
gravity theories, MGTs, (in special for $R^{2}$ models and cosmology), our
former results on black holes and geometric flows of anisotropic
cosmological solutions in \cite{kounnas3,bakaslust} and can be considered as
extension of the works \cite{muen01,vvrfthbh}.

The Hamilton-Perelman theory of Ricci flows provides fundamental results on
topology of three dimensional spaces which can be related to the geometric
and topological structure of our Universe. The definition of Perelman's
functionals is of crucial importance in order to elaborate the proof of such
results in modern mathematics and suggests a number of new ideas on
analogous statistical thermodynamics for gravitational fields, on possible
connections to string theory \cite%
{friedan1,friedan2,friedan3,nitta,tsey,oliy,carfora} and applications in
geometric mechanics and modern commutative and noncommutative gravity \cite%
{vnhrf,vnrflnc,vrfijmpa,vacvis,vmedit}.

The main problem for further applications of the geometric flow theory in
modern gravity and quantum field theory is that existing rigorous
mathematical methods are based on specific techniques elaborated in
geometric analysis for flows of Riemannian metrics of Euclidean signature.
We can consider various formal modifications of Perelman functionals with
nonholonomic, commutative and noncommutative variables. Such values and
derived geometric evolution equations do not have a standard nonlinear
diffusion and entropy like interpretation for pseudo-Euclidean signatures
and generalized connections. We can investigate possible physical
implications of such generalized functionals and "relativistic evolution"
equations if certain classes of exact solutions can be constructed in
general forms. Positively, modified Perelman functionals and generalized
Hamilton Ricci flow equations for non--Riemannian geometries do have
physical importance in modern gravity and cosmology. This follows at least
from the fact that we can always consider non-relativistic limits with
corresponding 3+1 splitting of geometric models and gravitational theories.
In order to construct exact solutions in explicit form, we work with
nonholonomic 2+2 splitting but a second 3+1 fibration can be always
considered if it is necessary for computing certain physical important
values in non-relativistic limits.

In addition to the problem on elaborating models of relativistic geometric
flows with applications in modern gravity and cosmology, we emphasize that
it is of fundamental importance to study physically viable models of
supergeometric flows and their connections to string theory. This is a very
difficult task because one has not been elaborated generally accepted by
mathematicians concepts of supermanifolds and relevant directions for
supergeometric analysis, nonlinear functional analysis etc. Using certain
recent results on viable effective scalar potentials in modified models of $%
R^{2}$ gravity and cosmology, we can formulate supersymmetic extensions of
Perelman functionals and study cosmological Ricci soliton configurations
related to MGTs. This work can be considered as a first step to $SU(1,1+k)$
supersymmetric modifications of $R^{2}$ gravity, geometric flow theory and
acceleration cosmology. We support our approach by constructing in explicit
form various types of inhomogeneous and locally anisotropic cosmological
solutions. Corresponding spacetime models encode nontrivial off--diagonal
vacuum configurations, running constants and polarizations of physical
constants depending on geometric flow parameter $\tau .$ The main conclusion
is that geometric evolution flow effects may result in running of physical
constants at classical level and as such, in general, off--diagonal effects
may be important in modern astroparticle physics relating to understanding
dark energy and dark matter.

In order to compare our results on geometric flows and modified Ricci
solitons to those developed in other MGTs, we consider effective mimetic
F(R) gravity models accompanied by a potential with supersymmetric
corrections $\ ^{E}V$ and a Lagrange multiplier. It is not surprising that
there is a strong analogy between nonholonomic geometric evolution models,
Ricci solitons, and dynamical (generalized mechanical) models with Lagrange
multipliers. Both methods complement each other, being important in
constructing exact solutions and in order to study possible physical and
observational implications. The presence of generalized sources determined
by modified potentials, generating functions, nonholonomic constraints and
Lagrange multiplies offer many possible ways and various classes of exact
solutions for realization of geometric flow and cosmological evolution. Our
particular interest is to understand if certain supersymmetric modifications
determining modified geometric flows and new classes of cosmological Ricci
solitons can produce cosmological spacetimes that are in concordance with
observations. Given a generalized cosmological geometric flow/ Ricci soliton
configuration, we can use a reconstruction procedure elaborated in the text
to find mimetic potentials and Lagrange multiplier which can be modelled
equivalently by a viable F(R) cosmology, in particular by supersymmetric
modifications of $R^{2}$ theory. The reconstructing method elaborated in
\cite{5,odints2,bamba,odintsrec}\ and developed for nonholonomic
configurations in \cite{vcosmsol1,vcosmsol2,vcosmsol3,vcosmsol4,vcosmsol4a,vcosmsol5}
can be applied for arbitrary MGTs and for geometric flows of generic
off--diagonal metrics and generalized connections. This way we can study
physical viability of various flow evolution/ modified gravity theory via an
unified description of late--time acceleration cosmology and existing
observational data.

Furthermore, in the framework of nonholonomic geometric flow theory it is
possible to study issues on stable and unstable (towards liner
configurations) cosmological configurations. In principle, there is a strong
dependence on MGT details, off--diagonal geometric evolution / interactions
scenarios etc. For any classes of such generalized solution, we can consider
at the end certain cosmological limits with running constant configurations
and study further if there are stable de Sitter vacua, for instance, as
final attractors of the trajectories of corresponding nonholonomic dynamical
systems and speculate on inflation and acceleration scenarios. Interestingly
enough, we can analyze if certain supergravity potentials $\ ^{E}V$ via
geometric flows and off--diagonal interactions lead to cosmological
spacetimes with appealing cosmological properties. Our methods were applied
in adapted form both for an effective Jordan frame $R^{2}$ theory, in order
to calculate the spectral index of primordial perturbations, and in the
Einstein frame for running of the spectral index. An effective two scalar
field formalism was developed in the slow--roll limit.

Finally, we note that models of geometric flow evolution and mimetic MGTs
offer many possibilities for realizing cosmological scenarios with various
classes of effective sources and different types of generating functions.
Computation of Perelman entropy may allow us to decide what model of
geometric flows and cosmological Ricci solitons are physically viable. Such
computations were provided in \cite{vvrfthbh,muen01} for certain locally
anisotropic cosmological models and black hole and solitonic configurations
in $R^{2}$ gravity. We hope to address such issues for geometric flows
related to superstring theory and classify respective cosmological Ricci
solitons in our future publications.

\section*{Acknowledgments}

S. V. research was partially supported by IDEI, PN-II-ID-PCE-2011-3-0256,
DAAD, and QGR -- Topanga, Los Angeles, California. The
manuscript contains results presented at GR21 NY and a seminar at CSU Fresno (host professor D. Singleton), and provides in parallel to \cite{muen01} further developments and discussions of black hole solutions
obtained in \cite{kounnas3,kehagias} (published versions of arXiv: 1409.7076
and 1502.04192; such papers are cited many times in this article, with
certain dubbing of necessary formulas and text).

\end{document}